\documentclass[%
 reprint,
 superscriptaddress,
 showpacs,
 preprintnumbers,
 nofootinbib,
 amsmath,amssymb,
 aps,
 prc,
]{revtex4-1}

\usepackage{graphicx}
\usepackage{dcolumn}
\usepackage{bm}
\usepackage{color}
\usepackage{multirow}

\usepackage{mathrsfs} 
\usepackage{amssymb}  

\begin{document}

\preprint{\large{\textit{Submitted to Physical Review C}}}

\title{Measurement of the neutron $\bm{\beta}$-asymmetry parameter
$\bm{A_0}$ with ultracold neutrons}

\author{B.~Plaster}
\affiliation{Department of Physics and Astronomy, University of Kentucky,
Lexington, Kentucky 40506, USA}
\affiliation{W.\ K.\ Kellogg Radiation Laboratory,
California Institute of Technology, Pasadena, California 91125, USA}
\author{R.~Rios}
\affiliation{Los Alamos National Laboratory, Los Alamos, New Mexico 87545, USA}
\affiliation{Department of Physics, Idaho State University, Pocatello, Idaho
83209, USA}
\author{H.~O.~Back}
\affiliation{Department of Physics, North Carolina State University,
Raleigh, North Carolina 27695, USA}
\affiliation{Triangle Universities Nuclear Laboratory, Durham,
North Carolina 27708, USA}
\author{T.~J.~Bowles}
\affiliation{Los Alamos National Laboratory, Los Alamos, New Mexico 87545, USA}
\author{L.~J.~Broussard}
\affiliation{Triangle Universities Nuclear Laboratory, Durham,
North Carolina 27708, USA}
\affiliation{Department of Physics, Duke University, Durham, North Carolina
27708, USA}
\author{R.~Carr}
\affiliation{W.\ K.\ Kellogg Radiation Laboratory,
California Institute of Technology, Pasadena, California 91125, USA}
\author{S.~Clayton}
\affiliation{Los Alamos National Laboratory, Los Alamos, New Mexico 87545, USA}
\author{S.~Currie}
\affiliation{Los Alamos National Laboratory, Los Alamos, New Mexico 87545, USA}
\author{B.~W.~Filippone}
\affiliation{W.\ K.\ Kellogg Radiation Laboratory,
California Institute of Technology, Pasadena, California 91125, USA}
\author{A.~Garc\'{i}a}
\affiliation{Department of Physics, University of Washington, Seattle,
Washington 98195, USA}
\author{P.~Geltenbort}
\affiliation{Institut Laue-Langevin, 38042 Grenoble Cedex 9, France}
\author{K.~P.~Hickerson}
\affiliation{W.\ K.\ Kellogg Radiation Laboratory,
California Institute of Technology, Pasadena, California 91125, USA}
\author{J.~Hoagland}
\affiliation{Department of Physics, North Carolina State University,
Raleigh, North Carolina 27695, USA}
\author{G.~E.~Hogan}
\affiliation{Los Alamos National Laboratory, Los Alamos, New Mexico 87545, USA}
\author{B.~Hona}
\affiliation{Department of Physics and Astronomy, University of Kentucky,
Lexington, Kentucky 40506, USA}
\author{A.~T.~Holley}
\affiliation{Department of Physics, North Carolina State University,
Raleigh, North Carolina 27695, USA}
\author{T.~M.~Ito}
\affiliation{W.\ K.\ Kellogg Radiation Laboratory,
California Institute of Technology, Pasadena, California 91125, USA}
\affiliation{Los Alamos National Laboratory, Los Alamos, New Mexico 87545, USA}
\author{C.-Y.~Liu}
\affiliation{Department of Physics, Indiana University, Bloomington, Indiana
47408, USA}
\author{J.~Liu}
\affiliation{W.\ K.\ Kellogg Radiation Laboratory,
California Institute of Technology, Pasadena, California 91125, USA}
\affiliation{Department of Physics, Shanghai Jiao Tong University,
Shanghai, 200240, China}
\author{M.~Makela}
\affiliation{Los Alamos National Laboratory, Los Alamos, New Mexico 87545, USA}
\author{R.~R.~Mammei}
\affiliation{Department of Physics, Virginia Tech, Blacksburg, Virginia
24061, USA}
\author{J.~W.~Martin}
\affiliation{W.\ K.\ Kellogg Radiation Laboratory,
California Institute of Technology, Pasadena, California 91125, USA}
\affiliation{Department of Physics, University of Winnipeg, Winnipeg,
MB R3B 2E9, Canada}
\author{D.~Melconian}
\affiliation{Cyclotron Institute, Texas A\&M University, College Station,
Texas 77843, USA}
\author{M.~P.~Mendenhall}
\affiliation{W.\ K.\ Kellogg Radiation Laboratory,
California Institute of Technology, Pasadena, California 91125, USA}
\author{C.~L.~Morris}
\affiliation{Los Alamos National Laboratory, Los Alamos, New Mexico 87545, USA}
\author{R.~Mortensen}
\affiliation{Los Alamos National Laboratory, Los Alamos, New Mexico 87545, USA}
\author{R.~W.~Pattie, Jr.}
\affiliation{Department of Physics, North Carolina State University,
Raleigh, North Carolina 27695, USA}
\affiliation{Triangle Universities Nuclear Laboratory, Durham,
North Carolina 27708, USA}
\author{A.~P\'{e}rez~Galv\'{a}n}
\affiliation{W.\ K.\ Kellogg Radiation Laboratory,
California Institute of Technology, Pasadena, California 91125, USA}
\author{M.~L.~Pitt}
\affiliation{Department of Physics, Virginia Tech, Blacksburg, Virginia
24061, USA}
\author{J.~C.~Ramsey}
\affiliation{Los Alamos National Laboratory, Los Alamos, New Mexico 87545, USA}
\author{R.~Russell}
\affiliation{W.\ K.\ Kellogg Radiation Laboratory,
California Institute of Technology, Pasadena, California 91125, USA}
\author{A.~Saunders}
\affiliation{Los Alamos National Laboratory, Los Alamos, New Mexico 87545, USA}
\author{R.~Schmid}
\affiliation{W.\ K.\ Kellogg Radiation Laboratory,
California Institute of Technology, Pasadena, California 91125, USA}
\author{S.~J.~Seestrom}
\affiliation{Los Alamos National Laboratory, Los Alamos, New Mexico 87545, USA}
\author{S.~Sjue}
\affiliation{Department of Physics, University of Washington, Seattle,
Washington 98195, USA}
\author{W.~E.~Sondheim}
\affiliation{Los Alamos National Laboratory, Los Alamos, New Mexico 87545, USA}
\author{E.~Tatar}
\affiliation{Department of Physics, Idaho State University, Pocatello, Idaho
83209, USA}
\author{B.~Tipton}
\affiliation{W.\ K.\ Kellogg Radiation Laboratory,
California Institute of Technology, Pasadena, California 91125, USA}
\author{R.~B.~Vogelaar}
\affiliation{Department of Physics, Virginia Tech, Blacksburg, Virginia
24061, USA}
\author{B.~VornDick}
\affiliation{Department of Physics, North Carolina State University,
Raleigh, North Carolina 27695, USA}
\author{C.~Wrede}
\affiliation{Department of Physics, University of Washington, Seattle,
Washington 98195, USA}
\author{Y.~P.~Xu}
\affiliation{Department of Physics, North Carolina State University,
Raleigh, North Carolina 27695, USA}
\author{H.~Yan}
\affiliation{Department of Physics and Astronomy, University of Kentucky,
Lexington, Kentucky 40506, USA}
\author{A.~R.~Young}
\affiliation{Department of Physics, North Carolina State University,
Raleigh, North Carolina 27695, USA}
\affiliation{Triangle Universities Nuclear Laboratory, Durham,
North Carolina 27708, USA}
\author{J.~Yuan}
\affiliation{W.\ K.\ Kellogg Radiation Laboratory,
California Institute of Technology, Pasadena, California 91125, USA}

\collaboration{UCNA Collaboration}

\date{\today}

\begin{abstract}
We present a detailed report of a measurement of the neutron
$\beta$-asymmetry parameter $A_0$, the parity-violating angular
correlation between the neutron spin and the decay electron momentum,
performed with polarized ultracold neutrons (UCN).  UCN were extracted
from a pulsed spallation solid deuterium source and polarized via
transport through a 7-T magnetic field.  The polarized UCN were then
transported through an adiabatic-fast-passage spin-flipper field
region, prior to storage in a cylindrical decay volume situated within
a 1-T $2 \times 2\pi$ solenoidal spectrometer.  The asymmetry was
extracted from measurements of the decay electrons in multiwire
proportional chamber and plastic scintillator detector packages
located on both ends of the spectrometer.  From an analysis of data
acquired during runs in 2008 and 2009, we report $A_0 = -0.11966 \pm
0.00089~_{-0.00140} ^{+0.00123}$, from which we extract a value for
the ratio of the weak axial-vector and vector coupling constants of
the nucleon, $\lambda = g_A/g_V = -1.27590 \pm
0.00239~_{-0.00377}^{+0.00331}$.  Complete details of the analysis are
presented.
\end{abstract}


\maketitle


\section{Introduction}
\label{sec:intro}


Precise measurements of neutron $\beta$-decay observables determine
fundamental parameters of the weak interaction and contribute to tests
of the Standard Model
\cite{nico05,severijns06,abele08,nico09,dubbers11,severijns11}.
Because the momentum transfer in neutron $\beta$-decay ($n \rightarrow
p + e^{-} + \overline{\nu}_e + 781.5$ keV) is small compared to the
$W^-$ mass, the decay can be modeled as a four-fermion contact
interaction with an amplitude under the Standard Model given by
\begin{equation}
\mathcal{M} = \frac{G_F V_{ud}}{\sqrt{2}} \langle p | J^\mu | n \rangle
L_\mu ,
\end{equation}
where $G_F$ is the Fermi weak coupling constant, $V_{ud}$ is the
weak-quark-mixing CKM matrix element, and $L_\mu = \overline{u}_e
\gamma_\mu (1 - \gamma_5) u_{\overline{\nu}}$ is the leptonic weak
vector and axial-vector current.  In its most general form, the
hadronic weak vector and axial-vector ($V-A$) current includes six form
factors \cite{goldberger58,weinberg58},
\begin{widetext}
\begin{equation}
\langle p | J^\mu | n \rangle = \overline{u}_p
\left[ g_V(q^2)\gamma^\mu - i \frac{g_{WM}(q^2)}{2M}\sigma^{\mu\nu}q_\nu +
\frac{g_S(q^2)}{2M}q^\mu +
g_A(q^2)\gamma^\mu\gamma^5 - i\frac{g_T(q^2)}{2M}\sigma^{\mu\nu}
\gamma_5 q_\nu + \frac{g_P(q^2)}{M}\gamma_5 q^\mu  \right]u_n ,
\end{equation}
\end{widetext}
where $q$ is the four-momentum transfer; $M$ is the nucleon mass; and
$g_V(q^2)$, $g_{WM}(q^2)$, $g_S(q^2)$, $g_A(q^2)$, $g_T(q^2)$, and
$g_P(q^2)$ are the vector, weak magnetism, induced scalar, axial
vector, induced tensor, and induced pseudoscalar form factors,
respectively.  In the limit of $q^2 \rightarrow 0$, the hadronic weak
current is dominated by the weak vector and axial vector coupling
constants of the nucleon, defined to be the values of the vector and
axial vector form factors at $q^2 = 0$, $g_A \equiv g_A(q^2 = 0)$ and
$g_V \equiv g_V(q^2 = 0)$.  Under the Conserved Vector Current (CVC)
hypothesis of the Standard Model and the assumption of
isospin symmetry, the vector coupling constant is $g_V = 1$
(independent of the nuclear medium).  Isospin-symmetry-breaking
effects to the value of $g_V$ in neutron $\beta$-decay have been
calculated in chiral perturbation theory, with the correction to $g_V$
found to be at a negligible $-4 \times 10^{-5}$ level \cite{kaiser01}.
Also per the CVC hypothesis, the weak magnetism coupling constant,
$g_{WM} \equiv g_{WM}(q^2=0)$, which appears at recoil order
in the vector current, is related to the proton and
neutron anomalous magnetic moments by
$g_{WM} = \kappa_p - \kappa_n$.

In contrast to the vector current, the axial-vector current is
renormalized by the strong interaction such that the value of $g_A$
must be determined experimentally and also by lattice Quantum
Chromodynamics (QCD) calculations.  Any contribution from the induced
pseudoscalar coupling constant, $g_P$, to neutron $\beta$-decay
observables is expected to be negligibly small, with the contribution
of $g_P$ to the energy spectrum calculated to be of order $m_e^2 / M
E_e \sim 10^{-4}$ \cite{holstein74}.

The two remaining terms, the induced scalar, $g_S(q^2)$, in the vector
current, and the induced tensor, $g_T(q^2)$, in the axial-vector
current, are termed second-class currents due to their
transformation properties under $G$-parity.  Under
the requirement of $G$-parity symmetry, both $g_S(q^2) = g_T(q^2) =
0$.  However, $G$-parity symmetry is violated within the Standard
Model due to differences in the $u$ and $d$ quarks' charges and masses
(i.e., isospin symmetry breaking effects).  An estimate for $g_T$
including SU(3) breaking effects suggested $|g_T|$ in neutron
$\beta$-decay to be $\alt 0.03$ \cite{donoghue82}, and an evaluation
of $g_T/g_A$ using QCD sum rules found $g_T/g_A = -0.0152(53)$
\cite{shiomi96}.  Finally, recent lattice QCD studies of SU(3)
breaking in semi-leptonic decays find small, $\mathcal{O}(0.1)$,
values for both $g_S(q^2)$ and $g_T(q^2)$ in neutron $\beta$-decay,
but the results are statistically limited and consistent with zero at
1--2 standard deviations \cite{sasaki09}.  However, despite these
hints for non-zero values of these second-class currents, their
contributions to neutron $\beta$-decay observables are again expected
to be negligibly small, as they also appear at order
$m_e^2 / M E_e \sim 10^{-4}$ in the energy spectrum
\cite{holstein74}.

Therefore, under the assumption that any such
contributions from $g_P$, $g_S$, and $g_T$ are negligibly small
relative to the current level of experimental precision, it is clear
that a description of neutron $\beta$-decay under the CVC hypothesis
of the Standard Model requires the specification of only two
parameters, $V_{ud}$ and $g_A$, given the high precision results for
$G_F$ achieved in muon decay \cite{webber11}.  Both $V_{ud}$ and $g_A$
can be accessed via measurements of two different types of neutron
$\beta$-decay observables: the lifetime, and angular correlation
coefficients in polarized and unpolarized $\beta$-decay.  The first of
these, the lifetime, as calculated from the amplitude and integration
over the allowed phase space, is of the form \cite{czarnecki04}
\begin{equation}
\frac{1}{\tau_n} = \frac{G_F^2 m_e^5}{2\pi^3} V_{ud}^2
\left(1 + 3\lambda^2\right) f \left(1 + \text{RC}\right),
\label{eq:lifetime-Vud}
\end{equation}
where $m_e$ is the electron mass and the parameter $\lambda$ is
defined to be the ratio of the axial vector and vector coupling
constants, $\lambda \equiv g_A / g_V$.  The numerical value for the
phase space factor of $f = 1.6887$ \cite{czarnecki04}
includes corrections for the Fermi function, the finite nucleon mass,
the finite nucleon radius, and the effect of recoil on the Fermi
function.  The factor $(1 + \text{RC})$ denotes the total effect of
all electroweak radiative corrections, including the
$\mathcal{O}(\alpha)$ outer (long-distance loop and bremsstrahlung
effects) and inner (short distance, including axial-vector-current,
loop effects) radiative corrections; an $\mathcal{O}(\alpha^2)$
correction resulting from factorization of the Fermi function; and
$\mathcal{O}(\alpha^2)$ leading-log and next-to-leading-log
corrections (for lepton and quark loop insertions in the photon
propagator) \cite{czarnecki04}.  The total electroweak radiative
correction has been calculated to be $(1 + \text{RC}) = 1.0390 \pm
0.0004$ \cite{marciano06} where the $\pm 0.0004$ uncertainty was
reduced by a factor of two (from its previous value of $\pm 0.0008$
\cite{czarnecki04}) after the development of a new method for
calculating hadronic effects in the matching of long- and
short-distance contributions to axial-vector current loop effects
(primarily from the $\gamma W$ box diagram).

The second type of observable, angular correlation
coefficients, parametrize the angular correlations between the momenta
of the decay products and the spin of the initial-state neutron.
In general, the directional
distribution of the electron and antineutrino momenta and the electron
energy in polarized $\beta$-decay is of the form \cite{jackson57}
\begin{eqnarray}
&&\displaystyle{\frac{d\Gamma}{dE_e d\Omega_e d\Omega_\nu}}
\propto p_e E_e (E_0 - E_e)^2
\nonumber \\
&&~~\times \left[1 + b\frac{m_e}{E_e} +
a\frac{\vec{p}_e \cdot \vec{p}_\nu}{E_e E_\nu} \right. \nonumber \\
&&~~~~~~~\left. +~\langle \vec{\sigma}_n \rangle \cdot
\left( A\frac{\vec{p}_e}{E_e} +
B\frac{\vec{p}_\nu}{E_\nu} +
D\frac{\vec{p}_e \times \vec{p}_\nu}{E_e E_\nu}\right) \right],
\label{eq:W_phase_space_distribution}
\end{eqnarray}
where $E_e$ ($E_\nu$) and $\vec{p}_e$ ($\vec{p}_\nu$) denote,
respectively, the electron's (antineutrino's) total energy and
momentum; $E_0$ ($= 781.5$ keV + $m_e$) is the
electron endpoint energy; and $\langle \vec{\sigma}_n \rangle$ is the
neutron polarization.  The angular correlation coefficients $a$
($e$-$\overline{\nu}_e$-asymmetry), $A$ ($\beta$-asymmetry), and $B$
($\overline{\nu}_e$-asymmetry) are, to lowest order, functions only of
$\lambda$ where, under a $\lambda < 0$ sign convention,
\begin{eqnarray}
a_0 = \frac{1 - \lambda^2}{1 + 3\lambda^2},~~~
A_0 = -2\frac{\lambda(\lambda + 1)}{1 + 3\lambda^2},~~~
B_0 = 2\frac{\lambda(\lambda - 1)}{1 + 3\lambda^2}. \nonumber \\
\label{eq:correlation_coefficients_lambda}
\end{eqnarray}

The contributions of terms in Eq.\
(\ref{eq:W_phase_space_distribution}) proportional to the Fierz
interference term $b$ and the time-reversal-odd
triple-correlation-coefficient $D$ are at recoil order for Standard
Model interactions \cite{holstein74,bhattacharya12,callan67,ando09},
and are negligible at the current level of experimental precision.
Note that to our knowledge, there are no published direct measurements
of $b$ in neutron $\beta$-decay.

As already noted, at lowest order $a_0$, $A_0$, and
$B_0$ are functions only of $\lambda$.  However,
recoil-order corrections, including the effects of
weak magnetism and $g_V$-$g_A$ interference, introduce energy-dependent
corrections to the asymmetry, and are of
$\mathcal{O}$(1\%) for $a$ and $A$
\cite{holstein74,wilkinson82,gardner01} and $\mathcal{O}$(0.1\%) for
$B$ \cite{gluck98}.  For $A$, the recoil-order corrections are of the
explicit functional form \cite{holstein74,wilkinson82,gardner01}
\begin{equation}
A = A_0 + A_1 \frac{\epsilon}{Rx} + A_2 R + A_3 Rx,
\label{eq:recoil_order_parametrization}
\end{equation}
where $R = E_0/M$, $\epsilon = (m_e/M)^2$, $x = E_e/E_0$, and the
$A_i$ ($i = 1,2,3$) coefficients are functions only of $\lambda$
and $g_{WM}$ (assuming $g_S = g_T = 0$, and negligible
contributions from $g_P$).
Under these assumptions, $A_1 = -0.3054$,
$A_2 = 0.7454$, and $A_3 = -3.0395$.
Note that the $q^2$ dependence of the form
factors does not appear until next-to-leading recoil order
\cite{gardner01}.

In addition to the above recoil-order corrections, there is a small
energy-dependent radiative correction (for virtual and bremsstrahlung
processes) to polarized asymmetries, resulting in a
$\mathcal{O}(0.1\%)$ correction to $A$ \cite{shann71,gluck92}.  After
application of these recoil-order and radiative corrections, a value
for $\lambda$ can be extracted from $a_0$, $A_0$, and $B_0$.  Note,
however, that for a given (relative) statistical precision, the
sensitivity of $A_0$ to $\lambda$ is slightly higher than that of
$a_0$, and a factor of $\sim 8$ greater than that of $B_0$, where at
leading order the relative uncertainties compare as
\begin{equation}
\frac{\delta |\lambda|}{|\lambda|} \approx
0.24\frac{\delta |A_0|}{|A_0|} \approx
0.27\frac{\delta |a_0|}{|a_0|} \approx 2.0\frac{\delta |B_0|}{|B_0|}.
\end{equation}

Thus, measurements of the angular correlation coefficients determine a
value for $\lambda$ (or $g_A$, assuming $g_V = 1$ under the CVC
hypothesis), a fundamental parameter in the nucleon weak current.  A
precise value for $g_A$ is also important in many other contexts.  In
hadronic physics studies of the spin structure of the nucleon
\cite{filippone02,bass05}, the Bjorken sum rule relates the difference
in the first moments of the proton and neutron spin-dependent $g_1$
structure functions (i.e., isovector channel), as probed in polarized
deep inelastic electron scattering, to $g_A$.  In QCD, the assumption
of a partially conserved axial-vector current (PCAC), valid in the
limit of a massless pion (identified as the Goldstone boson of the
spontaneously broken chiral symmetry), leads to the Goldberger-Treiman
relation \cite{goldberger58}, relating the value of $g_A$ to the pion
decay constant $f_\pi$, the weak pion-nucleon-nucleon coupling
constant $g_{\pi NN}$, and the nucleon mass.  The value of $g_A$ is
also important in astrophysical processes, including calculations of
solar fusion cross sections and rates, in particular, of the $pp$
fusion reaction, impacting the solar neutrino flux for this process
\cite{adelberger11}.  High-precision experimental results for $g_A$
also serve as an important benchmark for theoretical calculations of
$g_A$, both in fundamental lattice QCD calculations \cite{yamazaki08}
and in relativistic constituent quark model calculations
\cite{choi10}.  A precise value for $g_A$ is also
important as a phenomenological input parameter (together with other
low energy constants, such as the pion decay constant
$f_\pi$, the
nucleon mass, etc.)  to effective field theory calculations involving
the axial vector current \cite{gockeler05}.

Although not a fundamental weak interaction parameter by itself, a
precise value for the lifetime is important for Big Bang
Nucleosynthesis calculations, impacting the neutron-to-proton ratio
and hence the primordial $^4$He abundance at the time of freeze-out,
when the weak reaction rate became less than the Hubble expansion
rate \cite{mathews05}.  The value of the lifetime is also important
for the interpretation of data from neutrino oscillation experiments
employing antineutrinos from reactors \cite{mention11}, which
typically search for the reaction $\overline{\nu}_e + p \rightarrow
e^+ + n$ in detectors.  The cross section for this reaction is
inversely proportional to the neutron lifetime; therefore, an accurate
and precise experimental value for the lifetime is needed for an
interpretation of measured detector antineutrino reaction rates in
terms of the underlying neutrino oscillation physics.

Measurements of the lifetime and a value for $\lambda$ from
measurements of angular correlation coefficients permit the extraction
of a value for $V_{ud}$ solely from neutron $\beta$-decay observables
according to Eq.~(\ref{eq:lifetime-Vud}).  Although a value for
$V_{ud}$ from neutron $\beta$-decay \cite{pdg} is not yet competitive
with the definitive value deduced from measurements of $ft$ values in
superallowed $0^+ \rightarrow 0^+$ nuclear $\beta$-decay
\cite{towner10}, the appeal of such an extraction is that it does not
require corrections for isospin-symmetry-breaking and
nuclear-structure effects.  Ultimately, when the
precision on a neutron-based value for $V_{ud}$ approaches the
precision of the $0^+ \rightarrow 0^+$ result, the two values must
agree in the absence of new physics.  However, given that the
neutron-based value for $V_{ud}$ is not yet competitive, one can treat
the $0^+ \rightarrow 0^+$ value for $V_{ud}$ as a fixed input
parameter, and instead perform a robust test of the consistency of the
various measured neutron $\beta$-decay observables under the Standard
Model.  In particular, results for $g_A$ extracted from correlation
coefficient measurements can then be directly compared with results
from measurements of the neutron lifetime $\tau_n$.

Finally, measurements of the angular correlation coefficients
themselves are sensitive to beyond-the-Standard-Model physics, such as
scalar and tensor interactions
\cite{gudkov06,konrad10,bhattacharya12}.  With the projected
improvements to the experimental precision in future years, neutron
$\beta$-decay measurements will be sensitive to any such sources of
new physics at energy scales rivaling those probed directly at the
Large Hadron Collider \cite{bhattacharya12}.

\begin{table*}
\caption{Summary of published measurements of the neutron
$\beta$-asymmetry parameter $A_0$.  The error on the average
has been increased by a factor of $\sqrt{\chi^2/\nu} = 2.47$.}
\begin{ruledtabular}
\begin{tabular}{lllccc}
Experiment& Years Published& Type& Polarization& $A_0$ Result& Notes \\ \hline
PERKEO \cite{bopp86}& 1986& cold neutron beam& $0.974 \pm 0.005$&
  $-0.1146 \pm 0.0019$& \footnotemark[1] \\
PNPI \cite{erozolimskii91,yerozolimsky97}& 1991, 1997& cold neutron beam&
  $0.770 \pm 0.007$& $-0.1135 \pm 0.0014$&
  \footnotemark[2] \\
ILL-TPC \cite{schreckenbach95,liaud97}& 1995, 1997& cold neutron beam&
  $0.981 \pm 0.003$& $-0.1160 \pm 0.0015$&
  \footnotemark[3] \\
PERKEO II \cite{abele97,abele02}& 1997, 2002& cold neutron beam&
  $0.989 \pm 0.003$& $-0.1189 \pm 0.0007$&
  \footnotemark[4] \\
UCNA \cite{pattie09,liu10}, this work& 2009, 2010& stored ultracold neutrons&
  $1.0^{+0}_{-0.0052}$& $-0.11966 \pm 0.00089 ^{+0.00123}_{-0.00140}$&
  \footnotemark[5] \\
PERKEO II \cite{mund12}& 2012& cold neutron beam&
  $0.997 \pm 0.001$& $-0.11996 \pm 0.00058$& \footnotemark[6] \\ \hline
\multicolumn{6}{c}{Current Average Value: $A_0 = -0.11846 \pm 0.00104$
  ($\chi^2/\nu = 24.35/4$)} \\
\end{tabular}
\end{ruledtabular}
\begin{minipage}{\textwidth}
\footnotetext[1]{Included a $\sim 10$\% correction to the asymmetry for
                 magnetic mirror effects.}
\footnotetext[2]{The result reported in \cite{yerozolimsky97} superseded
                 that reported in \cite{erozolimskii91} of
                 $A_0 = -0.1116(14)$, on the basis of a revised value for the
                 polarization.}
\footnotetext[3]{The final result reported in \cite{liaud97} was identical to
                 a first result reported in \cite{schreckenbach95}.}
\footnotetext[4]{The final result of $A_0 = -0.1189(7)$ was the
                 combined result of $-0.1189(12)$ reported in \cite{abele97}
                 and $-0.1189(8)$ reported in \cite{abele02}.}
\footnotetext[5]{The result of $A_0 = -0.1138(46)(21)$ reported in
                 \cite{pattie09} was from a proof-of-principle
                 measurement and was not included in the result reported
                 in \cite{liu10}.}
\footnotetext[6]{Accounting for correlated systematic errors in
                 \cite{abele02,mund12}, the combined
                 PERKEO II result is $A_0 = -0.11951 \pm 0.00050$.}
\end{minipage}
\label{tab:summary_measurements}
\end{table*}

The remainder of this article is organized as follows.  In Section
\ref{sec:status_measurements}, we summarize the current status of
measurements of $A_0$.
We then outline the
experimental motivation for a measurement of $A_0$ with ultracold
neutrons in Section \ref{sec:ucna_experiment}, and then present a
detailed description of the UCNA (``Ultracold Neutron Asymmetry'')
Experiment \cite{pattie09,liu10} at the Los Alamos National
Laboratory.  Our measurement procedures and experimental geometrical
configurations are reported in Section \ref{sec:measurements}.
Results from our calibration and analysis procedures are discussed in
Section \ref{sec:analysis}.  Details of our procedure for the
extraction of asymmetries are presented in Section
\ref{sec:asymmetry}, and the corrections to measured asymmetries for
various systematic effects are discussed in Section
\ref{sec:corrections}.  Systematic uncertainties are summarized in
Section \ref{sec:uncertainties}, and our final results for $A_0$ are
then reported in Section \ref{sec:summary_final_results}.  We then
conclude with a brief summary of the physics impact of our work in
Section \ref{sec:conclusions}.  The data presented here were obtained
during data-taking runs in 2008--2009 and published rapidly in 2010
\cite{liu10}; in this article we provide a more detailed account of
the experiment and analysis procedures.

\section{Status of Measurements of $\bm{A_0}$}
\label{sec:status_measurements}

\begin{figure}
\includegraphics[scale=0.45,clip=]{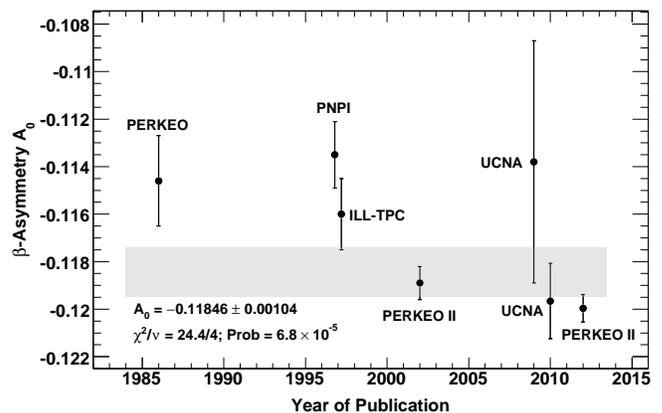}
\caption{Results from measurements of the $\beta$-asymmetry parameter
$A_0$ \cite{bopp86, erozolimskii91, yerozolimsky97, schreckenbach95,
liaud97, abele97, abele02, pattie09, liu10}.  The band ($\pm 1
\sigma$) indicates the current world average value of
$A_0 = -0.11846 \pm 0.00104$.}
\label{fig:status_beta_asymmetry}
\end{figure}

The current status of published results \cite{bopp86, erozolimskii91,
yerozolimsky97, schreckenbach95, liaud97, abele97, abele02, pattie09,
liu10,mund12} for the neutron $\beta$-asymmetry parameter $A_0$ is
summarized in Table \ref{tab:summary_measurements} and shown in Fig.\
\ref{fig:status_beta_asymmetry}.  Other than the UCNA Experiment, all
of the experiments have been performed with beams of polarized cold
neutrons, with reported values for the polarization ranging from
$0.770 \pm 0.007$ \cite{erozolimskii91,yerozolimsky97} to $0.997 \pm
0.001$ \cite{mund12}.  Magnetic solenoidal spectrometers providing $2
\times 2\pi$ solid angle acceptance for detection of the decay
electrons were employed in the PERKEO \cite{bopp86} and PERKEO II
\cite{abele97,abele02,mund12} experiments at the Institut
Laue-Langevin (ILL).  In contrast, the solid angle was defined by the
geometric acceptance in an experiment at the Petersburg Nuclear
Physics Institute (PNPI) \cite{erozolimskii91,yerozolimsky97} in which
the decay electrons and protons were detected in coincidence in
detectors surrounding the beam decay region, and in an experiment at
the ILL \cite{schreckenbach95,liaud97} which utilized a time
projection chamber for reconstruction of the electron track.

The current world average value for $A_0 = -0.11846 \pm 0.00104$
includes the most recent PERKEO II result\footnote{Note that in
computing the world average, we employed the combined PERKEO II result
of $-0.11951 \pm 0.00050$ reported in \cite{mund12} which accounted
for correlations of systematic errors in the two separately published
PERKEO II results \cite{abele02,mund12}.}~\cite{mund12}, but excludes
the UCNA proof-of-principle result \cite{pattie09}.  Note that the
current error bar of $\pm 0.00104$ includes the Particle Data Group's
$\sqrt{\chi^2/\nu}$ scaling \cite{pdg}.  The need for this expanded
error bar suggests an incomplete assessment of the systematic errors
in one or more of the cold-neutron-based experiments.

\section{UCNA Experiment}
\label{sec:ucna_experiment}

\subsection{Overview of Experiment}
\label{sec:experiment_overview}

The UCNA Experiment, installed in Area B of the Los
Alamos Neutron Science Center (LANSCE) at the Los Alamos National
Laboratory (LANL), was designed to perform the first-ever measurement
of the neutron $\beta$-asymmetry parameter with ultracold neutrons
(UCN), and to-date is the only experimental measurement of any neutron
$\beta$-decay angular correlation coefficient performed with ultracold
neutrons (UCN).  UCN are defined to be neutrons with kinetic energies
sufficiently low ($\alt 335$ neV, corresponding to speeds $\alt 8$ m
s$^{-1}$) such that they undergo total external reflection at any
angle of incidence from an effective potential barrier (a volume
average of Fermi potentials $V_{\text{Fermi}}$) at the surfaces of
certain materials \cite{ucn_book}.  Thus, UCN can be stored in
material-walled vessels, whereas cold neutrons (kinetic energies
0.05--25 meV, speeds 100--2200 m s$^{-1}$) must be transported along
neutron guides at reflection angles less than the guide critical
angle, resulting in short residency times in an apparatus.

\begin{figure}[b]
\includegraphics[angle=270,scale=0.65,bb = 12 190 585 550,clip=]
{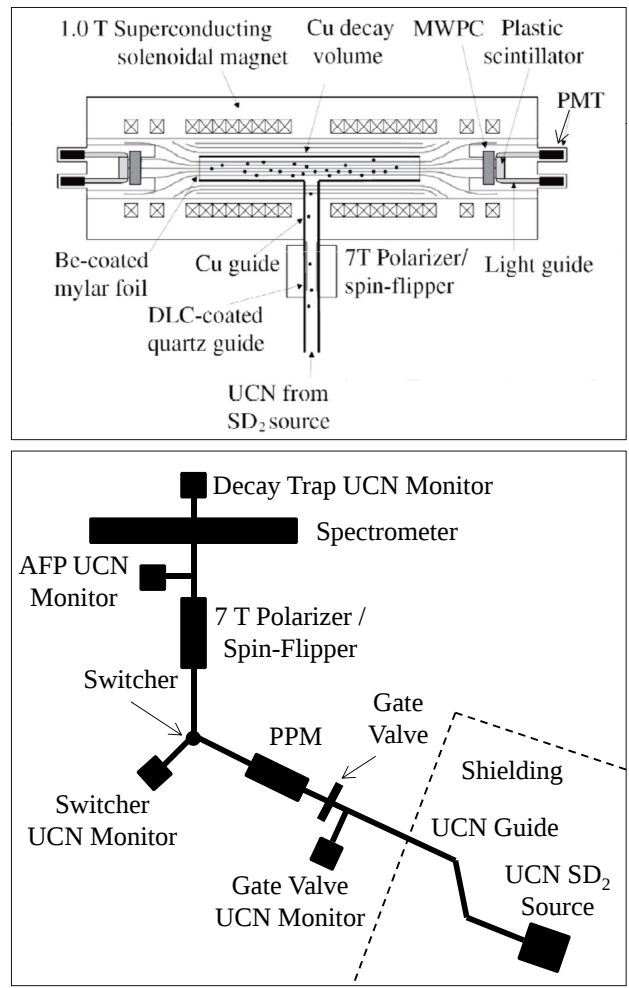}
\caption{Schematic diagram (not to scale) of the primary components of
the UCNA Experiment's $\beta$-asymmetry measurement.  The top panel
depicts the layout of the 7.0-Tesla polarizing magnet and AFP
spin-flipper, the 1.0-Tesla spectrometer, the decay trap, and the
electron detectors.  The bottom panel depicts the UCN source and
shielding, the layout of the UCN transport guides, the location of the
UCN gate valve, the location of the UCN switcher, the locations of all
the magnets, and the locations of all the UCN monitor detectors.  All
of these components are described in detail later in the text.  [Note
that during our data-taking runs in 2009, the (bare) Cu UCN guides
located between the spin-flipper and the entrance to the decay trap
(as depicted in the top panel) were replaced with diamondlike
carbon-coated Cu guides; see Section \ref{sec:measurements_geometries}
for details.]}
\label{fig:ucna_schematic}
\end{figure}

A schematic diagram of the UCNA Experiment is shown in Fig.\
\ref{fig:ucna_schematic}, and the basic principle of the experiment is
as follows.  Spallation neutrons resulting from the interaction of a
pulsed (typically 0.2 Hz) 800 MeV proton beam with a tungsten target
were moderated in cold polyethylene to the cold neutron regime, and
then downscattered to the UCN regime in a solid deuterium (SD$_2$)
crystal.  The UCN were then transported along a series of UCN guides
through a 7.0-Tesla solenoidal polarizing magnet,
where the spin-dependent $-\vec{\mu} \cdot \vec{B}$ potential ($\pm
60$ neV T$^{-1}$) served as a spin-state selector for magnetic moments
$\vec{\mu}$ oriented parallel to the direction of the longitudinal
magnetic field $\vec{B}$.  The polarized UCN were then transported
along non-magnetic UCN guides through an adiabatic-fast-passage
(AFP)
spin-flipper 1.0-Tesla field region, used to prepare UCN with spins
either parallel or anti-parallel to the magnetic field.  The UCN were
then directed to the center of a 12.4-cm diameter, 3-m long cylidrical
decay storage volume located within the warm bore of a 1.0-T
solenoidal spectrometer.  Emitted $\beta$-decay electrons then
spiraled (with a maximum Larmor diameter of 7.76 mm for 782 keV
endpoint electrons emitted perpendicular to the 1.0-T field) along the
magnetic field lines towards one of two electron detectors located on
both ends of the spectrometer.

In principle, the $\beta$-asymmetry $A$ can be extracted from
measurement of the $W(\theta) \propto (1 + P_nA\beta\cos\theta)$
angular distribution by forming an energy-dependent ``measured
asymmetry'', $A_{\text{meas}}(E_e)$, of the detectors' 
(background-subtracted) count rates,
\begin{equation}
A_{\text{meas}}(E_e) = \frac{r_1(E_e) - r_2(E_e)}{r_1(E_e) + r_2(E_e)} =
P_n A \beta \langle\cos\theta\rangle,
\label{eq:measured_asymmetry}
\end{equation}
where $r_{1(2)}(E_e)$ denote the energy-dependent count rates observed
in the two detectors, $P_n$ denotes the neutron polarization, $\beta$
denotes the electron velocity in units of $c$, and
$\langle\cos\theta\rangle$ is the average value of $\cos\theta$
integrated over the detectors' angular acceptance for that particular
value of $E_e$.  Note that for nominal values of $P_n \sim 1$, $A \sim
-0.12$, $\beta \sim 0.75$, and $\langle\cos\theta\rangle \sim 1/2$,
the experimental measured asymmetry is of order $|A_{\text{meas}}|
\sim 0.04$.

In practice, the asymmetry is extracted from ratios of the two
detectors' energy-dependent count rates for the two neutron spin
states with polarizations oriented parallel and anti-parallel to the
magnetic field via a ``super-ratio'' technique.  Here, the
super-ratio, $R$, is defined in terms of the measured
energy-dependent detector count rates for the two spin states,
$r_{1(2)}^{+(-)}(E_e)$, to be
\begin{equation}
R = \frac{r_1^-(E_e) \cdot r_2^+(E_e)}
{r_1^+(E_e) \cdot r_2^-(E_e)},
\end{equation}
with the energy-dependent measured asymmetry, $A_{\text{meas}}(E_e)$,
then calculated from the super-ratio according to
\begin{equation}
A_{\text{meas}}(E_e) = \frac{1-\sqrt{R}}{1+\sqrt{R}} =
P_n A \beta \langle\cos\theta\rangle.
\label{eq:individual_super_ratio}
\end{equation}
The merit of this super-ratio technique is that effects due to
differences in the two detectors' efficiencies and spin-dependent
differences in the efficiencies for transport of the two UCN spin
states into the spectrometer cancel to first order.
In a binned analysis, energy-dependent detection
efficiencies also largely cancel in the super-ratio, and are
negligible for the energy bin sizes used in this work.

The motivation for the development of the UCNA Experiment was
several-fold.  First, the use of UCN in a neutron $\beta$-asymmetry
experiment controls key neutron-related systematic corrections and
uncertainties, including the neutron polarization and
neutron-generated backgrounds.  As discussed in detail later in this
article, the polarization has been demonstrated to be $>99.48$\% at
the 68\% C.L., with the precision, at present, limited only by
statistics.  Further, neutron-generated backgrounds have been
constrained to be negligible, a direct result of the relatively small
number of neutrons present in the apparatus at any time, the small
probability for their capture and subsequent generation of
accompanying irreducible gamma ray backgrounds, and the fact that
nearly all of the neutrons present in the apparatus are located within
the spectrometer's decay volume.  In the
UCNA Experiment,
a relatively large fraction, $\sim 1/40$, of the UCN stored in the
decay volume contribute to the measured decay rate, whereas in cold
neutron beam experiments typically only $\sim 1/10^7$ of the neutrons
passing through the apparatus contribute to the decay rate
\cite{abele97}.  Therefore, control of neutron-generated backgrounds
is expected to be intrinsically more challenging in cold neutron beam
experiments.

Second, as described in detail elsewhere
\cite{ito07,plaster08}, the electron detector system developed for the
UCNA Experiment, consisting of a low-pressure multiwire proportional
chamber (MWPC) backed by a plastic scintillator, provides position
sensitivity, suppresses ambient gamma-ray backgrounds, and permits the
reconstruction of low-energy-deposition electron backscattering
events.  The UCNA Experiment is the first neutron
$\beta$-asymmetry experiment to employ a MWPC, providing the
experiment with two critical advantages.  First, the position
information permits the definition of a fiducial volume on an
event-by-event basis.  Second, the position information also permits
for an event-by-event correction for the scintillator's
position-dependent energy response.

We now provide a more detailed description of the primary components
of the UCNA Experiment.

\subsection{UCN Source and Guide Transport System}
\label{sec:experiment_source_guides}

A detailed description of the design principles and performance of the
LANL SD$_2$ UCN source is given elsewhere
\cite{morris02,saunders04,saunders11}; therefore, we provide only a
brief description here.  Protons from the 800 MeV LANSCE accelerator
were delivered in pulsed mode\footnote{Each proton
beam pulse consisted of five 625 $\mu$s beam bursts separated by 0.05
s, with 5.2~s between each pulse's leading edge burst.} at
a repetition rate of 0.2~Hz to a tungsten
spallation target, which was surrounded by a room-temperature
beryllium reflector.  With the spallation target operated in this
pulsed mode, prompt beam related backgrounds can be eliminated with
simple timing cuts, with negligible loss of duty factor for the
$\beta$-decay measurements performed with the UCN stored in the
electron spectrometer.  The spallation neutrons were moderated in
cold-helium-gas-cooled polyethylene (maintained at a temperature of
$\sim 150$ K for time-averaged proton beam currents of 5.8 $\mu$A)
located between the tungsten target and the beryllium reflector.  The
moderated cold neutrons were then downscattered to the UCN regime in a
$\sim 2$ L cylindrical volume of ortho-state SD$_2$ \cite{liu00,liu03}
embedded at the bottom of a vertically-oriented cylindrical
liquid-helium-cooled aluminum cryostat coated with $^{58}$Ni,
presenting a nominal effective potential of 342 neV
to the emerging UCN flux.  The SD$_2$ was maintained at temperatures
$< 10$ K during the proton beam pulses on the spallation target.  A
butterfly-style ``flapper'' valve coated with $^{58}$Ni was located
immediately above the SD$_2$ volume.  This ``flapper'' valve opened
and subsequently closed (with opening and closing response times of
about 0.1 s) with each proton pulse, in order to increase the storage
lifetime of the UCN in the volume of the source above the SD$_2$
volume.  A typical UCN lifetime with the flapper open was $9.6 \pm
0.2$~s, whereas the lifetime with the flapper closed was $39.4 \pm
0.1$~s.  The flapper leads to a corresponding increase in the UCN
density.

The UCN were then extracted from the source along
horizontally-oriented 10.16-cm diameter stainless steel guides
(presenting a nominal
potential of 189 neV) through the biological shielding
surrounding the source and out into the experimental area.  As shown
in Fig.\ \ref{fig:ucna_schematic}, this system of guides through the
biological shielding included two $45^\circ$ bends to eliminate
neutrons with kinetic energies above the stainless steel guide
potential.  The maximum UCN density at the biological shield exit
that we have obtained is $52 \pm 9$ cm$^{-3}$
\cite{saunders11},
but for this work (runs in 2008--2009) typical densities were
$\sim 35 \pm 6$ cm$^{-3}$.  After
exiting the biological shield, the UCN were transported along
stainless steel guides through a gate valve, which served to separate
the UCN source from the experiment, thus permitting measurements of
backgrounds in the electron spectrometer detectors with the
proton beam still operating in its normal pulsed mode,
but no accompanying UCN transport to the spectrometer.  A 6.0-T
superconducting solenoidal pre-polarizing magnet (PPM) was located
immediately downstream of this gate valve.  The PPM was included in
the experiment design in order to minimize UCN transport losses
through a thin (0.0254-mm thick) Zr foil which served to separate the
vacuum in the SD$_2$ source from the downstream vacuum in the
remainder of the experiment.  Note that the UCN population was
polarized after transport through the PPM's longitudinal magnetic
field.

To preserve this initial polarization, 10.16-cm diameter
electropolished Cu guides (nominal
potential of 168 neV) were installed
downstream of the PPM magnet.  The UCN were then transported along
these guides through a ``switcher'' valve, which allowed the
downstream guides comprising the $\beta$-asymmetry measurement to be
connected to either the upstream guides from the UCN source, or to a
$^3$He UCN detector \cite{morris09} used, as described later, for
measurements of the depolarized population.  These electropolished Cu
guides then transported the UCN through the primary 7.0-T polarizing
magnet (called the AFP magnet).  A 100-cm long quartz
guide section coated with diamondlike carbon (DLC)
\cite{mammei-thesis} passed through the center of a resonant (1.0-T)
``bird-cage'' r.f.\ cavity \cite{holley_afp}, used for adiabatic fast
passage (AFP) spin-flipping of the UCN.

Downstream of this DLC-coated quartz guide, another section of
10.16-cm diameter Cu guide transitioned to a 7-cm (vertical) $\times$
4-cm (horizontal) rectangular Cu guide, which transported the UCN
through a horizontal penetration in the 1.0-T solenoidal electron
spectrometer coil into the decay trap.  Permanent
magnets were attached to the outer surfaces of the rectangular guide
in order to suppress Majorana spin-reorientations
\cite{vladimirski61} of neutrons passing through ``field
zeros'' in the 1.0-T solenoidal spectrometer's field.


The UCN rate along the transport guide system was monitored with
$^3$He UCN detectors \cite{morris09} at two key locations: at the gate
valve (for monitoring of the SD$_2$ source performance), and slightly
downstream of the APF spin flipper (for monitoring of the AFP
spin-flipping efficiency).  These $^3$He UCN detectors were coupled to
the guide system via small (0.64-cm diameter) holes in the bottom of
the guides.  Note that the UCN density in the spectrometer for the
spin-flipped state was smaller than that for the unflipped state,
because of losses (after the 2.0-T $\vec{\mu} \cdot \vec{B}$ energy
boost associated with the spin flip) in the transport guides located
between the AFP spin-flip region and the electron spectrometer.
The measured $\beta$-decay rates for the spin-flipped
state were $\sim 25$\% smaller than those for the non-spin-flipped
state.

The maximum neutron $\beta$-decay rates measured in the spectrometer
during the 2008--2009 runs correspond to a stored density of
approximately 1~cm$^{-3}$ in the decay trap. Major
sources of loss are transport through the high field regions in the
PPM and in the AFP magnet.  The transport though the PPM is about
25\%.  Approximately half of the loss is due to polarization of the
neutrons, and the other half is due to UCN absorption in the Zr foil
and non-specular scattering on the UCN guide walls in the high field
region of the magnet which leads to enhanced wall losses. There is an
approximate 15\% loss in UCN density in the transition from the
stainless steel to the copper guides because of the lower Fermi
potential of the copper. Transmission through the AFP magnet is about
60\%, again due to non-specular scattering in the high field
region. There is a 50\% loss in density in loading the decay trap
because the loading time (which is determined by the
aperture of the above-described 7-cm $\times$ 4-cm rectangular guide)
and decay trap lifetime are nearly the same.  Finally, there is an
approximate factor of two loss in the transport from the biological
shield exit to the decay trap due to guide losses.  (The typical loss
per bounce in the guide system is $3 \times 10^{-4}$ which is
dominated by gaps in the guide couplings.)  Thus, all of these factors
combined account for the reduction in the UCN density from its intial
value of $\sim 35$ cm$^{-3}$ at the biological shield
exist to $\sim 1$ cm$^{-3}$ in the spectrometer decay trap
during runs in 2008--2009.

\subsection{Decay Trap Geometry}
\label{sec:experiment_decay_trap}

The decay trap consisted of a 300-cm long, 12.4-cm diameter
electropolished Cu tube situated along the warm bore axis of the 1.0-T
solenoidal electron spectrometer.  The vacuum pressure in the UCN
guides downstream of the Zr foil in the PPM and in the decay trap was
typically $\sim 10^{-5}$ Torr.  The ends of the decay trap were closed
off with variable thickness mylar end-cap foils, whose inside surfaces
were coated with 300 nm of Be (nominal 252 neV
potential) which served to increase the UCN storage time in the decay
trap (and, hence, the $\beta$-decay rate).  An additional important
feature of the end-cap foils is that they eliminated the possibility
for neutron $\beta$-decay events in the region of the spectrometer
where the field is expanded from 1.0-T to 0.6-T (discussed later in
Section \ref{sec:experiment_electron_spectrometer_SCS}).
Collimators with inner radii of 5.84 cm mounted on the
two ends of the decay trap suppressed events originating near the
decay trap walls and also functioned as mounts for the end-cap foils.
As discussed in more detail later, the thicknesses of the mylar
end-cap foils were varied from 0.7 $\mu$m to 13.2 $\mu$m to study key
systematics related to electron energy loss in, and backscattering
from, these foils.  

The UCN density in the decay trap was monitored by a $^3$He UCN
detector coupled to a small (0.64-cm diameter) hole in the bottom of
the decay trap center, as indicated schematically in Fig.\
\ref{fig:ucna_schematic}.

\subsection{Polarization and Spin Flipping System}
\label{sec:experiment_spin_flipping}

A detailed description of the 7.0~T polarizing magnet and the AFP
spin-flip system is given elsewhere \cite{holley_afp}; therefore, we
provide only a brief description of this system here.

The solenoidal superconducting magnet which serves as the primary UCN
polarizer (the AFP magnet) and provides the requisite environment for
an adiabatic fast-passage (AFP) spin flipper was designed by American
Magnetics using a cryostat supplied by Ability Engineering. It
possesses a 194.9~cm long, 12.7~cm diameter warm bore and provides
both a peak field of 7.0-T near the entrance as well as a 44.5-cm long
precision gradient spin-flip region with an average
field of 1.0 T, chosen to reduce neutron spectral
differences between the flipped and unflipped spin states in the
electron spectrometer decay trap volume. When energized to 96.45~A,
the main coil of this magnet produces both the maximum polarizing
field as well as the 1~T field, with an average
gradient of $6 \times 10^{-5}$ T cm$^{-1}$ through the latter. Ten
superconducting shim coils centered on the uniform field region and
spaced every 5.1~cm provide the ability to further tailor the uniform
field in order to optimize performance of the spin flipper.

Due to the high field in the spin flip region, the spin flipper was
constructed using an efficient high-pass birdcage resonant cavity
geometry \cite{holley_afp}.  For the UCNA Experiment, this
configuration was realized with eight Cu tubes (rungs) arranged in a
cylindrical geometry and connected at the top by 820~pF American
Technical Ceramics chip capacitors and at the bottom by a sliding Cu
tuning ring whose position determined the inductance presented by the
Cu tubes. When excited by an r.f.\ signal such a geometry is resonant,
with the fundamental mode corresponding to a discretized sinusoidal
distribution of current in the rungs. This current distribution
provides a transverse r.f.\ field, one rotating component of which is
utilized for AFP spin flipping, providing efficient spin reversal over
a wide band of neutron speeds \cite{holley_afp}.  The specific
operation frequency was adjusted by moving the tuning ring, and the
cavity formed after tuning the spin flipper to operate at $\sim
28$~MHz was $\sim 15$~cm long (8.74~cm diameter), coaxial with a 7~cm
diameter DLC-coated quartz UCN guide.

\begin{figure*}
\includegraphics[angle=270,scale=0.60,clip=]{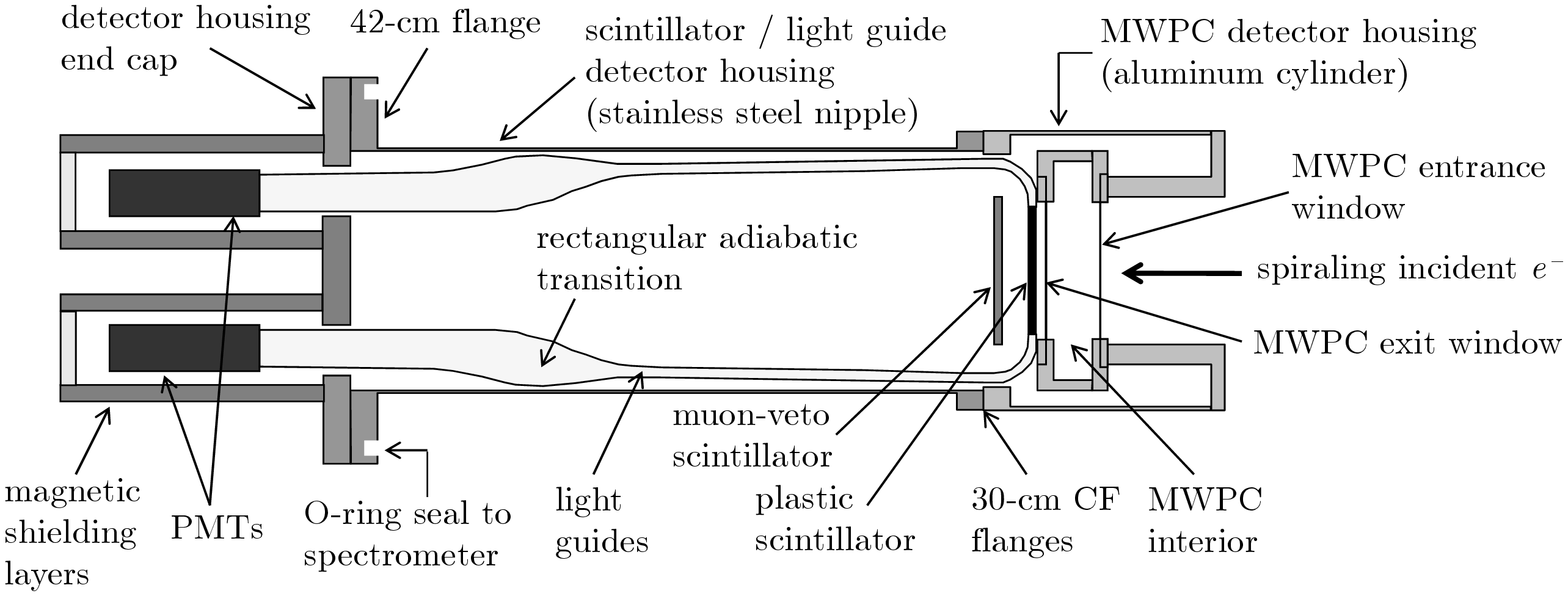}
\caption{Schematic diagram (not to scale) of the MWPC and plastic
scintillator electron detector package.}
\label{fig:mwpc_scintillator_schematic}
\end{figure*}

The UCNA spin flipper was typically operated with 40~W of input power,
which necessitated an impedance-matching system comprised of a
calculated length of drive line and three Jennings vacuum variable
capacitors. Water cooling was also required, and was accomplished by
flowing chilled, filtered tap water serially through the rungs. In
order to provide for stable electrical operation and to prevent r.f.\
radiation from inducing noise elsewhere in the experiment, the
birdcage cavity was driven in a balanced mode and electrically
shielded by a grounded Al cylindrical enclosure which also provided a
vacuum seal around the DLC-coated quartz guide. The interior of this
enclosure connected through four bellows to the outside of the AFP
magnet so that the actual r.f.\ cavity remained at atmosphere while
the AFP magnet bore and the guide system were under vacuum. This
arrangement also provided feedthroughs for the r.f.\ drive line, water
cooling lines, an RTD temperature sensor, and an r.f.\ field sensor
loop.

Initial characterization of the spin flipper was performed in a
crossed polarizer analyzer geometry as described in \cite{holley_afp},
which determined the average spin flip efficiency to be $0.9985 \pm
0.0004$. During the actual running of the UCNA experiment during the
years 2008--2009, tuning of the spin flipper, as well as run-to-run
monitoring of its performance, was accomplished using a $^3$He UCN
monitor located just downstream of the AFP magnet, $\sim 1$~m below a
$\sim 0.64$~cm hole in the bottom of the guide (the location of this
UCN monitor is indicated schematically in Fig.\
\ref{fig:ucna_schematic}).  This detector had a magnetized Fe foil
covering the detector acceptance which provided for spin state
selection.

\subsection{Electron Spectrometer System}
\label{sec:experiment_electron_spectrometer}

The electron spectrometer system, consisting of a 1.0-T
superconducting solenoidal magnet and a multiwire proportional chamber
(MWPC) and plastic scintillator electron detector package, is
described in detail elsewhere \cite{ito07,plaster08}.  Nevertheless,
for completeness, we discuss the primary components of this system
here.  As described below, the electron spectrometer system was
designed both to suppress the total electron backscattering fraction
and to reconstruct low-energy-deposition backscattering events.  The
primary components of the two identical MWPC and plastic scintillator
detector packages are shown in Fig.\
\ref{fig:mwpc_scintillator_schematic}.

\subsubsection{1.0-T Superconducting Solenoidal Magnet}
\label{sec:experiment_electron_spectrometer_SCS}

The spectrometer magnet \cite{plaster08} is a warm-bore 35-cm
diameter, 4.5-m long superconducting solenoid (hereafter, SCS magnet).
The coil, which was designed and fabricated by American Magnetics,
Inc., consists of a main coil winding with a single persistence heater
switch, 28 shim coil windings (each with individual persistence heater
switches), and three rectangular 7-cm $\times$ 4-cm radial
penetrations (two providing horizontal access, and one providing
vertical access, to the warm bore).  These penetrations are located at
the center of the coil.  The magnet's 1600-L capacity liquid helium
cryostat was designed and fabricated by Meyer Tool and Manufacturing,
Inc.
The magnet's full energized field strength of 1.0 T requires a current
of 124~A in the main coil winding.  Note that the magnet's shim coils
were not energized during the data-taking runs reported in this
article.

An important feature of the SCS magnet design was that the field is
expanded, as indicated schematically in Fig.\
\ref{fig:ucna_schematic}, from 1.0 T in the decay trap region to 0.6 T
at the location of the MWPC and plastic scintillator electron
detectors, to suppress large-pitch-angle backscattering.  In
particular, the field-expansion ratio of 0.6 maps pitch angles of
$90^\circ$ in the 1.0-T region to pitch angles of $51^\circ$ in the
0.6-T region.  Another important feature of the magnet design
concerned the field uniformity in the decay trap region.  Electrons
emitted with momentum $p_0 = (p_{\perp,0}^2 +
p_{\parallel,0}^2)^{1/2}$, with $p_{\perp,0}$ ($p_{\parallel,0}$) the
initial transverse (longitudinal) momentum component, in some local
field $B_0$ will be reflected from field regions $B$ if $B >
B_{\text{crit}} \equiv (p_0^2/p_{\perp,0}^2) B_0$, thus contributing
to a false asymmetry.  By this same process, electrons emitted at
large pitch angles in the vicinity of a local field minimum will be
trapped.

\begin{figure}
\includegraphics[scale=0.45,clip=]{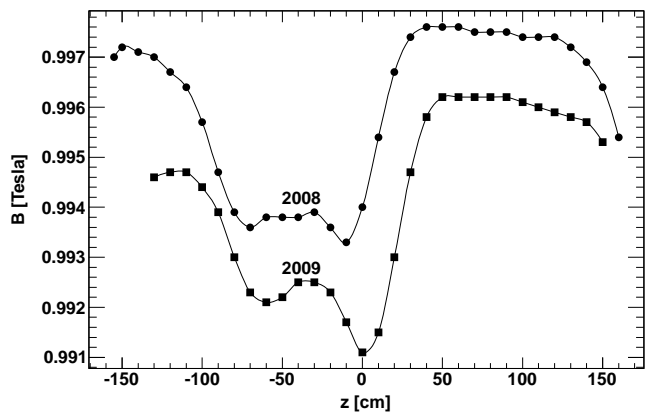}
\caption{Measured SCS field profile during 2008 (circles) and 2009
(squares).  The two data sets have been offset for purposes of
clarity.  The coordinate system is such that the 300-cm long decay
trap is centered at $z = 0$.}
\label{fig:scs_field_uniformity}
\end{figure}

The SCS field profile measured during the data-taking period reported in
this article is shown in Fig.\ \ref{fig:scs_field_uniformity}.  As can
seen there, the field was uniform to the level of $5 \times 10^{-3}$
over the decay trap length, but included a $\sim 0.005$~T ``dip'' near
the center of the decay trap.  Note that the field uniformity shown
here was degraded from that originally published in \cite{plaster08};
this was the result of damage to the shim coil persistence heater
switches during multiple magnet quenches.  The impact of this field
non-uniformity on the measured asymmetry is discussed later in this
article.

\subsubsection{Multiwire Proportional Chamber}
\label{sec:experiment_electron_spectrometer_MWPC}

Some of the most important features of the MWPC \cite{ito07} are as
follows.  First, because an MWPC is relatively insensitive to
gamma rays, requiring a coincidence between the MWPC and the
scintillator greatly suppressed gamma-ray backgrounds, a dominant
background source in previous experiments (see, e.g., background
spectrum in \cite{abele97}).

Second, the MWPC permitted reconstruction of an event's transverse
$(x,y)$ position.  This permitted the definition of a fiducial volume
and the subsequent rejection of events occurring near the decay trap
walls.  Such electrons can scatter from the decay trap walls, leading
to a distortion in the energy spectrum and/or a bias to the asymmetry.
The $(x,y)$ position information from the MWPC also permitted a
characterization of the scintillator's position-dependent response, as
the scintillator was viewed by four photomultiplier tubes (discussed
in the next section).  The 64-wire anode plane was
strung with 10-$\mu$m diameter gold-plated tungsten wires, and the
two cathode planes (oriented at $90^\circ$ relative to each other)
were each strung with 64 50-$\mu$m diameter gold-plated aluminum
wires.  The wire spacing on both the anode and cathode planes was 2.54
mm, yielding an active area of $16.3 \times 16.3$ cm$^2$.  This area
in the 0.6-T field-expansion region mapped to a $\sqrt{0.6}(16.3
\times 16.3)$ cm$^2$ = $12.6 \times 12.6$ cm$^2$ square in the 1.0-T
region, thereby providing full coverage of the 12.4-cm diameter decay
trap volume.  As demonstrated previously \cite{ito07,plaster08}, the
center of the event (i.e., the center of the charge cloud resulting
from the electron's Larmor spiral in the MWPC gas) could be
reconstructed with an accuracy of better than 2 mm, sufficient for the
definition of a fiducial volume.

Third, to suppress ``missed backscattering events'' (i.e., those
events depositing no energy above threshold in any detector element
along the electron's trajectory prior to backscattering), the entrance
window separating the MWPC fill gas from the spectrometer vacuum was
designed to be as thin as possible.  
Fourth, because of this thin
entrance window requirement, the fill gas pressure was required to be
as low as possible.  The chosen fill gas, C$_5$H$_{12}$
(2,2-Dimethylpropane, or ``neopentane''), a low-$Z$ heavy hydrocarbon,
was shown to yield sufficient gain at a pressure of 100 Torr and a
bias voltage of 2700 V.  At this pressure, the minimum window
thickness (over the MWPC's 15-cm diameter entrance and exit windows)
shown to withstand this 100 Torr pressure differential with minimal
leaks from pinholes was 6 $\mu$m of aluminized mylar.  Note that the
front window was further reinforced by Kevlar fibers.

\subsubsection{Plastic Scintillator Detector}
\label{sec:experiment_electron_spectrometer_scintillator}

The plastic scintillator detector was a 15-cm diameter, 3.5-mm thick
disk of Eljen Technology EJ-204 scintillator.  This 15-cm diameter
mapped to a 11.6-cm diameter disc in the 1.0-T decay trap region,
providing nearly full coverage of the decay trap volume.  The range of
an endpoint energy electron in the plastic was 3.1 mm; therefore, the
3.5-mm thickness was sufficient for a measurement of the full
$\beta$-decay energy spectrum, while minimizing the ambient gamma
ray background rate.

With the scintillator located in the 0.6-T field-expansion region
at a distance of 2.2 m from the center of the SCS magnet,
light from the disc was transported over a distance of $\sim 1$ m
along a series of UVT light guides to photomultiplier tubes which were
mounted in a region where the magnetic field was $\sim 0.03$ T.  The
light guide system, shown schematically in Fig.\
\ref{fig:mwpc_scintillator_schematic}, consisted of twelve rectangular
strips (39-mm wide $\times$ 10-mm thick UVT) coupled to the edge of
the scintillator disc with optical grease.  These twelve rectangular
strips were then bent through $90^\circ$ over a 35-mm radius,
transported over a distance of $\sim 1$ m away from the scintillator,
and then adiabatically transformed into four $39 \times 30$ mm$^2$
rectangular clusters, with 5.08-cm diameter Burle 8850 photomultiplier
tubes (PMTs) glued to each of these four rectangular clusters.
Therefore, each PMT effectively viewed one $\pi/2$ quadrant of the
scintillator face.

The magnetic shielding for each of the PMTs consisted of an array of
active and passive components, including (moving from the outside to
inside) steel and medium-carbon-steel shields, a bucking solenoidal
coil wound on the surface of a thin $\mu$-metal foil, and a
$\mu$-metal cylinder.  Magnetic end caps were not required.

The vacuum housing enclosing the scintillator, light guides, and PMTs
was maintained at $\sim 95$ Torr of nitrogen, and was separated from
the MWPC volume (with its 100 Torr of neopentane gas) by the MWPC exit
window.  The nitrogen volume pressure was maintained at a somewhat
lower pressure than the MWPC pressure to ensure that the MWPC exit
window bowed out, or away, from the MWPC interior, to avoid contact
with the MWPC wire planes.

\subsection{Scintillator Calibration and PMT Gain Monitoring}
\label{sec:experiment_calibration_gain}

The scintillators were calibrated periodically with conversion
electron sources, including commercially-available $^{109}$Cd (63 keV,
84 keV), $^{139}$Ce (127 keV, 160 keV), $^{113}$Sn (364 keV, 388 keV),
$^{85}$Sr (499 keV), and $^{207}$Bi (481 keV, 975 keV, 1047 keV)
conversion-electron sources, and a custom-prepared $^{114m}$In (162
keV, 186 keV, 189 keV, 190 keV) conversion-electron source (via
implantation of $^{113}$In onto an Al substrate and subsequent
irradiation in a reactor \cite{wrede11}).  These calibrations were
conducted \textit{in-situ} using a vacuum load-lock source insertion
system which permitted insertion and removal of calibration sources
with the electron spectrometer under vacuum.  The insertion point for
these sources was through one of the superconducting solenoid magnet's
horizontal rectangular penetrations at the center of the coil.  Note
that this source insertion system permitted the sources to be
positioned only along the horizontal axis of the decay trap's circular
geometry; however, as described later, the position dependence of the
energy calibration over the full circular geometry was achieved by
comparing the reconstructed neutron $\beta$-decay endpoint in a large
number of binned positions over the scintillator face.

The PMT gains were monitored on an approximate daily basis with a
$^{113}$Sn source using this source insertion system.  Fits to the
minimum-ionizing peak of cosmic-ray muons served as a run-to-run gain
monitor.

\subsection{Cosmic-Ray Muon Veto System}
\label{sec:experiment_muon_veto}

The electron spectrometer was surrounded with a cosmic-ray muon veto
system which consisted of the following components.  First, as shown
in Fig.\ \ref{fig:mwpc_scintillator_schematic}, a 15-cm diameter,
25-mm thick plastic scintillator (the ``backing veto'') was located
immediately behind each of the spectrometer scintillators.  Second, a
large-scale plastic scintillator and sealed drift tube veto counters
\cite{rios11} surrounded the electron spectrometer magnet.

\subsection{Electronics and Data Acquisition}
\label{sec:experiment_electronics}


\begin{figure}
\includegraphics[angle=270,scale=0.36,clip=]{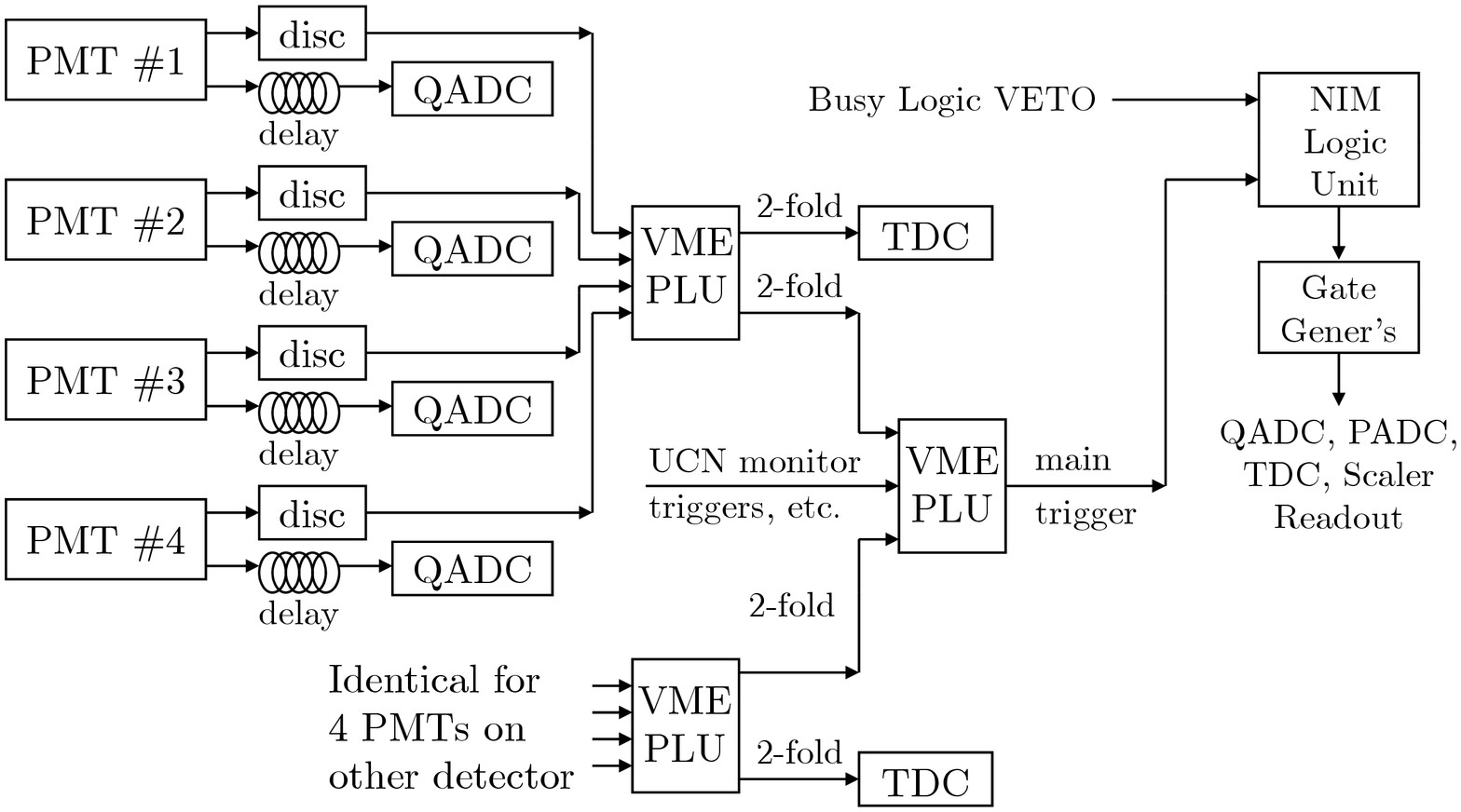}
\caption{Simplified schematic diagram of the data acquisition event
trigger logic.}
\label{fig:trigger_logic_schematic}
\end{figure}

The frontend electronics for the experiment consisted of a VME-based
system for the event trigger logic (via discriminators and
programmable logic units (PLUs)) and for the readout of scalers,
analog-to-digital convertors (ADCs) and time-to-digital convertors
(TDCs).  A NIM-based system coupled to the VME system was employed for
the implementation of a ``busy logic'', which served to veto event
triggers arriving during ADC/TDC conversion times (i.e., during these
modules' busy states).  This busy logic also prevented re-triggering
by correlated scintillator afterpulses (mostly occuring over a
$\sim 1$ $\mu$s window \cite{junhua_thesis}), and was
implemented with a LeCroy 222 gate generator in latch mode.

A simplified schematic diagram of the trigger logic is shown in Fig.\
\ref{fig:trigger_logic_schematic}.  For each detector package, a
trigger was defined by a two-fold PMT coincidence trigger above the
discriminator threshold for each PMT (nominally, set at 0.5
photoelectrons).  The resulting two-fold PMT trigger rate in each
scintillator was $\sim 50$~s$^{-1}$ (primarily from
low-energy background gamma rays); the singles rates in each PMT as
determined by counting in scalers were typically $\sim 500$--1000~s$^{-1}$
(from both dark noise and low-energy backgrounds).
The main event trigger was then defined to be the \texttt{OR} of the
two detectors' two-fold PMT triggers and other experiment triggers
(e.g., from the $^3$He UCN monitor detectors).  The logic for the
two-fold PMT coincidence triggers and the main event trigger was
performed with CAEN V495 Dual PLUs.  Those main event triggers not
vetoed by the busy logic then triggered gate/delay generators for the
readout of the ADCs, TDCs, and scaler modules.  The total
number of two-fold PMT coincidence triggers were counted in scalers as
a monitor of the DAQ dead time.

CAEN V775 TDC modules were used for the relative measurement of the
time-of-flight between the two detectors' two-fold PMT coincidence
triggers.  This relative timing information provided for the
identification of the detector with the earlier arriving trigger,
important, as discussed later, for the assignment of the initial
direction of incidence for electron backscattering events triggering
both scintillators.  These TDCs were also used to record the timing
information from the plastic-scintillator-based muon veto detectors.
A global event-by-event time stamp was defined by the counting of a 1
MHz clock in a CAEN V830 scaler.

CAEN V792 charge-integrating ADC (QADC) modules, triggered for readout
by a $\sim 140$ ns gate from a CAEN V486 gate/delay generator,
provided a measurement of the total charge measured in each PMT. 
The analog signals from the cosmic-ray muon backing vetos were also
read out by these QADC modules.  Peak-sensing CAEN V785 ADC (PADC)
modules, triggered for readout by a $\sim 12$ $\mu$s gate from a CAEN
V462 gate generator, digitized the MWPC anode and cathode-plane
signals.  Note that the anode signal that was read out was the
summation (i.e., single channel per anode plane) of the signals on all
64 of the wires comprising the anode plane.  The 64 wires on each of
the two cathode planes were read-out in groups of four (i.e., 16
channels per cathode plane); hereafter, we will simply call each of
these four-wire groups a ``wire''.  Analog signals from the $^3$He UCN
monitors and the drift tube cosmic-ray muon veto counters were also
read out with PADCs.


The data acquisition (DAQ) system was based on the \texttt{MIDAS}
package \cite{midas}, with a dedicated Linux-based workstation for
implementation of the frontend electronics acquisition code and a
separate dedicated Linux-based workstation for run control and online
analysis.  The frontend acquisition code accessed the VME crate via a
Struck PCI/VME interface.  The \texttt{MIDAS} raw data banks were
subsequently decoded into \texttt{CERNLIB} \texttt{PAW} \cite{cernlib}
and \texttt{ROOT} \cite{root} file formats for data analysis.

A separate data acquisition system, based on the
\texttt{PCDAQ} software package \cite{hogan99} was used to monitor the
proton beam charge incident upon the UCN source's tungsten spallation
target and to asynchronously monitor environmental variables in the
experimental area.  The incident proton flux was measured using an
integrating current toroid mounted around the proton beam line 8~m
upstream of the tungsten target, just before the proton beam entered
the biological shield.  As noted earlier in Section
\ref{sec:experiment_source_guides}, each proton beam pulse consisted
of five 625 $\mu$s beam bursts separated by 0.05~s, with 5.2~s between
each pulse's leading edge burst.  The proton charge integrating system
measured only the charge of the first of the five beam bursts in each
pulse; the resulting value was then scaled by five to yield
the total proton charge delivered during each of the 0.2 Hz beam pulses.

The environmental monitoring system asynchronously
read and stored up to 96 variables, on a typical time scale of 0.2 s
to 1.0 s between readings.  Variables measured included cryogenic
temperatures in the UCN source (read by Lakeshore 218 temperature
monitors), pressures in the different segments of the UCN guide system
(read by capacitance manometer, thermocouple, and cold-cathode ion
vacuum gauges), ambient temperature in the experimental area, and
liquid helium levels and gas pressures throughout the cryogenic systems.
The environmental data were time-stamped for later comparison to the
$\beta$-decay data acquired with the main data acquisition system.

\section{Measurements, Experimental Geometries, and Polarization}
\label{sec:measurements}

In this section we provide a detailed description of our measurement
procedures for $\beta$-decay and ambient background runs; the various
geometrical configurations of the experiment during our $\beta$-decay
runs; and our procedures for, and results from, measurements of the
neutron polarization.

\subsection{$\bm{\beta}$-Decay Run Cycle}
\label{sec:measurements_run_cycle}

\subsubsection{Octet Data-Taking Structure}
\label{sec:measurements_run_cycle_octet}

The data taking during normal $\beta$-decay production running was
organized into octets, each consisting of A- and B-type quartet run
sequences.  The structure of these quartet and octet run sequences,
shown in Table \ref{tab:octet_structure}, was such that the neutron
spin state (hereafter designated $+$ or $-$, with $+$($-$)
corresponding to the loading of UCN with AFP-spin-flipper-on (-off)
spin states into the electron spectrometer) was toggled according to a
$-++-+--+$ spin-sequence (for octets in which A-type runs preceded
B-type runs) or a $+--+-++-$ (i.e., complement) spin-sequence, with
the order of $\beta$-decay and ambient background run pairs toggled
for a particular spin state within each A-type or B-type run sequence.
Within each octet, the decision for whether the A-type runs would
precede or follow the B-type runs was made randomly.  The notation in
Table \ref{tab:octet_structure} is such that B$^{+(-)}$ and
$\beta^{+(-)}$ denote, respectively, ambient
background and $\beta$-decay runs for the two spin
states.  The notation for depolarization runs is such that D$^+$, for
example, denotes a measurement of the depolarized spin-state
population for which the spin-state was polarized in the $+$
spin-state during the preceding $\beta$-decay run.

\begin{table}
\caption{Run structure for the octet data taking sequence, consisting
of A- and B-type quartets.  See text for details.}
\begin{ruledtabular}
\begin{tabular}{ccccccccccccc}
A1& A2& A3& A4& A5& A6& & A7& A8& A9& A10& A11& A12 \\
B$^-$& $\beta^-$& D$^{-}$& B$^+$& $\beta^+$& D$^{+}$& &
  $\beta^+$& D$^{+}$& B$^+$& $\beta^-$& D$^{-}$& B$^-$ \\ \hline
B1& B2& B3& B4& B5& B6& & B7& B8& B9& B10& B11& B12 \\
B$^+$& $\beta^+$& D$^{+}$& B$^-$& $\beta^-$& D$^{-}$& &
  $\beta^-$& D$^{-}$& B$^-$& $\beta^+$& D$^{+}$& B$^+$ \\
\end{tabular}
\end{ruledtabular}
\label{tab:octet_structure}
\end{table}

As described in detail later, the $\beta$-decay yields were ultimately
obtained from background subtraction.  Although such a procedure is
potentially subject to systematic bias from time-varying backgrounds,
the merit of this octet data-taking structure is that linear
background drifts cancel to all orders (provided that the durations of
the background and $\beta$-decay runs do not change during the octet)
in the definition of asymmetries based on complete octet-structure
data sets.  Linear drifts in detector efficiency which
might affect background subtraction also cancel under the octet
structure.


\subsubsection{Run Cycle Procedure}
\label{sec:measurements_run_cycle_procedure}

\begin{table*}
\caption{Foil thicknesses for the different decay trap end-cap window
and MWPC window Geometries and the number of
$\beta$-decay events collected in each Geometry passing all analysis
cuts.}
\begin{ruledtabular}
\begin{tabular}{cccc}
& Decay Trap End-Cap& MWPC Window&
  Number of \\
Geometry (Year)& Window Thickness [$\mu$m]& Thickness [$\mu$m]&
  $\beta$-Decay Events \\ \hline
A (2008)& 0.7 (mylar) + 0.3 (Be)& 25&
  $5.2 \times 10^6$ \\
B (2008)& 13.2 (mylar) + 0.3 (Be)& 25&
  $5.3 \times 10^6$\\
C (2008)& 0.7 (mylar) + 0.3 (Be)& 6&
  $2.4 \times 10^6$ \\
D (2009)& 0.7 (mylar) + 0.3 (Be)& 6&
  $1.8 \times 10^6$ \\
\end{tabular}
\end{ruledtabular}
\label{tab:geometries}
\end{table*}

As noted previously in Section \ref{sec:experiment_source_guides}
and shown in Fig.\ \ref{fig:ucna_schematic}, a gate
valve separated the UCN source from the $\beta$-asymmetry experiment.
Measurements of the ambient backgrounds (runs A1/B1, A4/B4, A9/B9, and
A12/B12) were performed with this gate valve closed (i.e., with no UCN
in the decay trap), but with the proton beam still
operating in its normal pulsed mode and the AFP spin-flipper in its
appropriate run-paired state (i.e., so as to properly account for
beam-related backgrounds and any noise/backgrounds associated with the
operation of the AFP spin-flipper).  These background runs were
nominally 0.2 hours in duration.  The $\beta$-decay runs (runs A2/B2,
A5/B5, A7/B7, A10/B10) were performed with the gate valve open and the
AFP spin-flipper in its appropriate state for the entire duration of
the run, nominally 1.0 hour in duration.

During the $\beta$-decay runs, an equilibrium density
of both correctly polarized and incorrectly polarized UCN developed in
the decay trap. With the spin flipper off, the incorrectly polarized
population was dominated by depolarization due to material
interactions between the UCN and the walls of the decay trap and
guides. When the spin flipper was active, this incorrectly polarized
population was increased as a result of spin flipper inefficiency. In
the spin flipper off case, the lifetime of correctly polarized UCN in
the decay trap, dominated by the decay trap exit aperture, was
$\sim 21$~s, and the lifetime of incorrectly polarized UCN trapped in
the experimental geometry by the 7~T polarizing field was $\sim 31$~s,
dominated by losses in the low-field region between the AFP magnet and
SCS. In the spin flipper on state, the lifetime of correctly polarized
UCN was $\sim 17$~s, and the lifetime of incorrectly polarized UCN was
$\sim 44$~s.

At the immediate conclusion of a $\beta$-decay run, a depolarization
run (runs A3/B3, A6/B6, A8/B8, A11/B11) was conducted to measure the
depolarized fraction of the UCN population in the decay trap via the
following procedure.  First, the gate valve was closed, the proton
beam was gated off, and the switcher valve was re-configured such that
the guides downstream of this valve (i.e., from the switcher valve all
the way through the decay trap) were connected to a $^3$He UCN
detector (see the discussion in Section
\ref{sec:experiment_source_guides}).  The state of the AFP
spin-flipper was unchanged from its state during the immediately
preceding $\beta$-decay run.  At this point, UCN of the ``correct''
spin state in the experimental volume could exit the geometry
through the 7-T polarizing field, which now served as a spin-state
analyzer, and were counted in the UCN detector located at the switcher
valve.  This cleaning phase lasted 25 s, and the number of counts
recorded in the UCN detector during this time interval was
proportional to the number of correctly polarized UCN present in the
experimental geometry

Following this cleaning phase, the state of the spin-flipper was
changed.  This then permitted those UCN of the ``wrong'' spin state
located downstream of the spin-flipper, which until now had been
trapped within this volume by the 7-T polarizing field, to exit this
volume through the 7-T field, and to be counted in the UCN detector.
This counting during the unloading phase was performed for $\sim 200$
s, which provided for a measurement of both the number of wrong
spin-state neutrons as well as a measurement of the UCN detector
background on a depolarization run-by-run basis.

\subsection{Experiment Geometries}
\label{sec:measurements_geometries}

During the data-taking runs in 2008--2009 for the results reported in
this article, the experiment was operated in four different geometries
with different decay trap end-cap and MWPC entrance and exit window
foil thicknesses, to study key systematic corrections and
uncertainties related to energy loss in, and backscattering from,
these foils.  The foil thicknesses for these four different
experimental geometries, termed Geometries A, B, C, and D, are given
in Table \ref{tab:geometries}.  The number of
$\beta$-decay events collected in each Geometry passing all of the
analysis cuts detailed later in this article are also listed there.

Note that although the foil thicknesses for Geometries C and D were
identical, we defined separate geometries for these data-taking
periods because the UCN transport guides in the region between the APF
spin-flip region and the decay trap (i.e., the circular and
rectangular guides, see Section \ref{sec:experiment_source_guides})
were upgraded from (bare) electropolished Cu in Geometry C to
DLC-coated electropolished Cu in Geometry D.  This change to a guide
system with a higher effective UCN potential in the region downstream
of the AFP spin-flip region resulted in a different velocity spectrum
for those UCN stored in the decay trap, the details of which were
important for the interpretation of measurements of the UCN
polarization, described below.

\subsection{Polarization Measurements}
\label{sec:measurements_polarization}

\begin{figure}
\begin{center}
\includegraphics[scale=1.17]{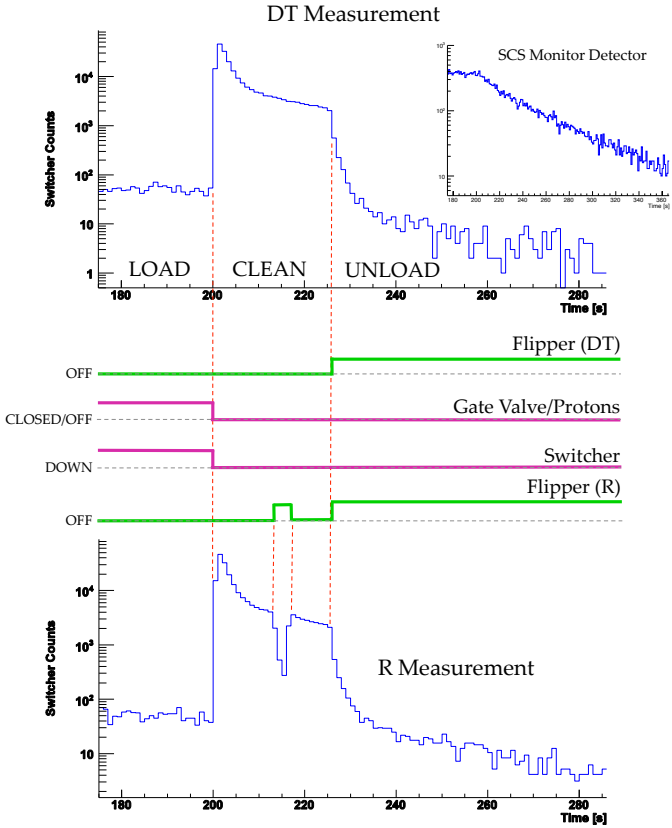}
\caption{(Color online) Arrival time
spectra from the switcher UCN monitor detector (two primary figures)
and the SCS monitor detector (top inset figure) characteristic of
depolarization trapping (DT) and reloaded background (R) measurements.
(The start time $t=0$ corresponds to the start of the
load interval or, conceptually, 200 s prior to the end of a
$\beta$-decay run.)  The states of the gate valve, switcher, and the
spin flipper during the load, clean, and unload intervals are
indicated schematically.  (With the switcher in the ``down'' state,
UCN located between the switcher and the decay trap are transported to
the switcher UCN monitor detector; in the opposite state, UCN are
transported from the SD$_2$ source to the electron spectrometer.)  The
timing and states depicted here are for the case of spin-flipper-off
loading. Flipper-on loading measurements reverse the state of the spin
flipper.}
\label{fig:polarimetry_figure}
\end{center}
\end{figure}

A pair of depolarization measurements for each spin state, i.e. a
D$^-$ run (following the loading of spin-flipper-off spin states
during the preceding $\beta$-decay run) and a D$^+$ run
(following the loading of spin-flipper-on states) as
described in Section \ref{sec:measurements_run_cycle_procedure},
provide, in principle, an \textit{in situ} measurement of the UCN
polarization at the end of the associated $\beta$-decay interval. This
pair of measurements automatically incorporates all depolarization
mechanisms including, in the case of flipper-on loading, spin flipper
inefficiency. Fig.\ \ref{fig:polarimetry_figure} depicts the
arrival time spectra in the switcher UCN monitor
detector and the decay trap UCN monitor detector (hereafter, SCS
monitor detector) characteristic of a depolarization measurement
during each of the ``load'', ``clean'', and ``unload'' intervals.  The
states of the gate valve, switcher, and the spin flipper during each
of these intervals are indicated schematically there.

Determination of the equilibrium polarization at the
end of the ``load" interval (corresponding to a time $t=200$ s in Fig.\
\ref{fig:polarimetry_figure}) was accomplished by using the switcher
UCN detector to count both the number of correctly polarized UCN in
the decay trap, $\mathscr{D}_p(t=200~\text{s})$, during the "clean"
interval, and then by changing the state of the spin flipper to count the
number of incorrectly polarized UCN in the decay trap,
$\mathscr{D}_d(t=225~\text{s})$, during the ``unload" interval. From
these signals, it is possible to extrapolate back to a depolarized
fraction $\xi \approx
\frac{\mathscr{D}_d(t=200~\text{s})}{\mathscr{D}_p(t=200~\text{s})}$
from which detection efficiencies cancel to first order, and which
provides the equilibrium polarization via $P=1 - 2 \xi$. The
extrapolation procedure requires knowledge of the appropriate storage
lifetimes for correctly and incorrectly polarized neutrons in the
system, obtained from the SCS monitor detector.

Models of the UCN transport confirm the intuitive expectation that the
time dependence of the ``clean" and ``unload" switcher detector
signals is characterized by double exponential behavior: the shorter
time constant is associated with emptying the guide system between the
switcher detector and the narrow rectangular guide to the decay trap,
while the longer time constant is associated with emptying the decay
trap through this rectangular guide. Analysis of the arrival time
spectra generated by the $D\pm$ measurements which formed part of the
beta asymmetry run cycle was accomplished by first fitting the
switcher detector timing spectrum during the clean interval to a
double exponential plus background (where the background was
determined from the last 100 s of the unload interval).  This
established the two amplitudes and associated time constants
$A^{(+/-)}_1, \tau^{(+/-)}_1$ and $A^{(+/-)}_2, \tau^{(+/-)}_2$ which
characterize the population of correctly polarized UCN in the system
at the end of the $\beta$-decay (loading) interval, where $-$
corresponds to flipper-off loading and $+$ corresponds to flipper-on
loading. Similarly, fitting the unload interval determined the
amplitude $A^{(+/-)}_<$ associated with the smaller time constant
$\tau^{(+/-)}_<$ and the amplitude $A^{(+/-)}_>$ associated with the
larger time constant $\tau^{(+/-)}_>$, which ideally characterize the
population of incorrectly polarized UCN present at the end of the
cleaning interval. In order to extrapolate this population back to the
end of the $\beta$-decay measurement interval, the storage lifetimes
$\tau_{+/-}$ of UCN trapped in the system due to their spin state
relative the state of the spin flipper must also be determined, where
here $+$ ($-$) corresponds to the storage lifetime of UCN whose spins
are parallel (anti-parallel) to the local magnetic field and are thus
trapped downstream of the spin flipper when it is off (on). This was
accomplished by fitting the unload interval of the SCS monitor timing
spectrum with a single exponential plus background. Note that $\tau_+$
is determined during a flipper-on loading (D$^+$) depolarization
measurement while $\tau_-$ is determined during a flipper-off loading
depolarization (D$^-$) measurement.
Monte Carlo
studies indicated that using these storage lifetimes to capture the
average behavior of the depolarized population during the clean
interval introduced no significant bias (at the current level of
precision) to the extrapolation of this depolarized population back to
the $\beta$-decay measurement.

In an ideal depolarization measurement, the cleaning interval is made
of sufficient length that contributions to the signal observed during
the unloading interval from correctly polarized UCN which are not
trapped when the spin flipper changes state are negligible. If the
number of free correctly polarized UCN is large and the depolarized
signal sufficiently small, however, waiting long enough for adequate
cleaning can reduce the incorrectly depolarized signal to levels below
the measurement threshold.  Since this was the case for the UCNA
geometries utilized in the 2008--2009 run period, the cleaning time
$\Delta$ was set to 25 s, just long enough to resolve $\tau_1$ and
$\tau_2$.  This enhanced the depolarized signal but necessitated
separate measurements to determine the correctly polarized background
in the unload timing spectrum, which for this clean interval was on
the same order as the depolarized signal.  In particular, depolarized
UCN coming from the decay volume are expected to appear as part of the
$A_<$ component, but the short cleaning time created a non-negligible
population of correctly polarized UCN in the guides between the spin
flipper and the polarizing field which are not trapped by the spin
flipper and which enter the decay trap before being detected in the
switcher detector, causing them to appear as part of $A_<$.

In order to correct for this \textit{reloaded} (R)
background, \textit{ex situ} measurements, denoted $R^{(+/-)}$, were
performed.
In these measurements, whose characteristic switcher
detector timing spectrum is shown in
Fig.~\ref{fig:polarimetry_figure}, thirteen seconds
prior to the start of the unloading phase the spin flipper state was
changed for three seconds in order to trap an additional reloaded
population, which then contributed to the amplitude
$\widetilde{A}^{(+/-)}_>$ determined from the unload phase of the
corresponding reload measurement. With this additional observable, the
reload-corrected polarizations were determined via
\begin{widetext}
\begin{eqnarray}
P^- &=& 1 - 2 \frac{e^{\Delta/\tau_{+}} \, (\tau^-_> - \tau^-_<) \,
\left\{ \zeta^-_1 \, (1-\mathfrak{r}^-) \, \left[A_< ^-- \mathscr{N}_-
\widetilde{A}_<^-\right] + A_<^- \right\}}{\zeta^-_2 \mathscr{D}_\mathrm{p}}
~~~~~\text{(flipper-off loading)} \notag
\label{eq:poleqs} \\
P^+ &=& 1 - 2 \frac{e^{\Delta/\tau_{-}} \, (\tau^+_> - \tau^+_<) \,
\left\{ \zeta^+_1 \, (1-\mathfrak{r}^+) \, \left[A_<^+ - \mathscr{N}_+
\widetilde{A}_<^+\right] + A_<^+\right\}}{\zeta^+_2 \mathscr{D}_\mathrm{p}}
~~~~~\text{(flipper-on loading),} \notag \\
\end{eqnarray}
\end{widetext}
where $\mathfrak{r}$ is a Monte Carlo calculated parameter on the
order of 0.60 needed to account for the presence of an extra
population between the spin flipper and the 7.0-T region trapped
by the three second flipper cycle,
\begin{equation}
\zeta_1^{(+/-)} = e^{-(\Delta_1 + \Delta_2)/\tau_2^{(+/-)}} \;
e^{\Delta_2/\tau_{(-/+)}}
\end{equation}
(where $\Delta_1$ is the length of the flipper cycle and $\Delta_2$ is
the interval between the end of the flipper cycle and the start of the
unload phase) is a scaling factor which accounts for the evolution of
the reloaded population trapped during the flipper cycle and corrects
for the larger population of correctly polarized UCN present to be
reloaded during the flipper cycle, $\mathscr{D}_\mathrm{p}$ is the
total number of background-subtracted counts recorded during the clean
interval, $\zeta_2$ is a factor which uses $A_1$,
$\tau_1$, $A_2$, and $\tau_2$ to extrapolate $\mathscr{D}_\mathrm{p}$
to the number of counts which would be observed for an infinitely long
cleaning period, and $\mathscr{N}$ is a normalization factor.
Values of $P^+$ and $P^-$ were obtained separately for the 2008 and
2009 data sets by summing all corresponding D and R runs and applying
Eq.~(\ref{eq:poleqs}). Since there was no statistically significant
difference at the 1$\sigma$ level between any of the four
measurements, a single reload-corrected value $P$ for the polarization
was obtained by performing a weighted average over the four
measurements.

Spin flipper inefficiency decreases the UCN polarization for
flipper-on loading, resulting in the expectation that $P^- > P^+$.
The resulting decreased polarization for the case of flipper-on
loading due to the spin flipper inefficiency is the actual
polarization of the UCN population stored in the decay trap, and no
further correction to the $P^+$ value is required.  However, the spin
flipper inefficiency also leads to a (smaller) increase in $P^-$ since
correctly polarized UCN which should remain trapped during the
unloading phase are freed when they are not flipped, adding to the
observed unload signal. Since this population is generated after the
$\beta$-decay measurement interval it requires a correction, which
will decrease the value of $P^-$. The accumulated data limited this
correction to be no larger than $\sim$0.15\% of the total
polarization, and error bars on $P$ were expanded accordingly. It is
also possible to have depolarized UCN populations whose storage
lifetime in the system is much shorter than the depolarization
measurement time, and which therefore have a low efficiency for
detection in a depolarization measurement. Neutrons with sufficient
energy to surmount the potential barrier presented by the 7.0-T
polarizing field (which may therefore enter the experiment in the
wrong spin state) and initially polarized UCN with energies higher
than the material potential of the decay trap walls (which can survive
in the system when confined to trajectories that sample the walls at
sufficiently oblique angles) are examples of such populations. Monte
Carlo calculations estimating the effect of these populations on the
neutron polarization indicated a negligible contribution at the
current level of precision, due largely to the short residency times
that such UCN posses. Expanding the error bars on $P$ to account also
for variations in the polarization due to pulsed loading (estimated to
be on the order of 0.04\% of $P$), a 1$\sigma$ lower limit of $P >
0.9948$ was determined.

\section{Calibration and Reconstruction}
\label{sec:analysis}

We now turn to a discussion of our data analysis and energy
calibration procedures.  We begin by defining the various possible
event types in the experiment, the selection rules for the observable
event types, and a desciption of our position recontruction algorithm
using the MWPC signals.  We also discuss our data ``blinding''
procedure, which ultimately resulted in ``blinded'' asymmetries which
were scaled by a randomly chosen scaling factor at the
$\mathcal{O}$(0--5\%) level.  Next, we discuss our energy calibration
procedures for the scintillator and the MWPC, and then compare for the
different event types our reconstructed energy spectra with simulated
Monte Carlo spectra.  Finally, we conclude this section with a
discussion of our procedure for the assignment of the initial energy
of the electron.

Hereafter, in our discussions of the data analysis of the detector
signals, we will refer to the two electron detectors as the ``East''
and the ``West'' detectors, corresponding to their actual physical
locations in the UCNA Experiment.

\subsection{Event Type Definitions}
\label{sec:analysis_event_type_definitions}

\begin{figure}
\includegraphics[angle=270,scale=0.36,bb = 53 61 448 739,clip=]
{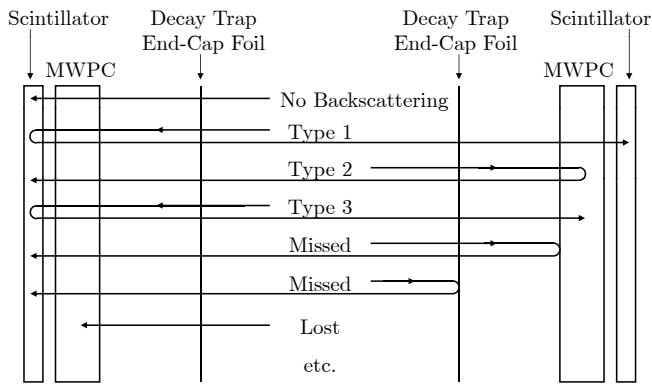}
\caption{Schematic diagram of the
various event types defined in the text.}
\label{fig:event_types_schematic}
\end{figure}

Measurement of the $\beta$-asymmetry requires an accurate
determination of the decay electron's initial direction of incidence.
This determination is complicated by backscattering effects, some of
which are not detectable.  We define the various classes of event
types, shown in Fig.\ \ref{fig:event_types_schematic}.

\begin{itemize}

\item
No backscattering events: Events in which an
electron, incident initially on one of the detectors, does not
backscatter from any element of that detector, and then generates a
two-fold PMT trigger in that side's scintillator.

\item
Type 1 backscattering events: Events in which an electron, incident
initially on one of the detectors, generates a two-fold PMT trigger in
that side's scintillator, backscatters from that scintillator, and
then generates a two-fold PMT trigger in the opposite side's
scintillator.  Note that a measurement of the relative
time-of-flight between the two scintillators' two-fold PMT triggers
determines the initial direction of incidence for Type 1
backscattering events.

\item
Type 2 backscattering events: Events in which an electron, incident
initially on one of the detectors, deposits energy above threshold in
that side's MWPC, backscatters from some element of that side's MWPC
(e.g., gas, wire planes, or the exit window) or the scintillator
without generating a two-fold PMT trigger (e.g., backscattering from
the scintillator's dead layer, or triggering only one PMT), and is
then detected in the opposite side's scintillator.  As can be seen in
Fig.\ \ref{fig:event_types_schematic}, the initial direction of
incidence of a Type 2 backscattering event would be misidentified
using only scintillator two-fold PMT trigger information.

\item
Type 3 backscattering events: Events in which an electron, incident
initially on one of the detectors, generates a two-fold PMT trigger in
that side's scintillator, backscatters from that side's scintillator,
deposits energy above threshold in the opposite side's MWPC, and is
then stopped in some element of the opposite side's MWPC or
scintillator without generating a two-fold PMT trigger (i.e., in the
dead layer, or triggering only one PMT).  Note that Type 2 and Type 3
backscattering events cannot be distinguished using only scintillator
two-fold PMT trigger information and a threshold cut on the MWPC
response (i.e., the Type 2 and Type 3 events depicted in Fig.\
\ref{fig:event_types_schematic} would not be distinguishable).

\item
Missed backscattering events: Events in which an electron, incident
initially on one of the detectors, backscatters from either the decay
trap end-cap foil or that side's MWPC without depositing energy above
threshold (e.g., from the entrance window, or from the gas in the
region between the cathode plane and the entrance window), and is then
detected in the opposite side's MWPC and scintillator.  Note that
missed backscattering events cannot be identified experimentally.

\item
Lost events: Events in which an electron, incident initially on one of
the detectors, deposits significant energy in a decay trap end-cap
foil and/or the MWPC, and does not generate a two-fold PMT trigger in
either of the scintillators.  Note that because these events do not
generate a DAQ event trigger, they cannot be identified
experimentally.

\end{itemize}

No backscattering events and Missed
backscattering events cannot be distinguished experimentally, and are,
hereafter, termed ``Type 0'' events.  Based on
scintillator information alone, Type 2 backscattering events cannot be
distinguished from Type 3 backscattering events.  Thus, we will refer
to these types of events as ``Type 2/3'' events.  Later in Section
\ref{sec:asymmetry_analysis_backscattering_choices_type23} we will
discuss the separation of Type 2/3 events using MWPC information and
simulation input.
Finally, Lost events cannot, of course, be reconstructed and can only
be corrected for in simulation.

\subsection{Run Selection}
\label{sec:analysis_run_selection}

Proton beam delivery constraints and other experimental issues
prevented on occasion the accumulation of complete octet data sets
during normal $\beta$-decay production running.  In the absence of a
complete octet, runs forming a quartet (i.e., runs A1--A12 or B1--B12
in Table \ref{tab:octet_structure}) or spin-pair (i.e., A1--A6,
A7--A12, B1--B6, or B7--B12) were retained for analysis.
Runs with clear detector issues (e.g., noisy channels associated with
the MWPC cathode planes) were discarded.

\subsection{Data Blinding}
\label{sec:analysis_blinding}

We performed a blinded analysis of our asymmetry data by applying
separate spin-dependent randomly chosen scaling factors to the two
detectors' count rates, thus effectively adding an unknown scaling
factor to the measured asymmetry.  This was implemented via the
following procedure.  First, note that the detector count rates were
based on a global event-by-event clock time which was defined, as
described earlier in Section \ref{sec:experiment_electronics}, by
counting a 1 MHz clock in a scaler.  Second, we generated two random
scale factors, $f_1$ and $f_2$, which were constrained to be between
$1.00 \pm (0.05 \times 0.04)$, where 0.04 represents the approximate
value of the measured asymmetry.

For runs with the AFP spin-flipper on ($+$ spin state), we then
scaled the east and west detector clock times, $t_E^+$ and
$t_W^+$, by these scale factors according to
\begin{equation}
t_E^+ = t \cdot f_1,~~~~~~~~~~
t_W^+ = t \cdot f_2,
\label{eq:blinded_time_stamps_1}
\end{equation}
where $t$ denotes the true global clock time.  Similarly, for those
runs with the AFP spin-flipper off ($-$ spin state),
\begin{equation}
t_E^- = t \cdot f_2,~~~~~~~~~~
t_W^- = t \cdot f_1.
\label{eq:blinded_time_stamps_2}
\end{equation}
In calculating the super-ratio for a spin-state run pair according to
Eq.\ (\ref{eq:individual_super_ratio}), the resulting blinded
measured asymmetry, $A_{\text{blind}}$, is then a function of
the blinded and true super-ratios, $R_{\text{blind}}$ and
$R_{\text{true}}$, according to
\begin{equation}
A_{\text{blind}} = \frac{1 - \sqrt{R_{\text{blind}}}}
{1 + \sqrt{R_{\text{blind}}}} =
\frac{1 - f\sqrt{R_{\text{true}}}}
{1 + f\sqrt{R_{\text{true}}}},
\end{equation}
where $f \equiv f_2/f_1$.

All of our analysis was performed with these blinded asymmetries.
Note that this did not impact our assessment of our energy
reconstruction algorithms (which do not, of course, depend on the
asymmetry), or our assessment of our systematic corrections for
backscattering and the $\cos\theta$-dependence of the acceptance, as
these were calculated (and subsequently benchmarked) in units of
$A_0$.

\subsection{Data Quality Cuts and Live Time Definition}
\label{sec:analysis_data_quality_cuts}

We subjected each run to a number of so-called ``global'' and
``event-by-event'' data quality cuts, resulting in the
removal of either a consecutive range of events or a single event,
respectively.  First, we note that there were sporadic corruptions to
the data stream resulting from malfunctioning DAQ electronics modules.
These electronics problems resulted in either the
corruption of all subsequent events following the occurence of the
problem, or the corruption of only a single isolated event.
Electronics problems resulting in the corruption of all subsequent
events included misalignments of the VME data banks (e.g., of the QADC
data bank relative to the PADC data bank) and sudden shifts in the TDC
channel peak positions of the two detectors' two-fold PMT self-trigger
timing peaks.  After the identification of either of these problems, a
global data quality cut was applied, resulting in the removal of all
subsequent events in that run.  Electronics problems which resulted in
the corruption of only a single event included corruptions to the
headers and/or footers of the PADC, QADC, or TDC event banks (e.g., an
event lacking a header or footer) and corruptions to the TDC bank
event counter relative to the \texttt{MIDAS} data acquisition event
counter.  An event-by-event data quality cut was then applied to those
events found to have either of these latter two types of electronics
problems.

A significant fraction of the data acquired during the
Geometries A and B running was discarded due to the above-described
electronics problems.  In particular, of the data acquired during the
second half of the Geometry A running and then during the entire
Geometry B running, $\sim 5$\% of the events lacked a header and/or
footer and up to $\sim 30$\% of the events suffered from the TDC bank
event counter problem.  In contrast, the fraction of events suffering
from electronics problems acquired during the Geometries C and D
running was small, with $<5$\% of the data affected by these
problems.

\begin{figure}
\includegraphics[scale=0.45,clip=]{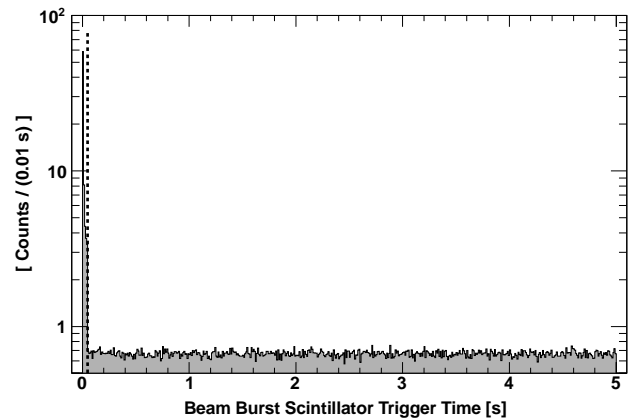}
\caption{Typical scintillator two-fold
PMT trigger rates during the 0.2 Hz proton pulse repetition
cycle.  For the clock timing shown here, each of the five
beam bursts comprising each proton pulse occur at $t =
0$, as the clock was reset to $t=0$ with each beam burst;
thus, under this reset scheme, all of the events
collected during the five beam bursts appear between 0 and 0.05 s (as
noted in Section \ref{sec:experiment_source_guides}, the time between
beam bursts was 0.05 s).  The dashed line indicates
the cut at 0.05~s.}
\label{fig:beam_burst_scintillator_trigger_time}
\end{figure}

Other global data quality cuts included the removal of all events
between 0.00 s to 0.05 s after each proton beam burst.  Typical
scintillator two-fold PMT trigger rates during the 0.2~Hz proton beam
beam pulse repetition cycle are shown in Fig.\
\ref{fig:beam_burst_scintillator_trigger_time}.  As can be seen there,
the peak scintillator trigger rates were up to a factor of $\sim 80$
higher during the proton beam bursts.  This figure illustrates one of
the merits of a pulsed-spallation-source of UCN, namely, that the
experiment can be performed in a low background environment (i.e.,
ambient backgrounds only) during the time between the proton beam
bursts.  Another global data quality cut included the
removal of events from $\beta$-decay runs occuring during time
periods when the rate on the $^3$He UCN monitor detector located near
the gate valve dropped below some threshold were vetoed, so as not to
degrade the signal-to-background ratio.

After application of the above-described global data
quality cuts, we then computed on a run-by-run basis a ``live time''
for each detector, defined to be the sum of the (blinded) clock times
of the run segments surviving the above-described global data-quality
cuts.  Specifically, if a run segment between event $M$ and event $N>M$
survived these global data-quality cuts, the corresponding east
and west detector live times, $T_E$ and $T_W$, for this run
segment were calculated as
\begin{equation}
T_E = t_E^N - t_E^M,~~~~~~~~~~T_W = t_W^N - t_W^M,
\end{equation}
where $t_E^M$ ($t_W^M$) and $t_E^N$ ($t_W^N$) denote the blinded time
stamps for the east (west) two-fold PMT trigger for events $M$ and
$N$, respectively, defined previously in Eqs.\
(\ref{eq:blinded_time_stamps_1}) and (\ref{eq:blinded_time_stamps_2}).


We note, however, an exception to the above-described procedure under
which we applied an event-by-event data quality cut to events with a
TDC event counter problem (with no subsequent correction to the live
time).  In particular, as was already noted, up to $\sim 30$\% of the
events collected during the Geometry B running suffered from this
problem.  Further, the fraction of corrupted events recorded by each
detector differed for the two neutron spin states, thus biasing the
extracted asymmetry.  Thus, it was necessary to correct the Geometry B
live times.  As discussed later in Section
\ref{sec:uncertainties_live_time}, we corrected for this Geometry B
live time problem using background gamma-ray events, which were
uncorrelated with the neutron $\beta$-decay events.  The correction
factors to the live times were then defined for each detector on a
run-by-run basis to be the ratio of the number of gamma-ray events
surviving the event-by-event TDC event counter cut to the total number
of recorded gamma-ray events.

\subsection{Event Reconstruction and Identification}
\label{sec:analysis_event_reconstruction}

Those events surviving the above data-quality checks were then
reconstructed and identified (i.e., tagged as a Type 0, Type 1, or
Type 2/3 event) based on the TDC measurement of the two detectors'
two-fold PMT coincidence trigger time-of-flight, the pulse height
(PADC) in the MWPCs, and the pulse height (PADC and QADC) and timing
(TDC) information from the various muon-veto detectors.  The available
detector information is described in more detail below.

\begin{figure}
\includegraphics[scale=0.46,clip=]{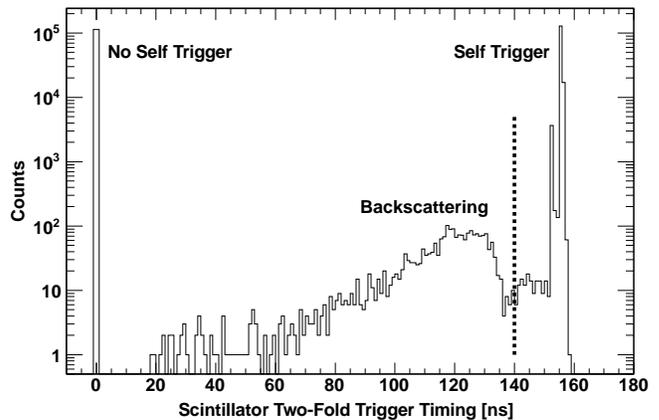}
\caption{Typical two-fold coincidence trigger
relative time-of-flight spectrum for one of the scintillator detectors
from a $\beta$-decay run.  See text for details.}
\label{fig:two_fold_time_of_flight}
\end{figure}

\subsubsection{Scintillator Timing Information}
\label{sec:analysis_event_reconstruction_scintillator_timing}

The logic outputs from the PLUs defining the two-fold PMT triggers for
the two detectors were routed to TDC channels, forming the
\texttt{START}s for their respective channels.  A copy of the main
event trigger was delayed by $\sim 155$ ns and was used to generate
the \texttt{COMMON STOP} for the TDC.  A typical two-fold coincidence
trigger time-of-flight spectrum for one of the detectors (for example,
of the West detector) is shown in Fig.\
\ref{fig:two_fold_time_of_flight}.  Main event triggers generated by a
West detector two-fold trigger appear at the self-trigger delay time
of $\sim 155$ ns, whereas main event triggers generated by a East
detector two-fold trigger, with no later arrival of a West detector
two-fold trigger, appear at the peak at 0 ns (i.e., a ``time-out'' for
that TDC channel).  Those events forming the broad secondary peak from
$\sim$ 20--140 ns, separated from the self-trigger peak by the dashed
line, correspond to main event triggers generated by a East detector
two-fold trigger, with the later arrival of a West detector two-fold
trigger.  That is, Type 1 backscattering events incident initially on
the (opposite-side) East detector comprise this broad peak.
The backscattering cut line (the dashed line in Fig.\
\ref{fig:two_fold_time_of_flight} at 140 ns) placed approximately 15 ns
before the self-timing peak is justified by the fact that the minimum
time-of-flight (i.e., straight-line trajectory) for an 800 keV
electron ($\beta$ = 0.919) traveling the 4.4 m distance between the
two scintillators is 15.9 ns.

Note that for the TDC dynamic range setting of $\sim
140$ ns shown in Fig.\ \ref{fig:two_fold_time_of_flight}, a ``true''
Type 1 event with a coincidence time-of-flight greater than $\sim 140$
ns would not appear in the opposite-side detector's coincidence timing
spectrum, but would instead appear in its ``time-out'' peak;
therefore, such a ``true'' Type 1 event would be misidentified in data
analysis as a Type 2/3 event.  However, as calculated in simulation,
the fraction of Type 1 events with a time-of-flight greater than $\sim
140$ ns is small, $\sim 2$\%,
and results in a negligible effect on the asymmetry.

\subsubsection{MWPC Spectra and Particle Identification}
\label{sec:analysis_event_reconstruction_mwpc_spectra}

\begin{figure}
\includegraphics[scale=0.90,clip=]{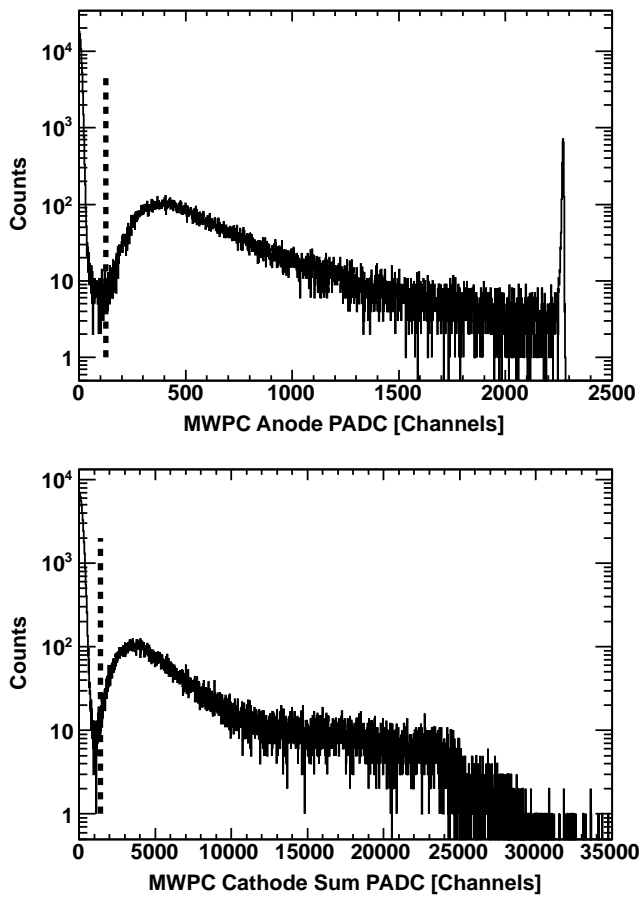}
\caption{ Typical MWPC anode and summed cathode pulse height spectra
from a $\beta$-decay run.  The dashed lines indicate the cut
positions.  The peak in the anode spectrum at channel $\sim 2250$ is
overflow.  Because the cathode sum is obtained via
summation over all of the individual cathode channels, channel
overflows are distributed throughout the summed spectrum.}
\label{fig:anode_cathode_spectra}
\end{figure}

As noted earlier, the MWPC anode and cathode pulse height signals were
read out on PADC channels during a $\sim 12$ $\mu$s window after each
event trigger.  Typical anode and cathode (summed over all of the
individual cathode channels) pulse height spectra during a
$\beta$-decay run are shown in Fig.\ \ref{fig:anode_cathode_spectra}.
For scintillator event triggers, the majority of the events appeared
in the pedestal, and were tagged as gamma ray events.
Those events satisfying a cut on either the anode PADC
channel number or the summed cathode PADC channel number, indicated
by the dashed lines there, were identified as charged particles
(electrons or muons).  Electron hits in a particular scintillator were
then further separated from muon hits by the requirement of no
coincident hits in any of the same-side muon veto detectors.
Note that the coincidence window for the
plastic scintillator (drift tube) muon veto detectors was defined by the
$\sim 140$~ns ($\sim 12$~$\mu$s) QADC (PADC) data acquisition gate.

\begin{figure}
\includegraphics[scale=0.47,clip=]{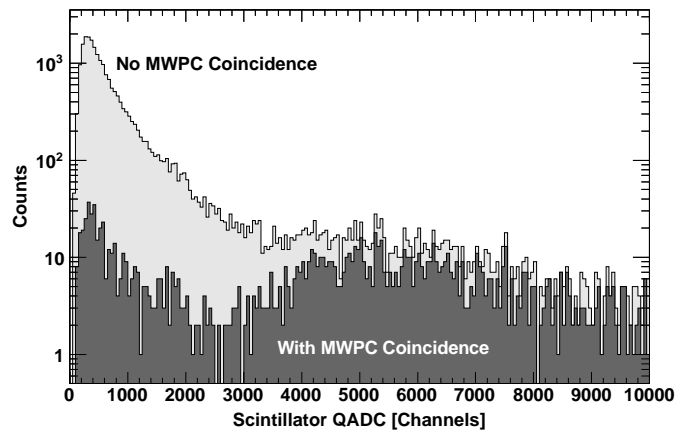}
\caption{Typical scintillator spectrum for a background run, without
(light shaded) and with (dark shaded) the requirement of a MWPC
coincidence cut.
See the text for a description of the features in the
resulting concidence spectrum.}
\label{fig:background_mwpc_coincidence}
\end{figure}

After removal of the events occuring during the pulsed proton beam
bursts, the requirement of a MWPC-scintillator coincidence cut further
reduced the total integrated ambient background rate in each
scintillator by a factor of $\sim 40$ from $\sim 50$
s$^{-1}$ to $\sim 1.2$ s$^{-1}$ over the range of QADC channels
corresponding to the neutron $\beta$-decay energy
spectrum\footnote{The total background rate without
the MWPC coincidence cut, and thus the background suppression factor
under the coincidence cut, varies strongly with the scintillator
threshold.  The numbers quoted above are from a typical background
run.}.  This is illustrated in Fig.\
\ref{fig:background_mwpc_coincidence}, where the (uncalibrated)
scintillator spectrum in QADC channels for a typical background run is
shown with and without the requirement of a MWPC-scintillator
coincidence for a typical background run.  The resulting broad peak
centered approximately at channel 5000 is from minimum-ionizing
cosmic-ray muons.  The origin of the surviving low-energy tail
(channels $\alt 2000$) was demonstrated in Monte Carlo studies to be
the result of cosmic-ray muon interactions with spectrometer materials
(e.g., in the SCS magnet's infrastructure) located near the MWPC and
scintillator detector, such as $\delta$-ray production and Compton
scattering of muon-induced gamma rays.

\subsubsection{Event Type Classification}
\label{sec:analysis_event_reconstruction_event_type_classification}

\begin{table}[t!]
\caption{Detector selection criteria (scintillator two-fold PMT TDC
trigger timing and MWPC pulse height) for the identification of Type
0, Type 1, and Type 2/3 events.  A $\surd$ indicates the requirement
of a valid detector hit, whereas a $\times$ indicates the requirement
of no detector hit.  The notation (*) indicates the requirement of an
earlier detector hit relative to the opposite-side detector hit.}
\begin{ruledtabular}
\begin{tabular}{lcccc}
\multirow{2}{*}{Event Type}& \multicolumn{2}{c}{East Detector}&
  \multicolumn{2}{c}{West Detector} \\
& TDC& MWPC& TDC& MWPC \\ \hline
Type 0 East Incidence& $\surd$& $\surd$& $\times$& $\times$ \\
Type 0 West Incidence& $\times$& $\times$& $\surd$& $\surd$ \\ \hline
Type 1 East Incidence& $\surd$ (*)& $\surd$& $\surd$& $\surd$ \\
Type 1 West Incidence& $\surd$& $\surd$& $\surd$ (*)& $\surd$ \\ \hline
Type 2/3 East Trigger& $\surd$& $\surd$& $\times$& $\surd$ \\
Type 2/3 West Trigger& $\times$& $\surd$& $\surd$& $\surd$ \\
\end{tabular}
\end{ruledtabular}
\label{tab:event_type_tagging}
\end{table}

Electron events were classified as Type 0, Type 1, or Type 2/3 events
according to the detector selection criteria listed in Table
\ref{tab:event_type_tagging}.  Electron events not satisfying any
of these criteria were discarded.

For Type 0 events, the assignment of the initial direction of
incidence (i.e., on the East or West detector) was, of course,
trivial.  For Type 1 events, the initial direction was determined from
the TDC two-fold trigger time-of-flight spectra, as discussed earlier
in Section
\ref{sec:analysis_event_reconstruction_scintillator_timing}.  At this
stage of the analysis, Type 2/3 events were identified as such but
were only tagged with the triggering scintillator side (i.e., not yet
assigned an initial direction of incidence).

\subsection{MWPC Position Reconstruction}
\label{sec:analysis_position}

\subsubsection{Algorithm}
\label{sec:analysis_position_algorithm}

\begin{figure*}
\includegraphics[scale=0.70,clip=]{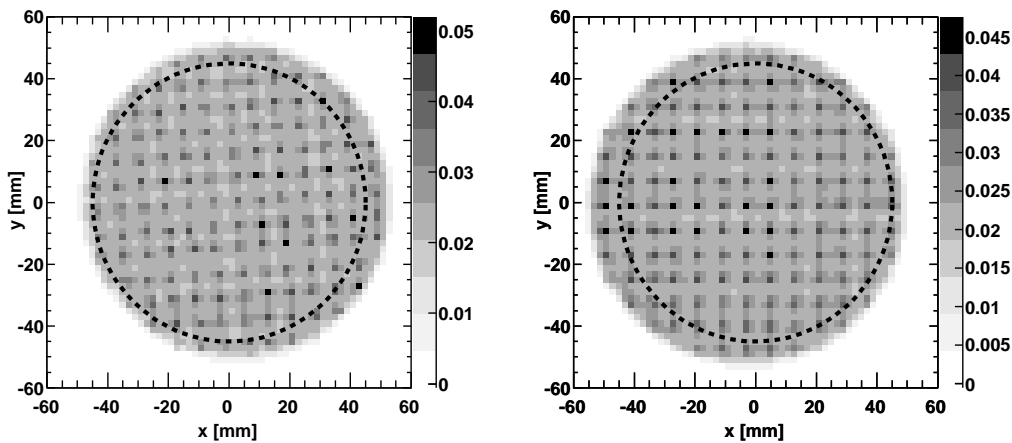}
\caption{Distributions of $(x,y)$ position distributions [for the
coordinate system choice (4) described in the text] for Type 0 neutron
$\beta$-decay events (left panel: East detector, right panel: West
detector).  The dashed circles denote the nominal 45 mm fiducial
volume radius cut.  The regularly-spaced ``spikes'' are an artifact of
the position reconstruction algorithm: for events in which two
adjacent cathode channels recorded overflow values, the algorithm
reconstructed the positions to sharp peaks halfway between the two
wires' positions (as discussed in Section
\ref{sec:analysis_position_algorithm}).}
\label{fig:east_west_xy_beta_decay}
\end{figure*}

The transverse $(x,y)$ position of an event was reconstructed from the
charge distribution on the MWPC cathode plane wires.  The position
reconstruction algorithm that was employed for the calculation of the
position from the signals that were digitized by the PADC modules
began by identifying, on each of the $x$- and $y$-planes, the wire
with the maximum PADC readout value above pedestal.  The position was
then reconstructed from the PADC reading of this ``maximum wire'' and
those of the immediately adjacent wires, with all other wires ignored,
by fitting these three wires' PADC readings as a function of their
$x$- or $y$-positions to a Gaussian shape\footnote{Although the
distribution of charge across the cathode planes is technically not
Gaussian (for the idealized case of a charge located
above an infinite grounded conducting cathode plane, the shape of the
induced charge distribution on the cathode plane would be equivalent
to that obtained via the method of images; a more realistic model for
the finite-length wires on the finite-size cathode plane would require
a finite-element analysis calculation), employing a Gaussian fit was
taken to be sufficient given that: (1) small-Larmor-radius calibration
source spots (e.g., $^{139}$Ce) reconstructed to correctly-sized
spots, and were highly repeatable; and (2) the distribution of
$\beta$-decay events, as can be seen in Fig.\
\ref{fig:east_west_xy_beta_decay}, is fairly uniform (as expected),
which then sets a limit on any positioning reconstruction errors.}.
The positions for $\sim 90$\% of the neutron $\beta$-decay events
could be reconstructed with this algorithm.

The positions for the remaining $\sim 10$\% of the events were
reconstructed under various special circumstances.  For example, if
there was only one
wire with a PADC value above pedestal, the position was
assigned to that particular wire's coordinate.  Or if only one of the
wires directly adjacent to the maximum wire recorded a PADC value
above pedestal, or if the maximum wire was located on one of the
cathode plane edges and the adjacent wire recorded a PADC value above
pedestal, the position was determined under the assumption that the
width of the charge distribution was 0.75 times the wire spacing (as
determined from the $\sim 90$\% of the events reconstructed with the
three-wire Gaussian fitting algorithm).  In the event two or more
wires recorded overflow PADC values, the positions were reconstructed
to sharp peaks halfway between the two overflow wires' coordinates.

Finally, if the $x$- or $y$-position could not be reconstructed (e.g.,
if none of the wires on one or both the cathode planes recorded a PADC
value above pedestal), the $x$- or $y$-position was defined to be 0.0.
This was potentially the source of a systematic bias, because as is
discussed in detail later in Section
\ref{sec:analysis_energy_calibration}, the scintillator energy
reconstruction was position dependent.  However, the fraction of
events identified as charged particles failing either the $x$- or
$y$-reconstruction was small, $< 10^{-3}$, and is
consistent with our later estimate (Section
\ref{sec:uncertainties_neutron_generated_backgrounds}) of the MWPC
efficiency.  The origin of these events may have been electronic noise
on the MWPC anode channel correlated with a scintillator trigger, thus
resulting in the identification of the event as a charged particle
with no corresponding signal on the cathode PADC channels.

Note that all of the $(x,y)$ position spectra shown hereafter have
been projected back to the spectrometer's 1.0-T field region
from the 0.6-T field-expansion region.

\subsubsection{Position Cuts, Fiducial Cut, and Coordinate Systems}
\label{sec:analysis_position_cuts}

The reconstructed positions were used to define a fiducial volume in
order to reject events originating near the decay trap
collimator
(inner radius of 58.4 mm).  Examples of such problematic events include both
$\beta$-decay electrons (subject to $\cos\theta$-dependent acceptance
and backscattering effects) and background electrons from the Compton
scattering of gamma rays in the decay trap material.

For Type 0 events, we required the reconstructed radius on the
triggering scintillator side to satisfy a conservative
$r_{\text{trigger}} < 45$ mm cut.  The maximum Larmor diameter for an
endpoint energy electron in the 1.0-T spectrometer field was 7.76 mm;
thus, this 45 mm cut was safely located $\sim 2$ Larmor diameters from
the decay trap wall.  For events identified as Type 1 or 2/3
backscattering events, in addition to this $r_{\text{trigger}} < 45$
mm trigger-side cut, we also made a further cut on the two sides'
$\vec{x}_E - \vec{x}_W$ vertex difference to eliminate accidental
backgrounds, with this cut defined to be $|\vec{x}_E - \vec{x}_W| <
25$ mm.

To study possible systematic effects associated with the position
reconstruction, we considered four different coordinate systems.  (1)
Coordinates defined by their nominal positions reconstructed from the
just-described Gaussian fits to the wires' PADC readings: $\vec{x}_E =
(x_E,y_E)$ and $\vec{x}_W = (x_W,y_W)$.  (2) West-side coordinates
defined by their nominal positions, $\vec{x}_W$, but transformed
East-side coordinates $\vec{x}'_E = (x'_E,y'_E)$ based on fits to Type
1 backscattering data which provided for a detector-to-detector
coordinate system based on these backscattering measurements of the
detector-pixel-to-detector-pixel magnetic field map.  These fits were
of the form $\vec{x}'_E = \mathcal{R}\vec{x}_E + \vec{c}$, where
$\mathcal{R}$ denotes a rotation matrix, and $\vec{c} = (c_x,c_y)$ is
a constant offset vector.  (3) Coordinates defined by the complement
of (2) (i.e., East-side coordinates defined by their nominal
positions, but West-side coordinates transformed according to the Type
1 backscattering detector-pixel-to-detector-pixel magnetic field map).
(4) A ``split'' coordinate system, in which the rotation matrix
$\mathcal{R}$ was applied to one set of coordinates, with the constant
offset vector $\vec{c}$ split between the two detectors, such that
$\vec{x}'_E = \mathcal{R}\vec{x}_E + \vec{c}/2$, and $\vec{x}'_W =
-\vec{c}/2$.  The latter was chosen as the default coordinate system
in the analysis presented hereafter.  We
assess the systematic uncertainty associated with
our choice of a coordinate system in
Section \ref{sec:uncertainties_fiducial_cut}.

Sample $(x,y)$ position distributions [calculated for the coordinate
system choice (4) described in the above paragraph] extracted from the
East and West MWPC for Type 0 neutron $\beta$-decay events (after
background subtraction)
are shown in Fig.\ \ref{fig:east_west_xy_beta_decay}.  The
distribution of background events was previously shown to be uniform
over the MWPCs' sensitive areas \cite{plaster08}.

\subsection{Monte Carlo Simulation Programs}
\label{sec:analysis_energy_monte_carlo}

We developed two independent Monte Carlo simulations of the
experimental acceptance, based on the \texttt{GEANT4} 
(version 9.2) \cite{geant4}
and \texttt{PENELOPE}
(version 3)
\cite{penelope} simulation codes.  These were employed
extensively as input to our energy calibration procedures (described
next) and for calculations of the systematic corrections for
backscattering and the $\cos\theta$-dependence of the acceptance
(e.g., from suppression of the acceptance at large angles), discussed
in detail in Section \ref{sec:corrections}.  The performance of these
simulation programs was benchmarked previously in a series of
measurements of backscattering from beryllium, plastic scintillator,
and silicon targets and found to be accurate to within 30\% in their
predictions for the angular and energy distributions of the
backscattered electrons
\cite{junhua_thesis,martin03,martin06,hoedl_thesis}.

Both of these simulations included detailed geometric models for: (1)
the 3-m long decay trap and its end-cap foil geometry; (2) the MWPCs,
including their entrance and exit windows, the Kevlar fiber support
for the entrance windows, the anode and cathode wire planes, and the
100 Torr neopentane fill gas; (3) the two dead regions in the MWPC
(i.e., the two regions between the cathode planes and the
entrance/exit windows); (4) the plastic scintillator disc; (5) the
measured scintillator dead layer of 3.0-$\mu$m thickness
\cite{junhua_thesis}; and (6) the magnetic field in the decay trap and
field expansion regions.  Note that depending on the study of
interest, the magnetic field was modeled either as a perfectly uniform
1.0-T field in the decay trap region analytically connected (subject
to the $\vec{\nabla} \cdot \vec{B} = 0$ requirement) to the 0.6-T
field in the field-expansion region, or via bicubic spline
interpolation of a three-dimensional grid of the field profile (e.g.,
such as for the measured field profile shown in Fig.\
\ref{fig:scs_field_uniformity}).  In both the \texttt{GEANT4} and
\texttt{PENELOPE} simulations, charged particles were transported
through the magnetic field via Runge-Kutta integration of the
equations of motion.

For simulation studies of calibration sources, events were generated
isotropically into $4\pi$ from a fixed point, with a model for the
source foil covers (assumed to be 3.6-$\mu$m mylar).  For neutron
$\beta$-decay simulations, events were generated uniformly over the
decay volume.  The \texttt{GEANT4} events were emitted isotropically
into $4\pi$ and then weighted with a $W(\theta) \propto (1 +
A\beta\cos\theta)$ weight factor, whereas the \texttt{PENELOPE} events
were sampled from the full phase-space distribution of Eq.\
(\ref{eq:W_phase_space_distribution}).

\subsection{Visible Energy Calibration}
\label{sec:analysis_energy_calibration}

\subsubsection{Overview}
\label{sec:analysis_energy_calibration_overview}

The overarching goal of the energy calibration procedure was to
calibrate the quantity of scintillation light produced by an electron
which deposited a certain amount of ``visible energy'' in the
scintillator and to calibrate the electron energy deposition in the
MWPC fill gas.  We define the scintillator visible energy,
$E_{\text{vis}}$, to be the total energy loss in the scintillator
active region (i.e., beyond the dead layer) measured by the
photomultiplier tubes.  The scintillator visible energy is, of course,
not equal to the initial energy of the emitted $\beta$-decay electron,
due to reconstructable energy loss in one or more of the MWPCs' active
regions, and non-reconstructable energy loss in one or more of the
decay trap end-cap foils, one or more of the MWPCs' non-active
elements (e.g., windows, wire planes, and gas region between the
cathode planes and the windows), and in one or more of the
scintillators' dead layers.  The relationship between the measured
visible energy in the scintillator, the measured energy deposition in
the MWPC, and the reconstructed ``true'' initial energy of the
$\beta$-decay electron, denoted $E_{\text{recon}}$, was determined
via comparison of conversion electron source
measurements to Monte Carlo studies (as discussed in Section
\ref{sec:analysis_initial_energy_reconstruction}).

As described earlier in Section
\ref{sec:experiment_electron_spectrometer_scintillator}, each PMT
effectively viewed one $\pi/2$ quadrant of the scintillator;
therefore, the response of each PMT was expected to vary as a function
of the $(x,y)$ position of the event.  Each PMT then produced a signal
which was a nonlinear function of the scintillation light it viewed,
with the nonlinearity the result of known physics (i.e., quenching
interactions in the scintillator \cite{birks51}) and also hardware
response issues.  The scintillation light was then ultimately
digitized by the QADC data acquisition modules.  Accounting for a
time-dependent gain of the PMT/QADC system, the total response of the
system to PMT $i$ in terms of digitized QADC channels was then modeled
as
\begin{equation}
\text{QADC}_i = g_i(t) \cdot f_i(\eta_i(x,y) \cdot E_Q),
\end{equation}
where $E_Q$ denotes the light produced in the scintillator,
$\eta_i(x,y)$ is the fraction of that light reaching PMT $i$ from the
position $(x,y)$ of the event, $f_i$ is the (possibly non-linear)
response function of PMT $i$, $g_i(t)$ denotes the time-dependent gain
fluctuation of PMT $i$, and $\text{QADC}_i$ is the digitized readout
for the event.  Note that we use the notation $E_Q$ for the produced
scintillation light because the produced light should be proportional
to the ``quenched energy'', which we define to be the true energy
deposition reduced by a Monte-Carlo-calculated quenching factor (based
on studies with a low-energy electron gun
\cite{junhua_thesis,yuan01}).  The QADC readout signal that is
measured also includes statistical fluctuations due to photoelectron
(PE) counting statistics, with the number of PEs proportional to the
scintillation light transported to the PMT, $N_i \propto
\eta_i(x,y)E_Q$, with the fluctuations in $N_i$ expected to follow a
Poisson distribution.

The scintillator energy calibration procedure was thus divided into
three primary tasks: (1) a determination of the linearity function
$f_i$ for each PMT; (2) a determination of the light-transport
efficiency $\eta_i(x,y)$ position response map for each PMT; and (3) a
determination of the time-dependent gain $g_i(t)$ for each PMT.
Detailed descriptions of each of these tasks follow below.  However,
note that the overall calibration procedure was highly iterative,
whereby previous approximate results from the other two tasks were
used to produce new, more refined results for the task in question.
Therefore, our discussions below of the individual tasks reference
input from the other two tasks.

\subsubsection{PMT Linearity Functions}
\label{sec:analysis_energy_calibration_linearity}

\begin{figure*}
\includegraphics[scale=0.85,clip=]{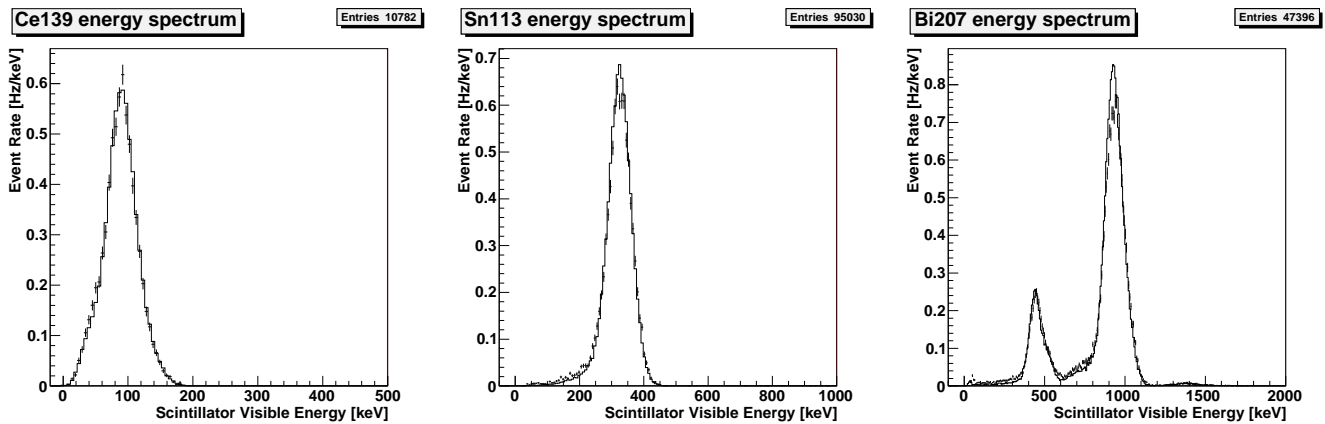}
\caption{Calibrated $E_{\text{vis}}$ visible energy spectra (data
points with error bars), compared with Monte Carlo calculations (solid
lines), for three of the calibration sources: $^{139}$Ce (left panel),
$^{113}$Sn (middle panel), and $^{207}$Bi (right panel).}
\label{fig:calibration_source_data}
\end{figure*}

The $f_i$ response functions for each of the PMTs were determined
each time a new set of calibrations was performed at
multiple $(x,y)$ positions with the conversion electron sources
using the remote source insertion system described previously in
Section \ref{sec:experiment_calibration_gain}.  Most of the useful
calibration data were obtained with the $^{139}$Ce, $^{113}$Sn, and
$^{207}$Bi sources; however, note that the $^{139}$Ce source was not
available during data taking for Geometries A and B.  The $^{109}$Cd
source was visible only in Geometries C and D
(configuration with 6-$\mu$m MWPC windows).

The QADC spectra for each PMT were then recorded for each $(x,y)$
source position.  The expected $E_Q$ spectrum for each PMT for each
$(x,y)$ source position was then determined from the Monte Carlo
simulation programs, with the Monte Carlo spectrum further smeared by
the PE counting statistics.  The smeared Monte Carlo spectrum was then
fit to a Gaussian (or, two Gaussians for the case of $^{207}$Bi) to
determine the $E_Q$ of the source conversion electron peaks.  The
measured QADC spectrum was also fit according to the same procedure,
thereby providing the measured QADC peak location for each conversion
electron peak.  Note, however, in order to prevent PMT nonlinearity
from shifting the fit position, the QADC spectrum was first linearized
using a linearity function $f_i$ determined from an earlier iteration
of this procedure.  The linearized spectrum of
$f_i^{-1}(\text{QADC}_i)$ was then fit, and the peak positions were
converted back to nonlinear QADC values via the $f_i$ functions.
Figure \ref{fig:calibration_source_data} shows sample data
compared with Monte Carlo spectra of the visible energy
$E_{\text{vis}}$ for three of the calibration sources ($^{139}$Ce,
$^{113}$Sn, and $^{207}$Bi).

\begin{figure*}
\includegraphics[scale=0.75,clip=]{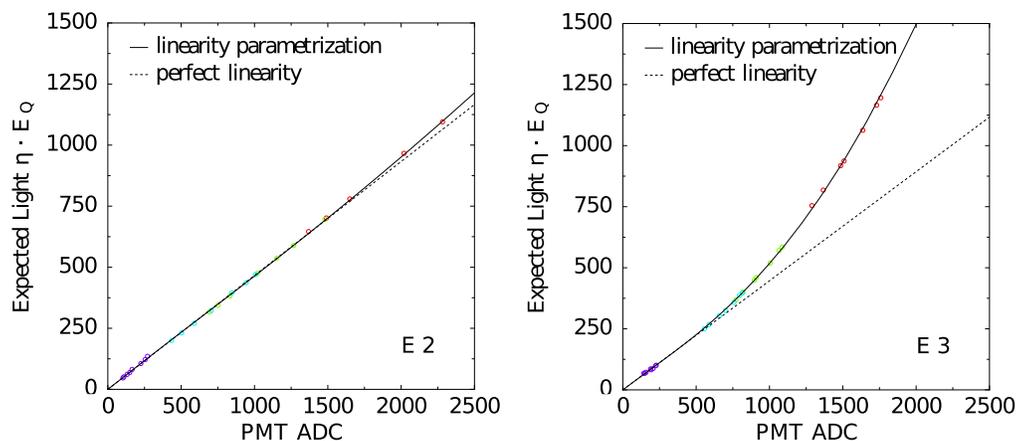}
\caption{Two examples of PMT linearity curves, for a PMT exhibiting a
nearly linear response (left panel) and a PMT exhibiting a highly
nonlinear response (right panel).  The curves are fits to the measured
QADC values (horizontal axis) of the peaks for the various calibration
sources at different $(x,y)$ positions versus the expected light
reaching each PMT.  The large nonlinearity observed in some
PMTs was traced to damaged bases and subsequently repaired for
runs after 2009.}
\label{fig:pmt_linearity_curves}
\end{figure*}

Using the position response map $\eta_i(x,y)$ for each PMT, the
expected light reaching each PMT $\eta_i(x,y) \cdot E_Q$ was plotted
against the observed QADC values of the peaks for the various sources
at the various $(x,y)$ positions.  The points on these plots thus
trace out the linearity curve for each PMT, with a fit to the
$\eta_i(x,y) \cdot E_Q$ values as a function of the QADC values
providing a parametrization of the linearity function $f_i^{-1}$ for
each PMT.  Examples of such linearity curves are shown in Fig.\
\ref{fig:pmt_linearity_curves} for two of the PMTs, with one of the
PMTs exhibiting a nearly linear response, and the other a highly
nonlinear response\footnote{The source of these
nonlinearities was later (in 2010) found to likely be the result of
problems with the PMT bases.  In particular, it was found that several
of the interstage capacitors were electrically shorted.  After this
problem was discovered, the PMTs and bases were replaced.}.  The
linearity curves were parametrized with a fit function which was
purely linear below some manually-determined transition point $x_0$.
Above this $x_0$, the nonlinearities were parametrized in the form
$y(x) = \exp[C_1 + \ln x + C_2[\ln(x/x_0)]^2 + C_3[\ln(x/x_0)]^3]$,
with $C_1$, $C_2$, and $C_3$ fitted constants.

\subsubsection{PMT Position Response Maps}
\label{sec:analysis_energy_calibration_position_response_map}

The overall principle for the determination of the light transport
efficiency to each PMT as a function of position, $\eta_i(x,y)$, was
to employ the measured neutron $\beta$-decay spectrum endpoint as a
``standard candle'' providing coverage over the entire detector
fiducial volume.  Assuming the linearity and time-dependent gain
functions, $f_i$ and $g_i(t)$, are known, the measured data provide a
measure of the amount of light reaching each PMT, $L_i(x,y) \equiv
\eta_i(x,y) \cdot E_Q = f_i^{-1}(\text{QADC}_i/g_i(t))$.  The spectrum
of $L_i(x,y)$ has the same shape as the spectrum of $E_Q$, but is
linearly ``stretched'' by the light-transport factor $\eta_i(x,y)$.
Determining the absolute light-transport efficiency would be quite
difficult, but, fortunately, is unnecessary.  Instead, measuring the
relative efficiency between different $(x,y)$ locations is sufficient,
normalizing $\eta_i$ by convention to $\eta_i(0,0) = 1$, and leaving
the unknown constant factor between the $\eta_i$ and the absolute
(unknown) light-transport efficiencies to be absorbed into the $f_i$.

In principle, the relative $\eta$ between two different
locations can be determined by
seeing how much the $L_i$ spectrum at one point needs to
be ``stretched'' to line up with the spectrum at another point.
In practice, the stretching factor $\eta_i(x,y)$ was
determined so that a Kurie plot of $\eta_i(x,y) \cdot L_i(x,y)$ was
aligned with a Kurie plot of a smeared Monte Carlo visible energy
spectrum.  To find the necessary $\eta_i(x,y)$, an iterative Kurie
plotting procedure was used.  Starting from an initial guess
$\eta_{i,0}(x,y)$ for each $\eta_i(x,y)$, a Kurie plot was then made
from the spectrum of $\eta_{i,0}(x,y) \cdot L_i(x,y)$.  This plot was
fit with a straight line over a visible energy range from 250--700
keV, which yielded some intercept $E_{\text{int}}$.  The position of
this fitted intercept relative to the expected intercept (from the
Monte Carlo visible energy spectrum smeared by the PE counting
statisics), $E_{\text{MC}}$, then provided an improved estimate for
the stretching factor, $\eta_{i,0}^\prime =
\frac{E_{\text{MC}}}{E_{\text{int}}} \eta_{i,0}$.  This procedure was
iterated several times (including iterations to the Monte Carlo
spectrum, to account for changes to the PE counting statistics
resulting from improved estimates for the light-transport efficiencies
at each $(x,y)$ position with each iteration), until the intercepts
from the Kurie plots for all of the points over the detector fiducial
volume were aligned with the smeared Monte Carlo spectrum intercept.
Note that the Monte Carlo spectrum of the visible energy varied with
the particular Geometry (i.e., A, B, or C/D, depending on the decay
trap end-cap foil and MWPC window thicknesses).  Thus,
even though the visible energy spectra varied with Geometry, we
emphasize that the Kurie fits to the visible energy were employed for
relative point-to-point $\eta_i(x,y)$ calibration purposes, and not
for absolute energy-scale calibration purposes.

\begin{figure}
\includegraphics[scale=0.38,clip=]{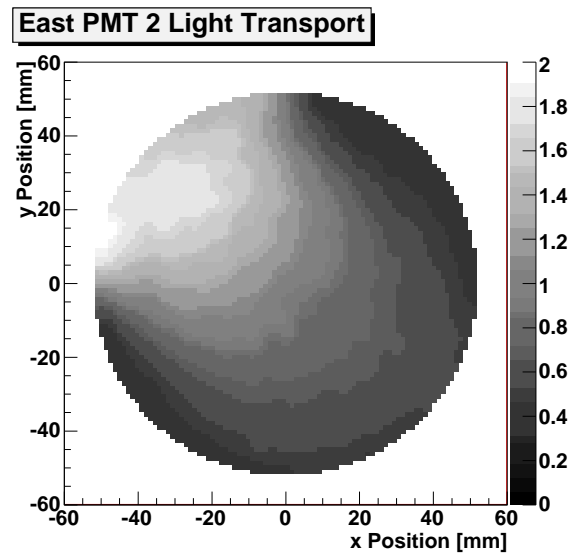}
\caption{Example of a $\eta(x,y)$ light-transport efficiency map for
a single PMT, nominally viewing the $(x<0,y>0)$ quadrant of the
scintillator disc.  By convention, $\eta(0,0)=1$.}
\label{fig:position_response_map}
\end{figure}

The above-described procedure for construction of these PMT position
response maps was implemented by combining $\beta$-decay data from
nearly the entire 2008--2009 dataset.  The detector face was divided
into 180 approximately-equal-sized pixels, and a background-subtracted
$L_i(x,y)$ spectrum for each PMT was generated for each of these
pixels.  The stretching factor $\eta_i(x,y)$ was then determined for
each of these pixels.  Having determined the $\eta_i(x,y)$ for each of
these discrete pixels, a continuous map of the light-transport
efficiency was then produced via bicubic spline interpolation in
polar coordinates.  Fig.\ \ref{fig:position_response_map} shows
an example of an interpolated position response map for one PMT,
exhibiting the expected strong dependence with position.

The primary source of uncertainty in the construction of these PMT
position response maps was the statistical uncertainty in the Kurie
fitting procedure, which was $\pm 1.0$\% at each of the 180 pixels and
independent of the $\eta_i(x,y)$ value determined for each pixel.  On
an event-by-event basis, there was an additional uncertainty from the
coupling between the varying $(x,y)$ position response, and the
uncertainty in the reconstruction of the position by the MWPC.  This
was determined to be $\sim \pm 1.5$\% per RMS mm uncertainty in
position reconstruction.  Note that the RMS uncertainty in the
position reconstruction is difficult to estimate, because on an
event-by-event basis, there will be a true variation due to the Larmor
spiral radius.  This RMS uncertainty is actually probably
significantly less than 2 mm, as the width observed with the
conversion electron calibration sources was mostly explained by their
Larmor radii.  Because the position response maps between the four
PMTs for each detector are significantly correlated, the uncertainty
of the combined four PMT response will be less than the $\sim
0.75$\%/mm expected under the assumption that the four PMT
uncertainties are independent.

\subsubsection{Gain Stabilization}
\label{sec:analysis_energy_calibration_gain_stabilization}

The gain of the PMT/QADC system drifted over time, typically on the
scale of a few percent over several hours, due to ambient temperature
fluctuations, etc.  Periodic $^{113}$Sn source calibrations conducted
every few days during production $\beta$-decay running provided the
first layer of gain stabilization.  For such a run, the $(x,y)$
position of the source and its QADC spectrum for each PMT was
measured.  The observed QADC spectrum and a Monte Carlo spectrum for
that $(x,y)$ position were then fit with the same procedure employed
for the determination of the PMT linearity functions, yielding a QADC
channel value for the $^{113}$Sn peak observed in each PMT $i$,
$\text{QADC}_{\text{Sn},i}$, and a Monte Carlo expected light yield,
$E_{Q,\text{Sn},i}$.  The PMT gain factor, $g_i$, was then set so that
the calibration curve placed the measured QADC value at its expected
$E_Q$,
\begin{equation}
\text{QADC}_{\text{Sn},i} = g_i \cdot f_i(\eta_i(x,y) \cdot L_{\text{Sn},i}).
\end{equation}
Note that the resolution of each PMT was also determined from the
fitted width of the $^{113}$Sn peak spectra.  As with the energy
calibration, the resolution was extracted from a comparison to Monte
Carlo spectra, in order to disentagle the effects of peak broadening
due to PE counting statistics from the multiple K, L, etc.\ conversion
electron lines.

All other runs were then further gain stabilized to match these
$^{113}$Sn calibration runs by comparing, on a run-by-run basis, the
shape of the measured QADC spectrum for events tagged as background
cosmic-ray muons by the backing veto.  Typically $\sim 5000$ such
events were identified in the $\sim 1$-hour long $\beta$-decay runs,
and the measured QADC spectrum was fitted to a Landau distribution.
If the QADC peak position for a muon event in PMT $i$ was
$\text{QADC}_{\mu,i}(0)$ during the $^{113}$Sn source calibration run
and then $\text{QADC}_{\mu,i}(t)$ at some later time $t$, this
gain shift was then corrected for by setting the time-dependent gain
correction factor to be
\begin{equation}
g_i(t) = \frac{f_i(\eta_i(x,y) \cdot E_{Q,\text{Sn},i})}
{\text{QADC}_{\text{Sn},i}} \cdot
\frac{\text{QADC}_{\mu,i}(0)}{\text{QADC}_{\mu,i}(t)}.
\end{equation}
These time-dependent gain corrections were typically on the order
of $\sim 5$\%.

The uncertainty in the $^{113}$Sn gain stabilization was dominated by
the position response uncertainty; since the source data were very
localized in position, they were subject to the statistics-limited
localized position map fluctuations of $\sim \pm 1$\% for each PMT.
The fit statistics for the cosmic-ray muon peaks
were typically
$\sim \pm 1.5$\% for each PMT on a run-by-run basis.  These two
uncertainties must then be combined in quadrature, with that
from the muon peak uncertainty contributing twice,
since calculation
of the $g_i(t)$ factors requires results from the
muon peak fits for both the $^{113}$Sn calibration run and the run
being calibrated.
Therefore, the total gain stabilization uncertainty
for each PMT was estimated to be $\sim 2.3$\%.  With the four PMTs
contributing approximately equally to a combined reconstruction of the
visible energy (discussed next), the gain stabilization fluctuations
were estimated to be $\sim 1.2$\%.

\subsection{Visible Energy Reconstruction and Resolution}
\label{sec:analysis_visible_energy_reconstruction}

\begin{figure*}
\includegraphics[scale=0.90,clip=]{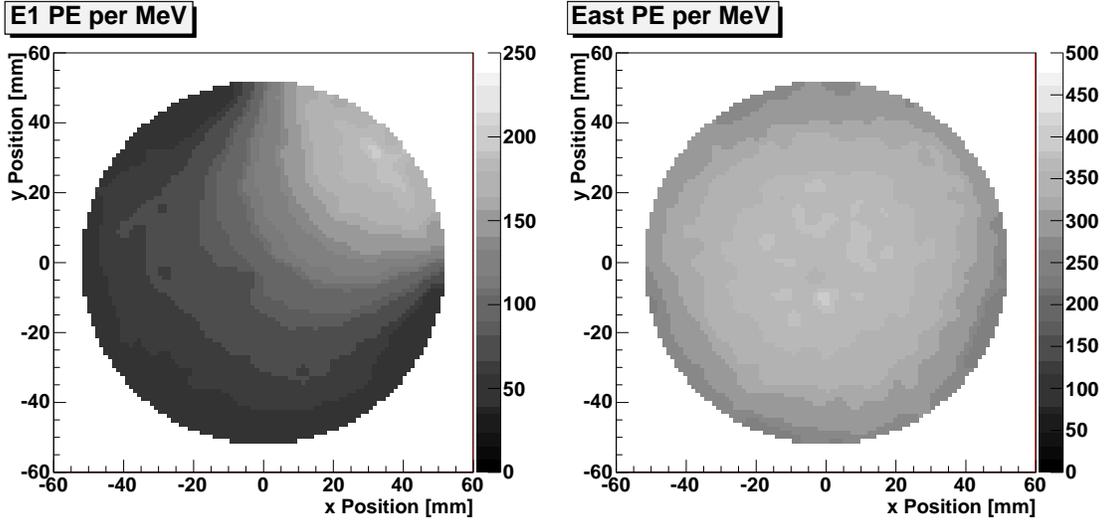}
\caption{Extracted energy response for one of the detectors in
photoelectrons per MeV as a function of $(x,y)$ position extracted
from one PMT (left panel) and the
statistically weighted combined response
from all four PMTs
(right panel), demonstrating the combined response exhibits a smoother
response as a function of the $(x,y)$ position.}
\label{fig:photoelectrons_position}
\end{figure*}

Thus far, the discussion has primarily focused on the calibration of
the individual PMTs.  The calibrated signals from the four PMTs for
each scintillator then provided, on an event-by-event basis, four
independent measurements of the visible energy.  These were then
statistically combined, with weighting according to their PE counting
statistics, to produce a single, more accurate result for the measured
visible energy.  Note that the dominant contribution to the
uncertainty, $\sigma_i$, in each individual PMT's measurement of the
visible energy was from Poisson counting statistics, as the previously
discussed position response map ($\sim 1.5$\%) gain stabilization
($\sim 2.3$\%) uncertainties are small compared to the order $\sim
10$\%-level individual-PMT PE counting statistics.

The motivation for extracting the event energy from a statistically
weighted average of the individual PMT energy measurements (as opposed
to a non-statistically-weighted sum of the individual PMT responses)
is as follows.  Consider an event with quenched energy $E_Q$ at
position $(x,y)$.  According to our model some fraction of the
produced scintillation light $L_i = \eta_i(x,y)E_Q$ will reach each
PMT, where it will be converted to $N_i \pm \sqrt{N_i}$ PEs according
to the quantum efficiency, $C_i$, of the PMT, such that $N_i = C_i
L_i$.  The QADC signal for PMT $i$ is then converted to an
individual-PMT estimate $E_i$, with estimated error $\sigma_i = E_i /
\sqrt{N_i}$.  Combining the four PMT estimates, with their respective
$1/\sigma_i^2$ statistical weights, and assuming the individual PMT
measurements are such that $E_i \approx E_Q$, we find
\begin{eqnarray}
E_Q &\approx& \frac{\sum_i \frac{N_i}{E_i^2}E_i}{\sum_i \frac{N_i}{E_i^2}} \pm
\frac{1}{\sqrt{\sum_i \frac{N_i}{E_i^2}}} \nonumber \\
&\approx& \frac{\sum_i \frac{N_i}{E_Q}}{\sum_i \frac{C_i L_i}{E_Q L_i/\eta_i}}
\pm \frac{1}{\sqrt{\frac{1}{E_Q}\sum_i \frac{C_i L_i}{L_i/\eta_i}}}
\nonumber \\
&=& \frac{\sum_i N_i}{\sum_i C_i \eta_i} \pm
\sqrt{\frac{E_Q}{\sum_i C_i \eta_i}},
\label{eq:PMT_weighted_mean}
\end{eqnarray}
which proves that the statistically weighted mean yields an estimate
for the energy which is the product of the sum of the total number of
photoelectrons, $N_{\text{tot}} = \sum_i N_i$, and the
position-dependent photoelectron-to-energy conversion factor,
$1/\sum_i C_i \eta_i(x,y)$.  Note that this form significantly
protects against errors in the reconstructed $(x,y)$ position, because
the position dependence of the individual PMT responses appears only
in the summed combination $\sum_i C_i \eta_i$, which is a smoother
function of $(x,y)$ than the individual maps.

The energy resolution of the detector was extracted from fits to the
measured $^{113}$Sn peak positions and widths, after accounting in
Monte Carlo for peak broadening from the K, L, etc.\ conversion
electron lines.  Figure \ref{fig:photoelectrons_position} shows plots
of the extracted number of photoelectrons as a function of position.
Averaged over the fiducial volume, the detector resolution was such
that $\sim 400$ PEs/MeV were observed in the East detector, and $\sim
500$ PEs/MeV in the West detector,
translating to a resolution of $\pm
9$\% at the $^{113}$Sn peak and $\pm 5$\% at the neutron $\beta$-decay
endpoint energy.

\begin{figure*}
\includegraphics[scale=0.99,clip=]{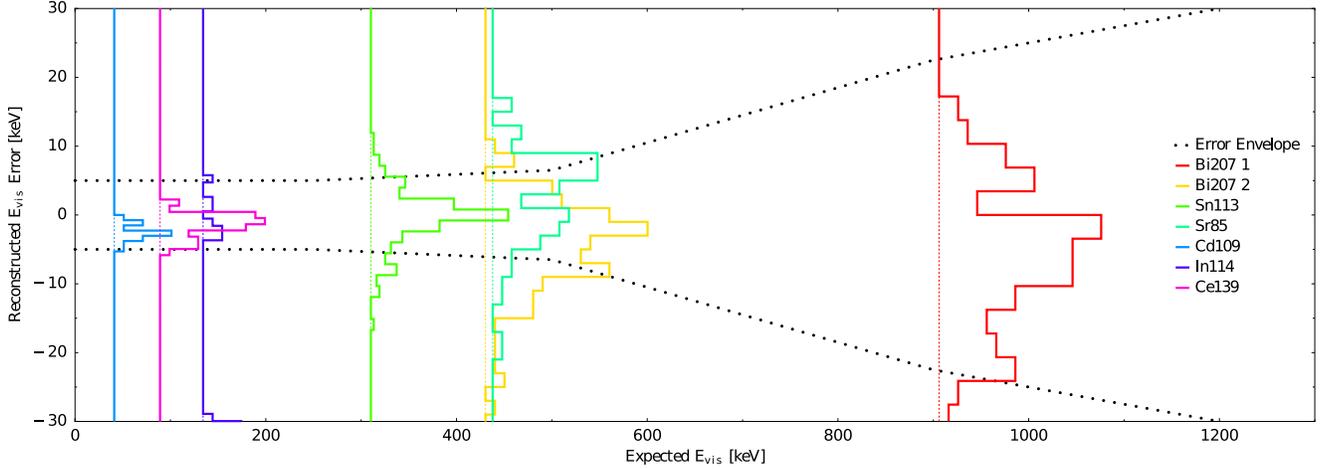}
\caption{(Color online) The vertical axes appearing at the locations
of the expected (i.e., as calculated in Monte Carlo) visible energies
(on the horizontal axis) for the various calibration sources show
histograms of errors (in keV) in the measured visible energies.  The
dashed lines then show our conservative estimate of an ``error
envelope'' of the total energy calibration uncertainty based on the
widths of the error distributions.}
\label{fig:energy_reconstruction_error_envelope}
\end{figure*}

\subsection{Visible Energy Reconstruction Uncertainty}
\label{sec:analysis_visible_energy_reconstruction_uncertainty}

The quality of the visible energy calibration procedure was checked by
comparing how closely the reconstructed visible energy spectra for the
conversion electron calibration sources were aligned with the Monte
Carlo predicted spectra.  The primary findings of this exercise were:

\begin{itemize}

\item
For each run, there was an overall $\pm 1.2$\% energy uncertainty from
gain stabilization.

\item
At low visible energies (50--100 keV), there was a $\pm 5$\% linearity
uncertainty, as deduced from a comparison of measured and simulated
$^{109}$Cd and $^{139}$Ce spectra.

\item
The linearity uncertainty was $\sim 0$ at the $^{113}$Sn peak energy,
since this served as an anchor point for the calibrations.  At this
visible energy range, the visible energy reconstruction error,
$\pm 1.7$\%, was primarily due to the errors in the
position response maps and the run-to-run gain stabilization
uncertainty.

\item
Residual nonlinearity induces a 1.3\% ($\sim 10$ keV) error at the
$\beta$-decay endpoint energy, based on fits over the visible energy
range of 300--700 keV.  This is then interpreted as the linearity
uncertainty around 500 keV.  The linearity uncertainty then increases
to $\pm 2.5$\% at the upper end of the visible energy range, $\sim
900$ keV, based on the $^{207}$Bi upper peak.

\end{itemize}

Figure \ref{fig:energy_reconstruction_error_envelope} shows a
histogram of the reconstruction errors in the source peak
energies from calibrations across all of the
Geometries (A, B, C, and D), together with an ``error envelope'' of
the total calibration uncertainty from the above described sources.
Note that this error envelope is a conservative estimate for the
uncertainty, based on the widths of the distributions
(and is especially conservative at energies below 200
keV, where the envelope is wider than the plotted $^{109}$Cd,
$^{139}$Ce, and $^{114m}$In distributions), as opposed to the
uncertainties in the means.  The impact of this uncertainty in the
visible energy calibration on the extraction of the $\beta$-asymmetry
is discussed later in Section
\ref{sec:uncertainties_energy_reconstruction}.

\subsection{Scintillator Trigger Efficiency Functions}
\label{sec:analysis_trigger_efficiency}

An extraction of the scintillator two-fold PMT trigger efficiencies as
a function of visible energy was important for comparisons of
Monte Carlo calculations with data (especially for backscattering
events which deposit small amounts of energy in the scintillator),
because the trigger efficiency ultimately impacts the reconstruction
of the event type.  We extracted these two-fold PMT trigger
efficiencies from the measured $\beta$-decay data as a function of the
measured visible energy according to the following procedure.  First,
we employed only electron events satisfying the MWPC position cuts
described in Section \ref{sec:analysis_position_cuts}.  Second, for
every possible pair of two PMTs that both triggered (i.e., generated a
signal above pedestal), we then incremented an $E_{\text{vis}}$
``total'' histogram for the other two PMTs.  For each of the other two
PMTs, if it triggered, we then incremented an $E_{\text{vis}}$
``trigger'' histogram.  Third, the trigger efficiency histogram for
each PMT, $\epsilon_i$, was then obtained by dividing the ``trigger''
histogram by the ``total'' histogram.  Finally, the total two-fold PMT
trigger efficiency for each scintillator as a function of the visible
energy was calculated as [$1 - (\text{probability no PMTs trigger}) -
(\text{probability only one PMT triggers})$], with the appropriate
combinatorics for each of the terms in this expression in terms of the
individual PMT $\epsilon_i$ efficiencies.

\begin{figure}
\includegraphics[scale=0.45,clip=]{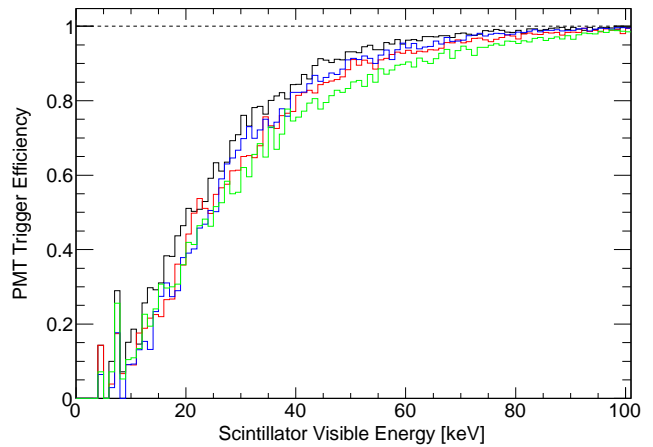}
\caption{(Color online) Individual PMT trigger efficiencies
$\epsilon_i$ for the four PMTs (indicated by the different colors) on
one of the detectors, as extracted from an analysis of Geometry C
data.}
\label{fig:pmt_efficiencies_geometry_c}
\end{figure}

The results from such an analysis of the individual PMT efficiencies
are shown for one of the detectors in Fig.\
\ref{fig:pmt_efficiencies_geometry_c} for one of the Geometries
(Geometry C).  The PMT efficiency curves for the other detector and
the other Geometries were similar.  These were incorporated in our
Monte Carlo simulation code.

\subsection{MWPC Energy Calibration}
\label{sec:analysis_mwpc_energy_calibration}

In addition to suppressing gamma-ray backgrounds and permitting the
definition of a fiducial volume, Monte Carlo studies suggested
that the measured energy deposition in the MWPC on the triggering
scintillator side could be used to separate Type 2/3 backscattering
events.  For example, in the event types schematic shown in Fig.\
\ref{fig:event_types_schematic}, the depicted Type 2/3 events would
otherwise appear identical simply in terms of detector signals above
threshold.  However, in the depicted Type 2 event, the electron
traverses the MWPC on the triggering-scintillator side only once,
whereas in the depicted Type 3 event, the electron traverses the MWPC
on the triggering-scintillator side twice.  Therefore, for the Type
2/3 events depicted there, the energy deposition in the MWPC on the
triggering-scintillator side would be expected to be greater for the
Type 3 event based on path length considerations.  Hereafter, we will
refer to the MWPC located on the earlier (and the only)
triggering-scintillator side as the ``Primary MWPC'', and the
opposite-side MWPC as the ``Secondary MWPC''.

An energy calibration of the MWPC response was performed according to
the following procedure.  First, we performed polynomial fits to Monte
Carlo data of energy deposition in the MWPC for Type 0 neutron
$\beta$-decay events (i.e., the calibration was based on Type 0
$\beta$-decay events).  Denote the resulting fit function
$f_{\text{MC}}(E)$.  Second, the MWPC detector face was divided into
10 mm $\times$ 10 mm$^{2}$ square bins (with 88 of these bins providing
coverage of the decay trap circular geometry).  In each of these 88
bins, we then fitted the function $f_{\text{MC}}(E(x))$, where $x$
denotes the MWPC anode PADC readout channel number, and $E(x)$ was a
function (taken to be polynomials) that coverted from channel number
to energy.  Lookup tables in binned $(x,y)$ positions and anode
channel numbers were then constructed for each detector for the
different Geometries (i.e., A, B, C, and D).

\begin{figure}
\includegraphics[scale=0.47,clip=]{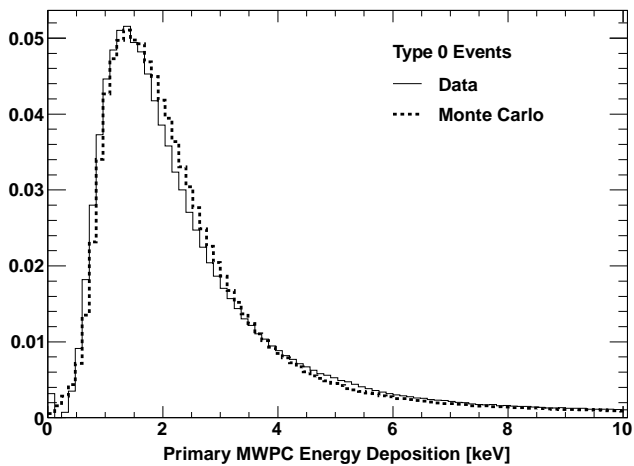}
\caption{Baseline calibrated Primary MWPC energy spectra for
Type 0 neutron $\beta$-decay events from Geometry B (solid line) compared
with Monte Carlo calculations (dotted line).  The histograms are normalized
to unity.}
\label{fig:mwpc_energy_calibration_geometry_b_1}
\end{figure}

\begin{figure*}[t]
\includegraphics[scale=0.73,clip=]{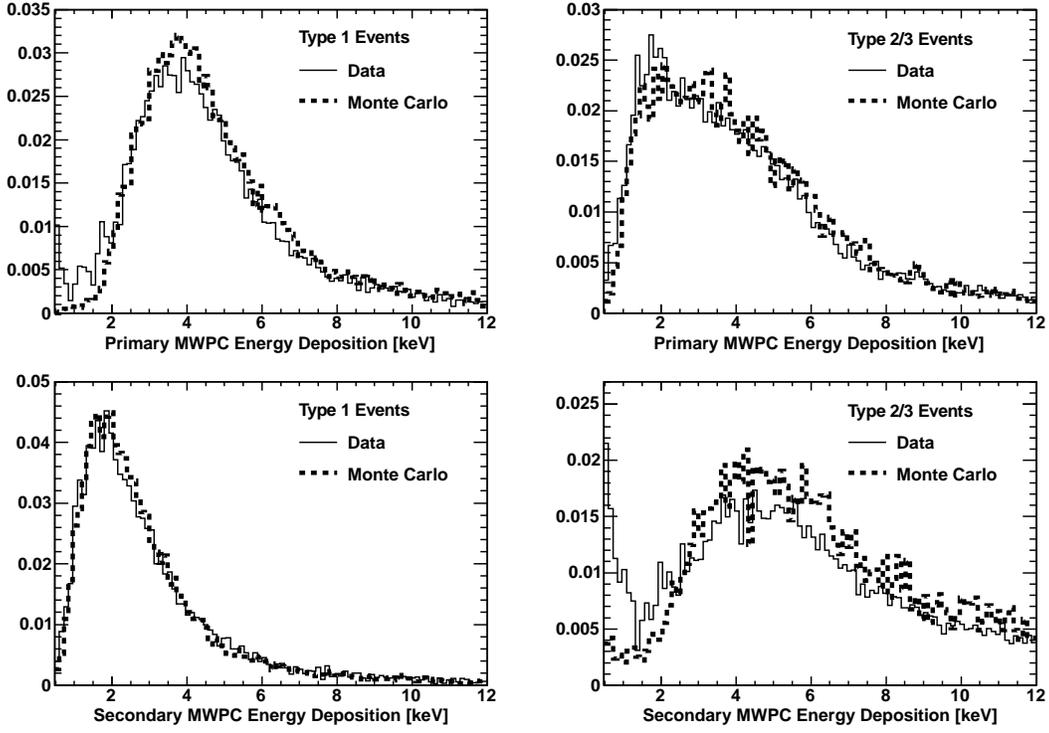}
\caption{Calibrated Primary (top panels) and Secondary (bottom panels)
MWPC energy spectra for Type 1 and Type 2/3 neutron $\beta$-decay
events from Geometry B (solid lines) compared with Monte Carlo calculations
(dotted lines).  The histograms are normalized to unity.}
\label{fig:mwpc_energy_calibration_geometry_b_2}
\end{figure*}

\begin{figure*}
\includegraphics[scale=0.80,clip=]{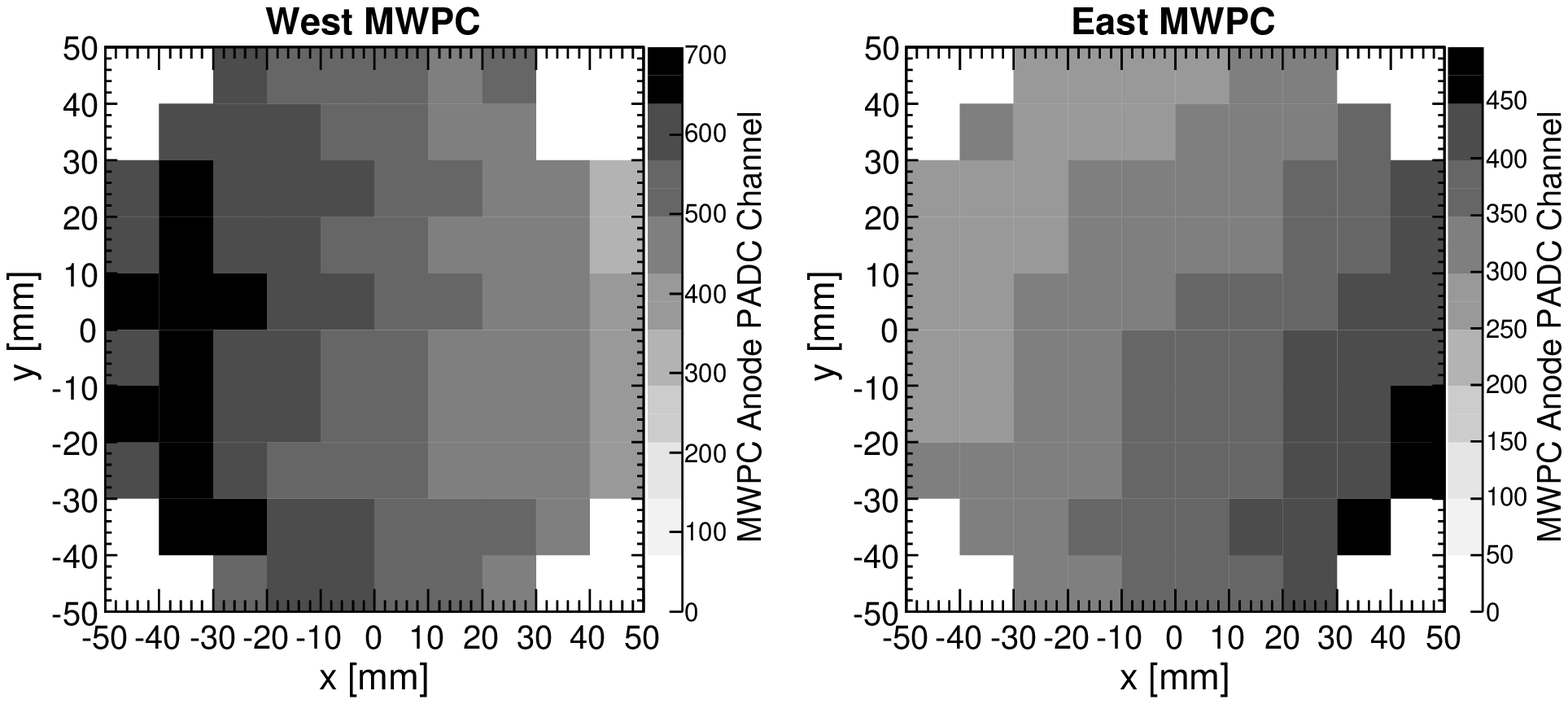}
\caption{$(x,y)$ position maps of the two MWPCs' anode PADC channel
numbers that correspond to a given calibrated energy (in this case,
4.14 keV).}
\label{fig:mwpc_calibration_position_dependence}
\end{figure*}

The quality of the MWPC energy calibration is shown in Figs.\
\ref{fig:mwpc_energy_calibration_geometry_b_1} and
\ref{fig:mwpc_energy_calibration_geometry_b_2}.  First, in Fig.\
\ref{fig:mwpc_energy_calibration_geometry_b_1} we compare a baseline
calibrated (and background-subtracted) Primary MWPC energy spectrum
for Type 0 neutron $\beta$-decay events from Geometry B with Monte
Carlo calculations for this Geometry.  Recall that the calibration was
based on polynomial fits to the Monte Carlo Type 0 (Primary MWPC)
spectra.  Second, in Fig.\
\ref{fig:mwpc_energy_calibration_geometry_b_2} we compare calibrated
(and background-subtracted) Primary and Secondary MWPC energy spectra
for Type 1 and Type 2/3 neutron $\beta$-decay events from Geometry B
with Monte Carlo calculations.  Note that the Type 1 Primary and
Secondary and the Type 2/3 Secondary spectra provide a
pseudo-independent check of the calibration, as the Monte Carlo
calculations were based on fits to the Type 0 MWPC Primary spectrum
(i.e., the spectrum on the earlier, or only, triggering-scintillator
side).  Note that the calibrated response exhibited a strong position
dependence, which can be seen in Fig.\
\ref{fig:mwpc_calibration_position_dependence}, where we show position
maps of the two MWPCs' anode channel numbers that correspond to a
particular fixed energy (for this plot, 4.14 keV, the relevance of
which for the separation of Type 2/3 events is discussed later in
Section \ref{sec:asymmetry_analysis_backscattering_choices_type23}).

\subsection{MWPC Position-Dependent Efficiency}
\label{sec:analysis_mwpc_efficiency}

\begin{figure*}
\includegraphics[scale=0.80,clip=]{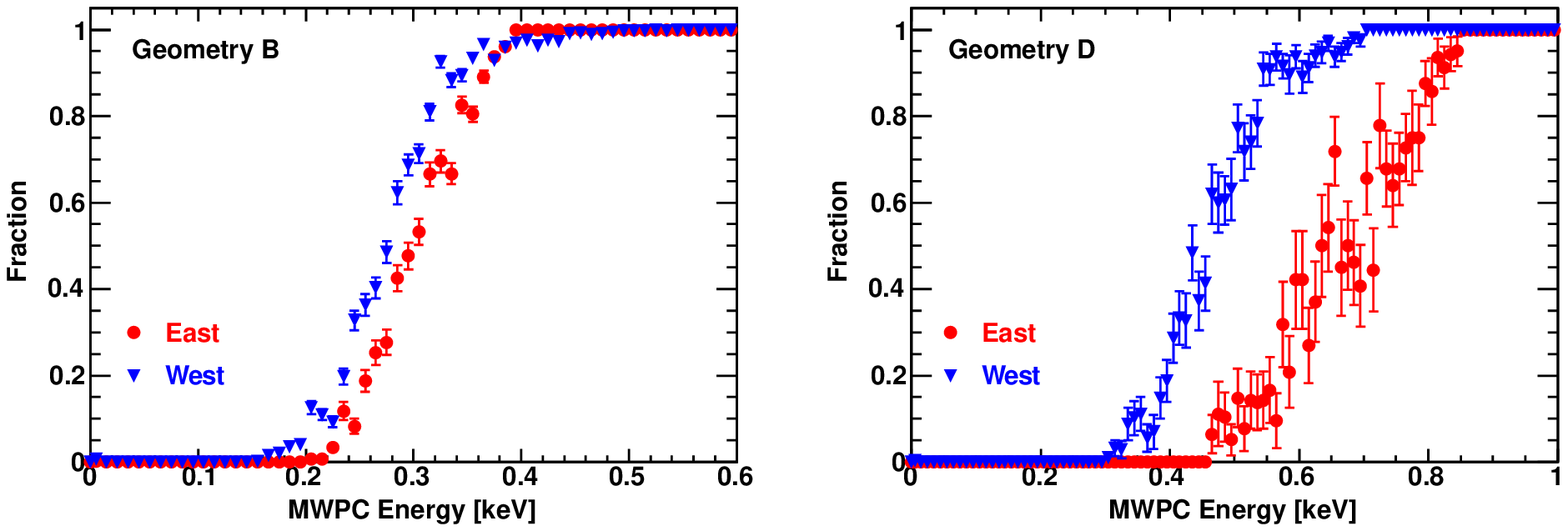}
\caption{(Color online) Efficiency for the identification of charged
particle events by the MWPC detectors (for a fixed PADC channel number
cut) plotted as a function of the measured MWPC energy deposition.
The non-step-function behavior of these efficiency curves is a result
of the MWPCs' position-dependent response.
Typical results for the East (red circles) and West (blue triangles)
detectors from Geometries B and D are shown in the left and right
panels, respectively.}
\label{fig:mwpc_efficiency_energy_deposition}
\end{figure*}

As discussed earlier in Section
\ref{sec:analysis_event_reconstruction_mwpc_spectra}, a fixed MWPC
PADC channel number cut was used to separate gamma ray and charged
particle events (i.e., this cut did not vary with the reconstructed
$(x,y)$ position).  If the MWPC response was
independent of position, a plot of the MWPC efficiency for the
identification of charged particles as a function of the measured MWPC
energy deposition would exhibit a step function at the cut energy (or
PADC channel number cut).  However, as just shown in Fig.\
\ref{fig:mwpc_calibration_position_dependence}, the MWPC response was
strongly position-dependent\footnote{We believe an
electronics issue (capacitances and/or inductances) on the MWPCs'
circuit boards may be the source of their observed position-dependent
response.}; therefore, for some given energy deposition in the MWPC,
the identification of an event as either a gamma ray event or a
charged particle event was subject to the MWPCs' position-dependent
response.

The extracted position-dependent efficiency for the identification of
charged particle events is illustrated in Fig.\
\ref{fig:mwpc_efficiency_energy_deposition}, where we have plotted the
fraction of events passing the standard gamma ray PADC channel number
cut as a function of the calibrated energy deposition in the two MWPCs
for two examples of typical data sets obtained during Geometries B and
D.  As can be seen there, over a particular range of energy
depositions the fraction of events passing this cut varies
monotonically from 0.0 to 1.0, as a result of the MWPCs'
position-dependent response.  These measured MWPC efficiency curves
were incorporated in our Monte Carlo simulation codes.  Note that
difference between the two MWPCs' efficiency curves was smallest
(greatest) for Geometry B (Geometry D), with the differences for
Geometries A and C in between those of Geometries B and D.

The possible impact of this position-dependent efficiency on the
identification of gamma ray events (as opposed to a cut on the MWPC's
calibrated energy response) and thus on the measured asymmetry is
discussed later in Section \ref{sec:uncertainties_mwpc_efficiency}.

\subsection{Initial Energy Reconstruction}
\label{sec:analysis_initial_energy_reconstruction}

The initial energy of the electron, hereafter denoted
$E_{\text{recon}}$, was reconstructed from the measured visible energy
in the scintillator based on the results of \texttt{GEANT4} Monte
Carlo simulations for the relation between the measured visible energy
in the scintillator, $E_{\text{vis}}$, and the actual initial energy
of the emitted $\beta$-decay electron.  Parametrizations relating
$E_{\text{recon}}$ to $E_{\text{vis}}$ were constructed for the
different event types (Type 0, Type 1, and Type 2/3) and for the
different Geometries (A, B, and C/D).  These parametrizations were
based on \texttt{GEANT4} simulations of conversion electron source
spectra, and were extracted from fits of the predicted mean value for
$E_{\text{vis}}$ to the true initial source energy, for the various
sources employed in the experiment.

Monte Carlo generated source electrons were separated into Type 0,
Type 1, and Type 2/3 events, according to the same selection rules as
applied in the data analysis.  For each Geometry and event type, two
different fits were constructed: ``Fit 1'' was based on the
scintillator visible energy, $E_{\text{vis}}$, only; whereas ``Fit 2''
included both the scintillator visible energy, $E_{\text{vis}}$, and
the calibrated MWPC energy, $E_{\text{MWPC}}$.  The input variables to
the fits were the mean values of: (1) the effective true source
energy, $E_{\text{true}}^{\text{source}}$, which accounted for the
difference between the Monte-Carlo-generated initial $\beta$-decay
energy and the electron's subsequent energy loss in the 3.6-$\mu$m
thick source enclosure foils; (2) the scintillator visible energy,
$E_{\text{vis}}$, summed over both scintillators, which accounted for
possible sub-trigger-threshold energy deposition in one of the
scintillators (e.g., in Type 2/3 events); and (3) the MWPC energy
deposition, $E_{\text{MWPC}}$, summed over both MWPCs.  For Type 0 and
Type 1 events, the functional forms for the fits of $E_{\text{recon}}
(\equiv E_{\text{true}}^{\text{source}})$ to $E_{\text{vis}}$ and
$E_{\text{MWPC}}$
\begin{equation}
E_{\text{recon}} = \left\{ \begin{array}{ll}
  f_Q E_{\text{vis}} + \Delta E,& \text{(Fit 1)} \\
  f_Q E_{\text{vis}} + \Delta E + E_{\text{MWPC}},& \text{(Fit 2)}
  \end{array} \right.
\end{equation}
and for Type 2/3 events
\begin{equation}
E_{\text{recon}} = \left\{ \begin{array}{ll}
  \epsilon E_{\text{vis}}^2 + f_Q E_{\text{vis}} + \Delta E,& \text{(Fit 1)} \\
  \epsilon E_{\text{vis}}^2 + f_Q E_{\text{vis}} + \Delta E + E_{\text{MWPC}}.&
  \text{(Fit 2)}
  \end{array} \right.
\end{equation}
For each Geometry, values were fitted for
$f_Q$, which represented a numerical
scintillator quenching factor; $\Delta E$, the energy loss in the
decay trap end-cap foils, MWPC windows, etc.; and $\epsilon$, a
second-order parameter which was used to control the fits for the Type
2/3 events.  The resulting fit parameters are given in Table
\ref{tab:Erecon_fit_parameters}, the point of which serves to set the
scale for the Monte Carlo-calculated $\Delta E$ energy loss for the
different Geometries.


\begin{figure}
\includegraphics[scale=0.47,clip=]{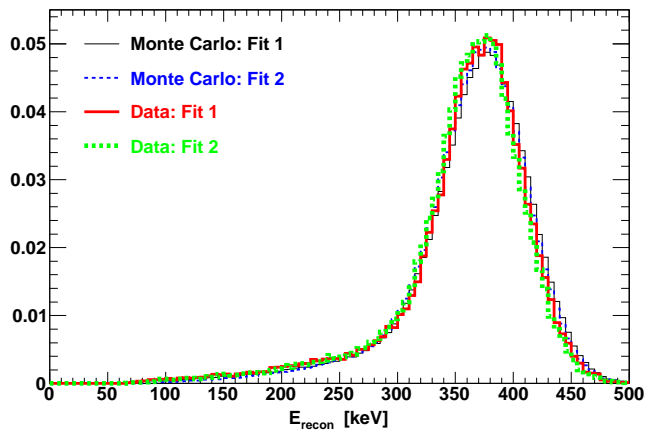}
\caption{(Color online) Reconstructions of $E_{\text{recon}}$ values
according to the Fit 1 (solid) and Fit 2 (dashed) parametrizations for
$^{113}$Sn source calibration Type 0 events from Monte Carlo (thin
lines) and experimental data (thick lines).  The histograms are
normalized to unity.}
\label{fig:Erecon_113Sn_Type0_data_MC_comparison}
\end{figure}

\begin{table*}
\caption{Values of fitted parameters for the $E_{\text{recon}}$
parametrizations.  See text for descriptions of parameters.  Note
that the uncertainties in the fitted parameters were highly correlated;
for brevity, we do not report the error matrix here.}
\begin{ruledtabular}
\begin{tabular}{lllll}
Event Type, Fit& Fit Parameter& Geometry A& Geometry B& Geometry C/D \\ \hline
\multirow{2}{*}{Type 0, Fit 1}& $\Delta E$ [keV]& 50.7(2)&
  63.0(2)& 33.3(1) \\
& $f_Q$& 1.0461(4)& 1.0459(4)& 1.0476(2) \\ \cline{2-5}
\multirow{2}{*}{Type 1, Fit 1}& $\Delta E$ [keV]& 129(1)& 149(1)& 73.5(6) \\
& $f_Q$& 1.031(2)& 1.040(3)& 1.035(1) \\ \cline{2-5}
\multirow{3}{*}{Type 2/3, Fit 1}& $\Delta E$ [keV]& 145(8)& 152(7)& 56(1) \\
& $f_Q$& 1.327(56)& 1.437(48)& 1.420(13) \\
& $\epsilon$ [keV$^{-1}$]& $-2.33(65) \times 10^{-4}$&
  $-3.60(58) \times 10^{-4}$& $-3.26(19) \times 10^{-4}$ \\ \hline
\multirow{2}{*}{Type 0, Fit 2}& $\Delta E$ [keV]& 45.1(2)&
  57.4(2)& 27.3(1) \\
& $f_Q$& 1.0486(4)& 1.0484(4)& 1.0510(2) \\ \cline{2-5}
\multirow{2}{*}{Type 1, Fit 2}& $\Delta E$ [keV]& 115(1)& 136(1)& 57.3(9) \\
& $f_Q$& 1.031(2)& 1.041(3)& 1.039(2) \\ \cline{2-5}
\multirow{3}{*}{Type 2/3, Fit 2}& $\Delta E$ [keV]& 121(6)& 129(7)& 41(1) \\
& $f_Q$& 1.276(40)& 1.396(53)& 1.274(15) \\
& $\epsilon$ [keV$^{-1}$]& $-1.88(47) \times 10^{-4}$&
  $-3.24(64) \times 10^{-4}$& $-1.92(22) \times 10^{-4}$ \\
\end{tabular}
\end{ruledtabular}
\label{tab:Erecon_fit_parameters}
\end{table*}

Figure \ref{fig:Erecon_113Sn_Type0_data_MC_comparison} compares
applications of these Fit 1 and Fit 2 parametrizations to
$E_{\text{recon}}$ reconstructions of $^{113}$Sn source calibration
Monte Carlo and experimental data for Type 0 events.  Good agreement
between Fits 1 and 2 is observed.  In the experimental data, the RMS
width of Fit 2 was slightly smaller (by $\sim 2$ keV), suggesting that
the inclusion of the MWPC energy slightly improved the resolution.
However, we ultimately chose to use Fit 1 instead of Fit 2 for our
asymmetry analysis because inclusion of the MWPC energy subjected the
value of $E_{\text{recon}}$ to the possibility of an overflow value
for the MWPC anode PADC readout (see Fig.\
\ref{fig:anode_cathode_spectra}), which occurred at a calibrated MWPC
energy of $\sim 10$ keV.  Because we could not reliably construct a
value for the MWPC energy in the event of an overflow readout, there
was a potential bias in the application of Fit 2 to the data.
Nevertheless, the possible impact of this small difference in the Fit
1 and Fit 2 results is discussed later in Section
\ref{sec:uncertainties_energy_reconstruction}.

\section{Asymmetry Analysis}
\label{sec:asymmetry}

In this section we outline our extraction of the asymmetries from the
calibrated data.  We begin by defining our different Analysis Choices
for the extraction of the asymmetries under which we included or
excluded the various backscattering event types.  Next, we describe
our procedure for the extraction of the binned (in energy)
background-subtracted event rates (and their statistical
uncertainties) from the $\beta$-decay and background runs, and proceed
to an extraction of the asymmetries under the octet data-taking
sequences via the super-ratio technique discussed earlier.  We then
show results from a number of basic data quality assessment checks,
including comparisons of the reconstructed energy spectra with Monte
Carlo results, assessments of the stability of the energy scale with
time (as quantified via Kurie fits to the $\beta$-decay endpoints),
and assessments of the statistical properties of the asymmetries under
the octet data-taking sequence.

\begin{table*}
\caption{Definitions of the different Analysis Choices for the
inclusion or exclusion of the various backscattering event types,
and the selection rules for the assignment of the electron's initial
direction.  ``Trigger Side'' refers to the scintillator generating the
trigger.}
\begin{ruledtabular}
\begin{tabular}{llll}
& \multicolumn{3}{l}{Assignment of Initial Direction for Backscattering
  Event Types} \\
Analysis Choice& Type 0& Type 1& Type 2/3 \\ \hline
1& Trigger Side& Earlier Trigger Side& Trigger Side \\
2& Trigger Side& Earlier Trigger Side& Omit \\
3 (default)& Trigger Side& Earlier Trigger Side& If Primary
  $E_{\text{MWPC}} > 4.14$ keV Trigger Side, Else Other Side \\
4& Trigger Side& Omit& Omit \\
5& Trigger Side& Earlier Trigger Side& Apply Likelihood Function
  Prob($E_{\text{MWPC}}$) for Assignment of Direction \\ \hline
6& Omit& Earlier Trigger Side& Omit \\
7& Omit& Omit& Trigger Side \\
8& Omit& Omit& If Primary
  $E_{\text{MWPC}} > 4.14$ keV Trigger Side, Else Other Side \\
9& Omit& Omit& Apply Likelihood Function
  Prob($E_{\text{MWPC}}$) for Assignment of Direction \\
\end{tabular}
\end{ruledtabular}
\label{tab:analysis_choices}
\end{table*}

\begin{figure}
\includegraphics[scale=0.90,clip=]{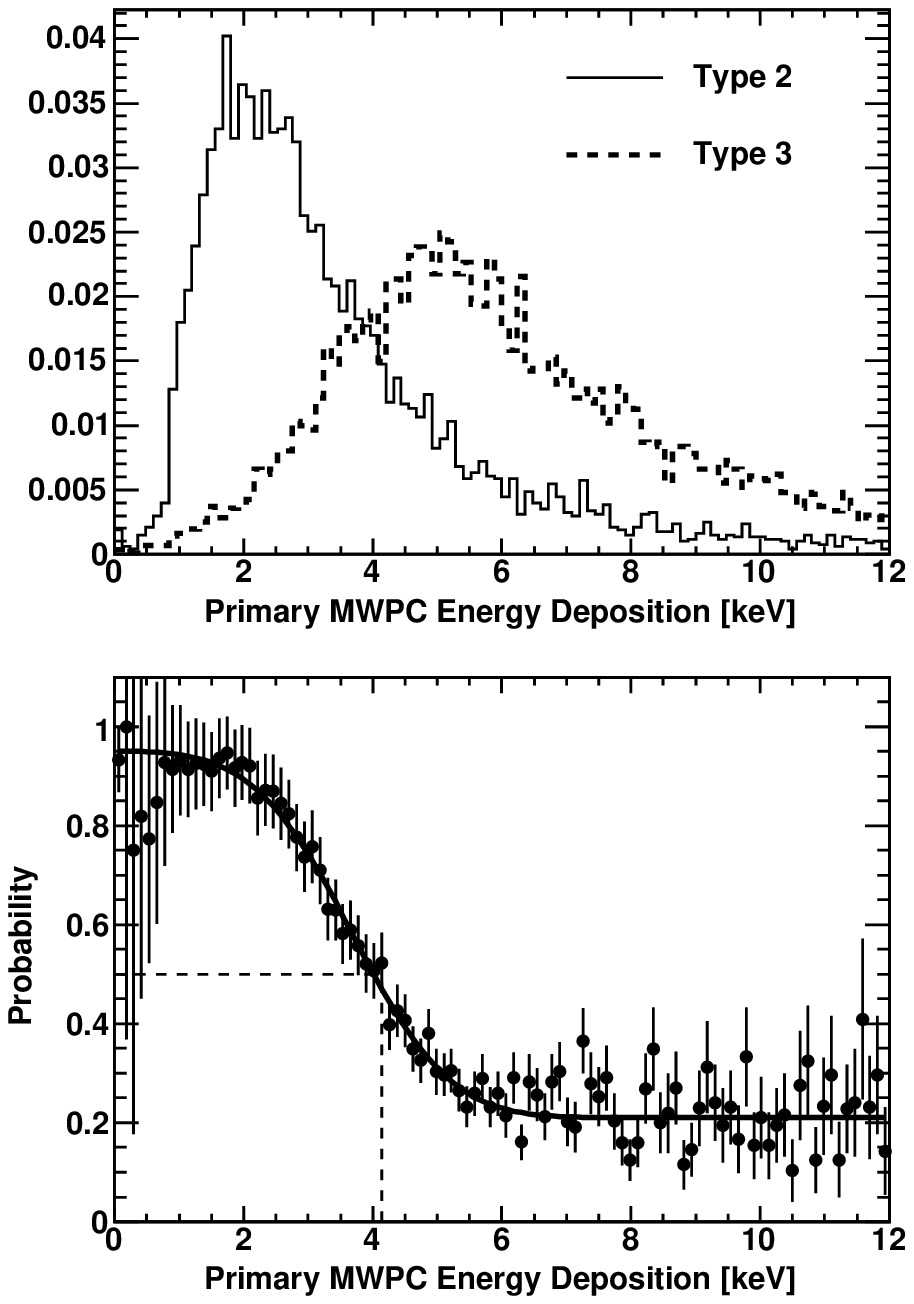}
\caption{Top panel: Monte Carlo calculations of the energy deposition
in the Primary MWPC for Type 2 and Type 3 neutron
$\beta$-decay events.  The histograms are normalized to unity.
Bottom panel: Monte Carlo calculations of the probability an event
identified in data analysis as a Type 2/3 event was a Type 2 event as
a function of the energy deposition in the Primary MWPC.  The dashed
lines indicate the threshold cut of $E_{\text{MWPC}} > 4.14$ keV for
the assignment of the initial direction to the triggering-scintillator
side in Analysis Choices 3 and 8 for the separation of Type 2/3
events.  The solid line is a fit to the calculations of the form
$\text{Prob}(E_{\text{MWPC}}) = -p_0 \text{erf}(p_1E_{\text{MWPC}} -
p_2) + p_3$, where $\text{erf}$ denotes the error function and the
$p_i$ are fit parameters.  This likelihood function was employed for
the separation of Type 2/3 events in Analysis Choices 5 and 9.}
\label{fig:monte_carlo_type2_type3}
\end{figure}

\subsection{Analysis Choices}
\label{sec:asymmetry_analysis_backscattering_choices}

\subsubsection{Definitions and Selection Rules}
\label{sec:asymmetry_analysis_backscattering_choices_definitions}

We extracted the asymmetries according to a number of different
Analysis Choices under which we included or excluded the various
backscattering event types and varied the selection rules for the
assignment of the electron's initial direction.  The motivation for
this study was to compare the variation of the measured asymmetry as a
function of the analysis choice with the variation predicted by the
Monte Carlo as a robust benchmark of our Monte-Carlo-calculated
corrections for backscattering and the $\cos\theta$-dependence of the
acceptance.

The selection rules for our various Analysis Choices,
numbered 1--9, are summarized in Table
\ref{tab:analysis_choices}.  Note that Analysis Choices 1--5 consider
Type 0 events, with different selections for the inclusion/exclusion
of Type 1 and 2/3 events, and further selection rules for the
identification of the initial direction of incidence for Type 2/3
events.  Analysis Choices 6--9 were included in order to study the
asymmetry from backscattering events for diagnostic purposes.

We employed Analysis Choice 3 as our default method for the extraction
of our final results for $A_0$, because this choice provided for
maximal use of detector information and minimized the magnitude of the
systematic corrections for backscattering and the
$\langle\cos\theta\rangle$ dependence of the acceptance.
Nevertheless, we show several results from the other Analysis Choices
below, as these results provide a powerful validation of our Monte
Carlo calculations.

\subsubsection{Treatment and Separation of Type 2 and Type 3 Events}
\label{sec:asymmetry_analysis_backscattering_choices_type23}

Note that in Analysis Choices 1 and 7 we assigned the initial
direction of incidence for Type 2/3 events simply to the
triggering-scintillator side.  However, we attempted in Analysis
Choices 3, 5, 8, and 9 to separate Type 2 and 3 events according to
selection rules on the energy response of the Primary MWPC (i.e.,
recall, the MWPC on the triggering scintillator side).  To illustrate,
the top panel of Fig.\ \ref{fig:monte_carlo_type2_type3} shows Monte
Carlo calculations of the energy deposition in the Primary MWPC for
Type 2 and Type 3 neutron $\beta$-decay events
separately (of course, such a direct separation was not possible in
data analysis).  The distinct spectra for these events suggested such
a separation could be performed via a probabilistic (likelihood)
approach, and the feasilibity of such was studied in Monte Carlo.
Results from this study are shown in the bottom panel of Fig.\
\ref{fig:monte_carlo_type2_type3}, where calculations of the
probability that an event identified in data analysis as a Type 2/3
event was actually a Type 2 event are plotted as a function of
$E_{\text{MWPC}}$.

The $E_{\text{MWPC}} > 4.14$ keV energy cut employed in Analysis
Choices 3 and 8 was chosen such that if a fixed Primary MWPC energy
cut was used to assign the initial direction of incidence for events
identified as Type 2/3 in data analysis to either the
triggering-scintillator side or the non-triggering-scintillator side,
the calculated probability for the incorrect assignment of the initial
direction to the triggering scintillator side was less than 50\%
(i.e., the calculated probability that the event was actually a Type 2
event).  This threshold cut is consistent with the conceptual
expectation (see, e.g., Fig.\ \ref{fig:event_types_schematic}) that an
observed Type 2/3 event which was actually a Type 2 event (i.e.,
initially incident on the non-triggering-scintillator side) would
deposit less energy in the Primary MWPC (one traversal) as compared to
a Type 3 event (two traversals).

\begin{figure}
\includegraphics[scale=0.47,clip=]{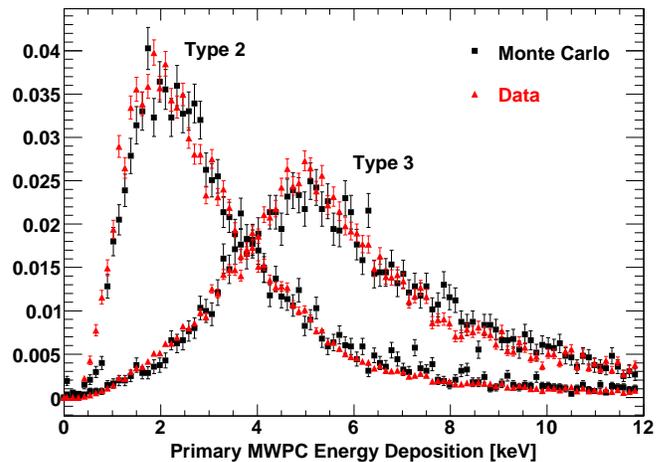}
\caption{(Color online) Comparison between data (red triangles) and
Monte Carlo (filled squares) for Primary MWPC energy spectra for
Geometry A Type 2 and Type 3 events, after separation of Type 2/3
events in data analysis according to the likelihood function
Prob($E_{\text{MWPC}}$).}
\label{fig:type2_type3_separation}
\end{figure}

In Analysis Choices 5 and 9 we then actually separated the Type 2/3
events according to the calculated likelihood function
Prob($E_{\text{MWPC}}$), such as that already shown in the bottom
panel of Fig.\ \ref{fig:monte_carlo_type2_type3}.  (Note that although
the curve shown there is from a Monte Carlo calculation for the
Geometry A configuration, the curves for the other Geometries were
similar.)  Fig.\ \ref{fig:type2_type3_separation} then demonstrates
the good agreement observed between the data and Monte Carlo for the
Primary MWPC energy spectra for Geometry A Type 2 and Type 3 events,
separated according to the likelihood function.\footnote{Note that
this good agreement was actually to be expected, because the
calibrated Primary MWPC energy spectra for Type 2/3 events agreed well
with Monte Carlo (as shown previously in Fig.\
\ref{fig:mwpc_energy_calibration_geometry_b_2}), and the Type 2 and
Type 3 separation follows from the Monte Carlo calculation of the
likelihood function.}

\subsection{Event Rates and Statistical Uncertainties}
\label{sec:asymmetry_rates_uncertainties}

For any given Analysis Choice, events were binned into 25 keV
$E_{\text{recon}}$ energy bins from 0--1200 keV, and then assigned an
initial direction of incidence according to the selection rules in
Table \ref{tab:analysis_choices}.  The measured event rate in each of
these $E_{\text{recon}}$ bins was then computed for each detector
according to
\begin{equation}
r_{\text{bin}} = \frac{N_{\text{bin}}}{T},
\end{equation}
where $N_{\text{bin}}$ denotes the number of events passing cuts in
that particular $E_{\text{recon}}$ bin, and $T$ denotes that
detector's live time (defined previously in Section
\ref{sec:analysis_data_quality_cuts}).

In energy bins containing $N < 20$ counts (e.g., in bins beyond the
$\beta$-decay endpoint for $\beta$-decay runs, or in nearly all bins
for the shorter background runs), where the assumption of
Gaussian errors approximated by Poisson
uncertainty is no longer valid,
we assigned a statistical uncertainty to the rate in
these bins according to the following procedure whose
key assumption was that the underlying background and neutron
$\beta$-decay spectral shapes did not change with time, even if the
run-by-run rates varied.

First, we generated high-statistics parent $\beta$-decay and
background spectra for each of the detectors (for a particular
Geometry and spin state) by combining many runs.  Second, on an
individual run-by-run basis, we then computed the measured rate for
each detector within an energy window,
$r^{\text{meas}}_{\text{window}}$, nominally 275--625 keV (see Section
\ref{sec:corrections_analysis_energy_window}), and compared this
measured rate with the rate for the parent spectrum,
$r^{\text{parent}}_{\text{window}}$, in this same energy window.
Third, the ratio of these two rates then defined a scaling factor $f
\equiv
r^{\text{meas}}_{\text{window}}/r^{\text{parent}}_{\text{window}}$ for
each of the detectors, which we then used to compute on a run-by-run
basis the expected rate for each energy bin,
$r^{\text{exp}}_{\text{bin}} = r^{\text{parent}}_{\text{bin}} \times
f$, for each detector.  Finally, the statistical uncertainty we then
assigned to the measured rate in each bin was then
\begin{equation}
\delta r_{\text{bin}} =
\sqrt{\frac{r^{\text{exp}}_{\text{bin}}}{T}},
\end{equation}
where $T$ again denotes that particular detector's (blinded) live
time.  For bins with $\geq 20$ counts, we employed the usual
$\sqrt{N}$ uncertainties.

For each $\beta$-decay and background run pair, we then subtracted on
a bin-by-bin basis the measured background rate from the measured
$\beta$-decay rate, with standard propagation of the statistical
errors calculated according to the above-described procedure for $N <
20$ or $N \geq 20$ counts.  We emphasize that this
procedure affected only the assignment of the statistical errors; the
actual measured counts were still employed for the calculation of the
run-by-run rates for each $\beta$-decay and background run pair.

\subsection{Asymmetry Extraction}
\label{sec:asymmetry_asymmetry_extraction}

We then extracted the experimental asymmetry on an
energy-bin-by-energy-bin basis for data grouped into either individual
spin-state run pairs, quartets, or octets (all of which were defined
previously in Section \ref{sec:measurements_run_cycle_octet}).

\subsubsection{Spin-State Pair Asymmetries}
\label{sec:asymmetry_asymmetry_extraction_spin_state}

For individual spin-state run pairings (i.e., A1--A6, A7--A12, B1--B6,
or B7--B12 run pairings in Table \ref{tab:octet_structure}), we
calculated the experimental measured asymmetries according to the
definition for the super-ratio of detector rates given previously in
Eq.\ (\ref{eq:individual_super_ratio}),
\begin{equation}
R = \frac{r_1^- \cdot r_2^+}{r_1^+ \cdot r_2^-},~~~~~
A_{\text{meas}} = \frac{1-\sqrt{R}}{1+\sqrt{R}}.
\end{equation}

\begin{figure*}
\includegraphics[scale=0.80,clip=]{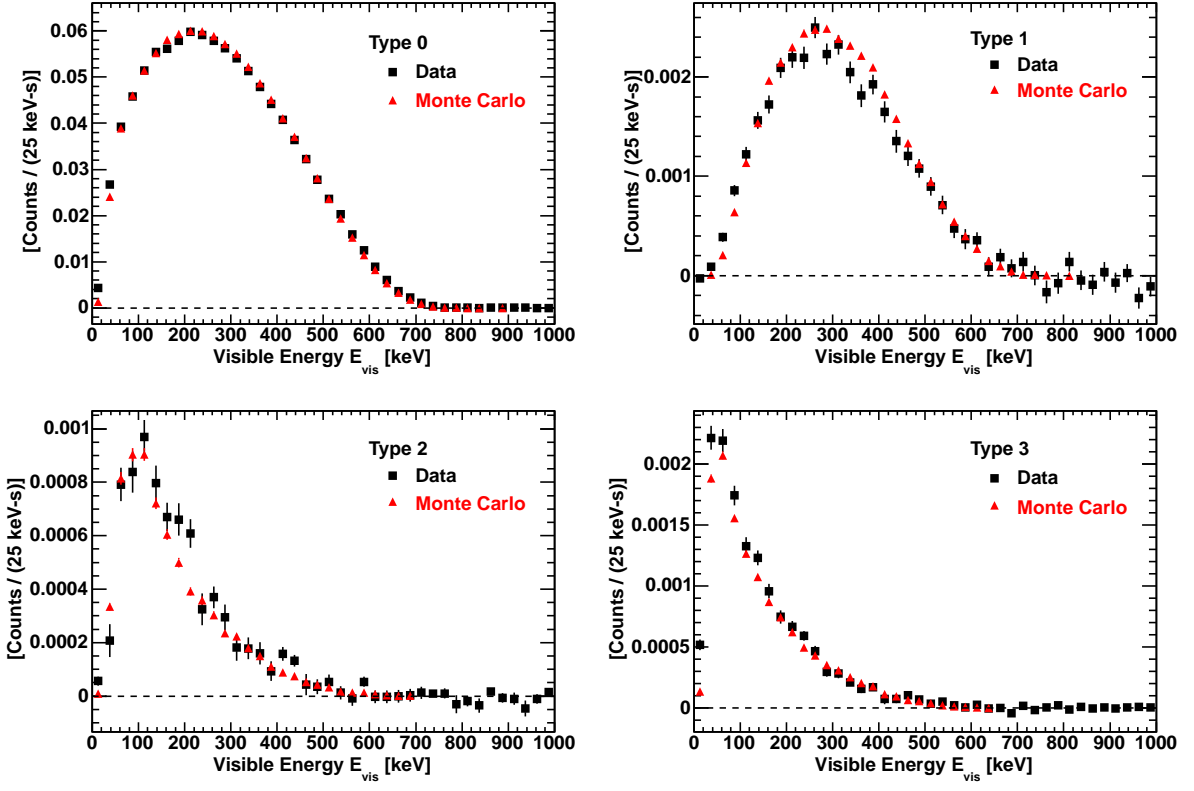}
\caption{(Color online) Comparisons of visible energy spectra for the
different event types extracted from data (black squares) and Monte
Carlo (red triangles) for the Geometry C configuration.  The same
selection rules for the event types were applied to both data and
Monte Carlo.  See text for details.  All error bars shown are
statistical; if not visible, the errors are smaller than the marker
size.}
\label{fig:visible_energy_data_mc_geometry_c}
\end{figure*}

\begin{figure*}
\includegraphics[scale=0.80,clip=]{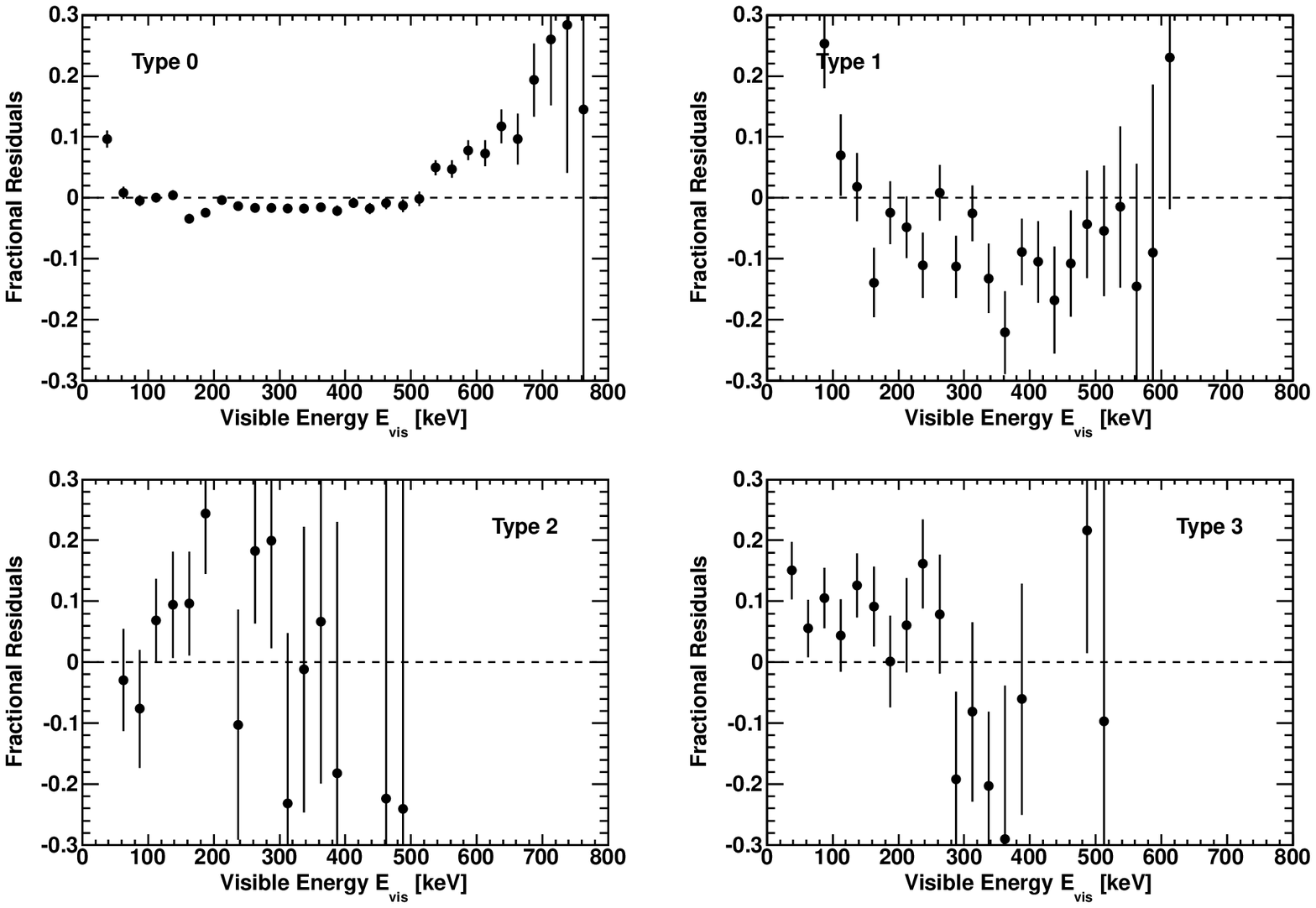}
\caption{Fractional residuals for the comparisons between the data and
Monte Carlo visible energy spectra shown in Fig.\
\ref{fig:visible_energy_data_mc_geometry_c}.  As can be seen in Fig.\
\ref{fig:visible_energy_data_mc_geometry_c}, there are very few Type 2
and Type 3 events with visible energies greater than $\sim 500$ keV.}
\label{fig:residuals_visible_energy_data_mc_geometry_c}
\end{figure*}

\subsubsection{Quartet Asymmetries}
\label{sec:asymmetry_asymmetry_extraction_quartet}

Conceptually, the asymmetry for a quartet run pairing (i.e., either
A-type A1--A12 or B-type B1--B12 runs in Table
\ref{tab:octet_structure}) can be computed from a ``summed super
ratio'' of detector rates,
\begin{equation}
R' = \frac{[r_1^{-(\text{A1,A2})} + r_1^{-(\text{A10,A12})}] \cdot
           [r_2^{+(\text{A4,A5})} + r_2^{+(\text{A7,A9})}]}
          {[r_1^{+(\text{A4,A5})} + r_1^{+(\text{A7,A9})}] \cdot
           [r_2^{-(\text{A1,A2})} + r_2^{-(\text{A10,A12})}]},
\end{equation}
where the notation is such that the run pairs within ($\cdots$)
parentheses indicate the $\beta$-decay run and ambient background run
background-subtraction pairing.  The expression for B-type runs is
similar.  The merit of this approach is that that linear background
and detector efficiency drifts cancel in this
definition of $R'$.  (This cancellation is exact provided that the
durations of the $\beta$-decay and background runs do not vary and
that any such linear drifts are constant in time.)

In practice, to properly account for the statistical uncertainty,
within each quartet we computed the statistically weighted mean of the
detector rates for each of the polarization states.
The resulting statistical uncertainties in the mean
rates were then propagated into the super ratio $R'$ and then into the
calculation of the asymmetry.  For example, we defined the detector
rate $r_1^+$ for an A-type quartet to be the statistically weighted
mean of the $r_1^{+(\text{A1,A2})}$ and $r_1^{+(\text{A10,A12})}$
rates.  The quartet-based asymmetry was then extracted from the super
ratio $R'$ of these quartet-averaged rates.

Note that a ``product super ratio'', $R''$, can also be
defined in terms of geometric means of detector rates for spin states as
\begin{equation}
R'' = \left[ \frac{[r_1^{-(\text{A1,A2})} r_1^{-(\text{A10,A12})}] \cdot
                   [r_2^{+(\text{A4,A5})} r_2^{+(\text{A7,A9})}]}
                  {[r_1^{+(\text{A4,A5})} r_1^{+(\text{A7,A9})}] \cdot
                   [r_2^{-(\text{A1,A2})} r_2^{-(\text{A10,A12})}]}
      \right]^{1/2}.
\end{equation}
We extracted asymmetries via both the summed and product super-ratio
approaches, and the central values from the two methods differed by
only 0.07\%.  However, the problem with the product super-ratio method
is that the resulting statistical uncertainty in $R''$ is dominated by
the rate from the run with the largest statistical uncertainty (e.g., if
one of the $\beta$-decay runs is significantly shorter in duration
than the others).  In contrast, the summed super-ratio method employs
inverse-square-uncertainty weighting.  Therefore, in our final
analysis we employed the summed super-ratio method.

\subsubsection{Octet Asymmetries}
\label{sec:asymmetry_asymmetry_extraction_octet}

Octet-based asymmetries were calculated similarly to quartet-based
asymmetries.  Now, for example, we defined the detector rate $r_1^+$
for a complete octet to be the statistically weighted mean of the
$r_1^{+(\text{A1,A2})}$, $r_1^{+(\text{A10,A12})}$,
$r_1^{+(\text{B4,B5})}$, and $r_1^{+(\text{B7,B9})}$ rates.  As with
the quartet-based asymmetry, the octet-based asymmetry was then
extracted from a summed super-ratio
$R'$ of these octet-averaged rates,
with propagation of the statistical uncertainties in
the octet-averaged rates through the super ratio and the asymmetry.
Again, the merit of the octet approach is that linear background
drifts cancel (subject to the same caveats as for the quartet
asymmetries).

\subsection{Comparison of Data and Monte Carlo Visible Energy Spectra}
\label{sec:asymmetry_data_monte_carlo_energy}

\subsubsection{Visible Energy Spectra}
\label{sec:asymmetry_data_monte_carlo_energy_spectra}

We now show an example (for the entire Geometry C data set) of
comparisons of the measured background-subtracted visible energy
$E_{\text{vis}}$ spectra with results from Monte Carlo
(\texttt{GEANT4} except where noted) calculations
(assuming the Particle Data Group value for $A_0$) in Fig.\
\ref{fig:visible_energy_data_mc_geometry_c} for the different event
types.\footnote{Note that comparing the measured $E_{\text{vis}}$
spectra with Monte Carlo is equivalent to comparing the measured
$E_{\text{recon}}$ spectra with Monte Carlo, because the mappings from
$E_{\text{vis}}$ values to $E_{\text{recon}}$ values were via the
parametrizations extracted from Monte Carlo calculations, discussed
previously in Section
\ref{sec:analysis_initial_energy_reconstruction}.}  The fractional
residuals, (Data $-$ Monte Carlo)/Data, are shown in Fig.\
\ref{fig:residuals_visible_energy_data_mc_geometry_c}.  Although not
shown explicitly in this article, we note that we achieved the same
level of agreement between data and Monte Carlo shown in Fig.\
\ref{sec:analysis_initial_energy_reconstruction} for all of the
Geometries and both detectors.

Both the measured and Monte Carlo data shown there were extracted from
events triggering one of the particular scintillators.  Type 2/3
events were separated in both data and Monte Carlo according to the
selection rules of Analysis Choice 3.  The Type 0, Type 2, and Type 3
$E_{\text{vis}}$ spectra shown there are the spectra observed in the
triggering scintillator, whereas the Type 1 spectra are summed over
both of the scintillators.

The spectra shown there are for one of the neutron spin states (in
particular, AFP spin-flipper off).  Note that the total measured
background-subtracted rate during runs when the neutron spin was
flipped with the AFP spin-flipper was $\sim 30$\%
less than that during runs for the unflipped spin state, due to UCN
losses along the transport guides between the AFP
spin-flip region and the SCS decay-trap volume after the 2-T
equivalent UCN energy increase following the spin flip.

\subsubsection{Backscattering Strengths and Fractions}
\label{sec:asymmetry_data_monte_carlo_energy_backscattering}

The normalization of the measured and Monte Carlo (\texttt{GEANT4})
spectra shown in Fig.\ \ref{fig:visible_energy_data_mc_geometry_c} was
performed according to their respective integral of the Type 0
spectrum over the complete visible energy range of 0--800 keV.  After
this relative normalization, the \texttt{GEANT4} spectra were further
internally normalized to account for a known deficit in the
\texttt{GEANT4} backscattering strength \cite{martin03,martin06}.
This was accomplished by applying two scale factors, $f_{\text{bulk}}$
(for backscattering from the scintillator bulk material) and
$f_{\text{thin}}$ (decay trap and MWPC thin windows), to the
\texttt{GEANT4} backscattering distributions such that the simulated
backscattering strengths matched the measured backscattering
strengths.  Specifically, we applied $f_{\text{bulk}}$ to Type 1 and
Type 3 events, whereas we applied $f_{\text{thin}}$ to Type 2 events,
because the former (latter) correspond to backscattering from the bulk
scintillator (MWPC windows, gaseous materials, etc.).  The numerical
values of these scale factors were $f_{\text{bulk}} = 1.3$ for all of
the Geometries, and $f_{\text{thin}} = 1.3$ (1.6) for Geometries A and
B (C and D).  The Type 1, Type 2, and Type 3 Monte Carlo spectra shown
in Fig.\ \ref{fig:visible_energy_data_mc_geometry_c} are scaled by
these scale factors.

For completeness, the measured event type fractions integrated over
the entire visible energy range from 0--800 keV for the different
Geometries are tabulated in Table \ref{tab:event_type_fractions}.

\begin{table}
\caption{Measured event type fractions for each Geometry integrated
over the entire visible energy range from 0--800 keV and averaged over
the detectors and the spin states.  Type 2/3 events were separated
according to Analysis Choice 3.}
\begin{ruledtabular}
\begin{tabular}{lcccc}
Geometry& Type 0& Type 1& Type 2& Type 3 \\ \hline
A& 0.947& 0.032& 0.011& 0.011 \\
B& 0.952& 0.030& 0.010& 0.008 \\
C& 0.930& 0.047& 0.009& 0.014 \\
D& 0.939& 0.039& 0.008& 0.014 \\
\end{tabular}
\end{ruledtabular}
\label{tab:event_type_fractions}
\end{table}

\subsection{Fitted Endpoint Distributions}
\label{sec:asymmetry_endpoint_distributions}

\begin{figure}
\includegraphics[scale=0.45,clip=]{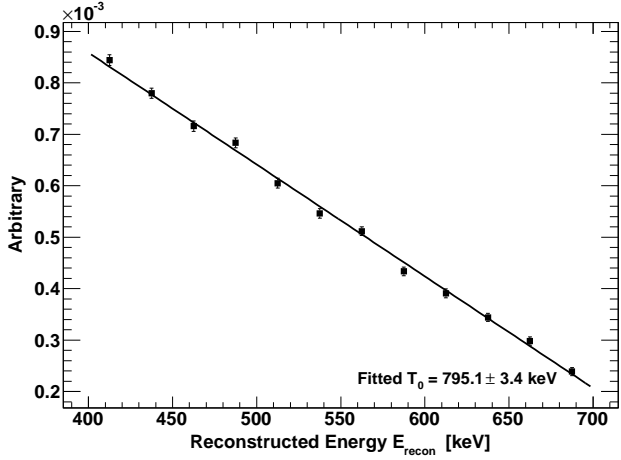}
\caption{Example of a Kurie fit to the (background-subtracted)
reconstructed energy $E_{\text{recon}}$ spectrum for a $\sim 1$-hour
long $\beta$-decay run.  The errors shown are statistical.}
\label{fig:kurie_fit}
\end{figure}

As a measure of the stability of the energy calibration with time,
background-subtracted $E_{\text{recon}}$ spectra for each
$\beta$-decay run were converted to Kurie plots and fitted over the
range of 400--700 keV.  An example of such a fit to the
$E_{\text{recon}}$ spectrum from a $\sim 1$-hour long $\beta$-decay
run is shown in Fig.\ \ref{fig:kurie_fit}.  There, we fitted the
measured rate, binned in energy, $dW/dE_e$, to the function
\begin{equation}
\frac{1}{p_e}\sqrt{\frac{dW}{dp_e}} =
\frac{1}{p_e}\sqrt{\frac{dW}{dE_e}\frac{dE_e}{dp_e}} \propto (T_0 - T_e),
\end{equation}
where $T_0$ denotes the fitted endpoint (kinetic) energy, $p_e =
\sqrt{T_e^2 + 2T_e m_e}$ is the magnitude of the electron momentum,
$T_e$ ($= E_{\text{recon}}$) is the kinetic energy, and $dE_e/dp_e =
p_e/E_e = p_e/(T_e + m_e)$.

Fits to Kurie plots of Monte Carlo spectra for the two spin states (in
order to account for the $\beta\cos\theta$ factor in the angular
distribution) which included the energy-dependent recoil order effects
and the finite scintillator energy resolution yielded fitted values
for the endpoint of $\sim 787$ keV (with a $\sim \pm 0.4$ keV
difference for the two spin states).

\begin{figure}
\includegraphics[scale=0.90,clip=]{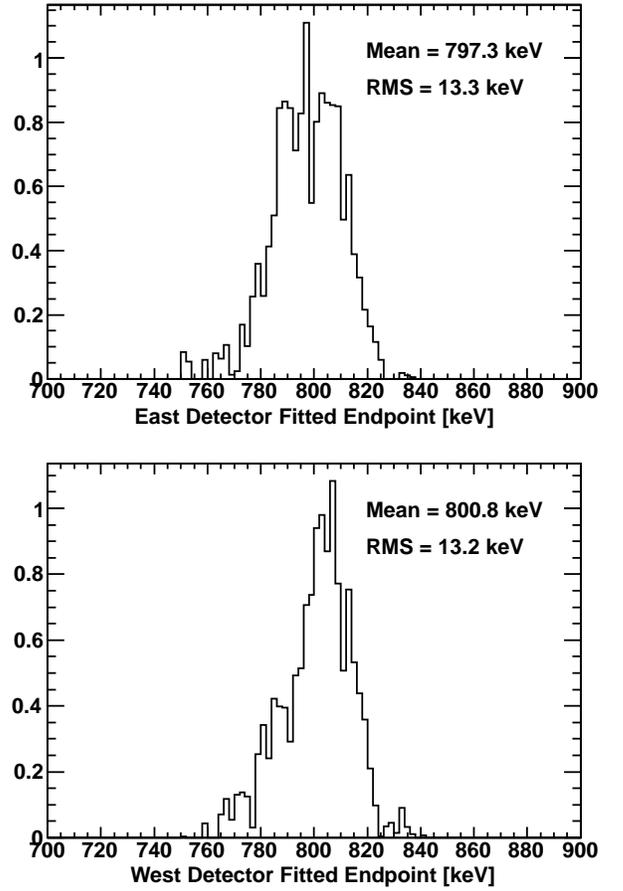}
\caption{Distributions of fitted endpoints for the two detectors,
averaged over Geometries and spin states.  The histogram contents
were weighted by the inverse square of their uncertainties.}
\label{fig:kurie_fit_distributions}
\end{figure}

Distributions of fitted endpoints for the two detectors (employing
``Fit 1'' for the $E_{\text{recon}}$ reconstruction), summed over all
Geometries and spin states and weighted by the inverse square of their
statistical uncertainties are shown in Fig.\
\ref{fig:kurie_fit_distributions}.  For both detectors, typical
run-to-run fluctuations (as characterized by the RMS) were less than
$\sim 13$ keV.  However, there was a systematic $\sim 10$--14 keV
difference between the mean fitted endpoints and the Monte Carlo
result of $\sim 787$ keV, which is addressed later in Section
\ref{sec:uncertainties_energy_reconstruction}.

\subsection{Statistical Properties of the Asymmetries}
\label{sec:asymmetry_tatistical_properties}

\begin{figure}
\includegraphics[scale=0.95,bb= 13 5 267 379]{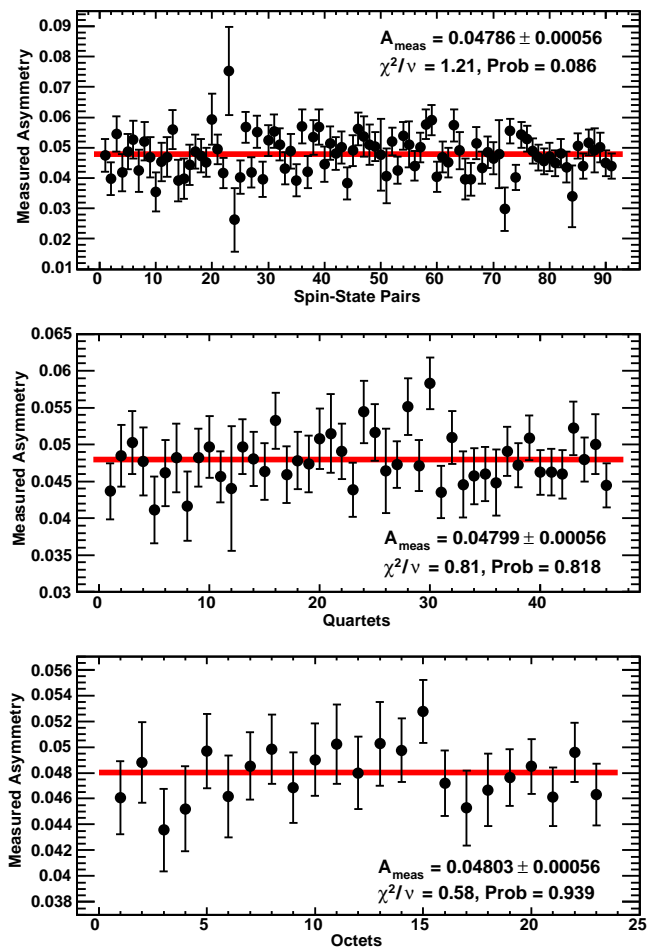}
\caption{Extracted values for the (blinded) measured asymmetries,
$A_{\text{meas}}$, from the Geometry B data set obtained under Analysis
Choice 3 for spin-state run pairings (top panel), quartet run pairings
(middle panel), and octet run pairings (bottom panel).}
\label{fig:asymmetries_pair_quartet_octet_geometry_b}
\end{figure}

As a demonstration of the utility of our octet data-taking procedure,
we extracted values for the measured blinded asymmetries, $A_{\text{meas}}$,
under Analysis Choice 3 integrated over a 225--675 keV energy window
for the three different run groupings discussed previously: spin-state
pairs, quartets, and octets.  Incomplete octets were not discarded;
instead, we retained individual spin-state pairs or quartets within
incomplete octets in our analysis.

Sample results from an analysis of the Geometry B data set are shown
in Fig.\ \ref{fig:asymmetries_pair_quartet_octet_geometry_b}.
{As can be seen there, the $\chi^2/\nu$ value for the
quartet analysis is improved over that for the spin-state pair
analysis.  Further, there is a slight shift in the central value of
the asymmetry between the spin-state pair analysis and the quartet
analysis.  If there were linear drifts in the backgrounds and/or
detector efficiencies, this is the expected result, as any such linear
drifts would bias the spin-state pair analysis, but would cancel in
the quartet analysis.  In comparing all of the Geometries, the
$\chi^2/\nu$ values for the octet analyses of Geometries A, B, C, and
D were 30.1/29, 12.8/22, 20.1/13, and 6.5/7, respectively (i.e., the
relatively small $\chi^2/\nu$ value for Geometry B was not
representative of the entire data set).

All of the asymmetry results presented hereafter were obtained under
the octet analysis.

\section{Systematic Corrections for $\bm{A_0}$ Extraction}
\label{sec:corrections}

In this section we discuss the systematic corrections for
backscattering, the $\cos\theta$-dependent acceptance (hereafter also
termed the ``angle effect''), and the Standard Model recoil-order and
radiative corrections that were applied to the measured asymmetries in
order to extract the desired $\beta$-asymmetry parameter $A_0$.  The
Monte Carlo corrections for the backscattering and angle effect
corrections presented here were based on our \texttt{GEANT4} Monte
Carlo simulation code.  However, as noted earlier, we also developed a
\texttt{PENELOPE} simulation code and, in general, we obtained good
agreement between the \texttt{GEANT4}- and
\texttt{PENELOPE}-calculated corrections.  The small differences
between the results from these two simulation programs are discussed
in the context of the systematic uncertainty for the backscattering
and $\cos\theta$-dependent acceptance corrections later in Section
\ref{sec:uncertainties_angle_effects_backscattering}.


\subsection{Monte Carlo Benchmark: Scaling of Asymmetries with Analysis Choice}
\label{sec:corrections_asymmetries_analysis_choices}

First, as a benchmark of the validity of our Monte Carlo treatment of
the angle and backscattering effects, we demonstrate that the scaling
of our (still-blinded) measured asymmetries, $A_{\text{meas}}$, with
the Analysis Choice (calculated according to the selection rules in
Table \ref{tab:analysis_choices}) is consistent with our Monte Carlo
predictions.  An example of this is shown in Fig.\
\ref{fig:asymmetries_analysis_choice_geometries}, where we have
plotted the (blinded) measured asymmetries $A_{\text{meas}}$ and both
the \texttt{GEANT4} and \texttt{PENELOPE} Monte Carlo predictions for
$A_{\text{meas}}$, integrated over an energy window of 225--675 keV,
as a function of the Analysis Choice for each of the Geometries.  Note
that the Monte Carlo predictions shown there were arbitrarily
scaled (for effectively an arbitrary $A_0$) in order
to match the data at Analysis Choice 3 (given that
the measured asymmetries shown there were blinded); the point of this
exercise was to demonstrate that the scaling of the Monte
Carlo-calculated asymmetries with the Analysis Choice matched that of
the measured asymmetries.  [An absolute comparison would have required
scaling each by the (unknown) unblinded values of $A_0$ for each
Geometry.]

\begin{figure*}
\includegraphics[scale=0.80,clip=]{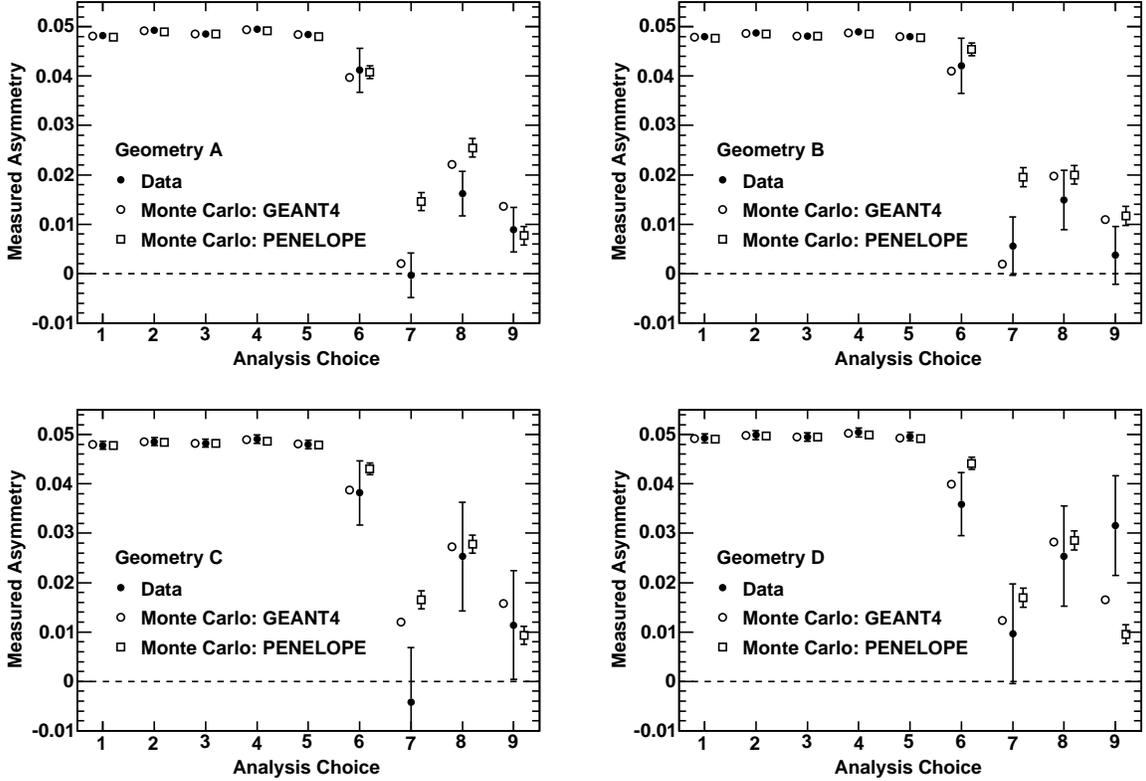}
\caption{Comparison of the scaling of
the (blinded) measured asymmetries extracted from the data,
$A_{\text{meas}}$, and the Monte Carlo-calculated
asymmetries (from both \texttt{GEANT4} and \texttt{PENELOPE}) with
the Analysis Choice for each of the Geometries.  The asymmetries shown
here were integrated over an energy window of 225--675 keV.  The
results from the Monte Carlo calculations were scaled for each
Geometry in order to match the data at Analysis Choice
3.  All errors are statistical, with the \texttt{GEANT4} statistical
errors smaller than the symbol size.}
\label{fig:asymmetries_analysis_choice_geometries}
\end{figure*}

This agreement between the scaling of our measured and
Monte-Carlo-calculated results for $A_{\text{meas}}$ with the Analysis
Choice (and, thus, upon the inclusion or exclusion of the different
backscattering event types, with different selection rules therein for
identification of the backscattering event types) provides a powerful
demonstration of the validity of our Monte Carlo-calculated
corrections for backscattering and the $\cos\theta$-dependence of the
acceptance, and also validates our application of the scale factors
$f_{\text{bulk}}$ and $f_{\text{thin}}$ to the \texttt{GEANT4}
backscattering distributions, discussed previously in Section
\ref{sec:asymmetry_data_monte_carlo_energy_backscattering}.

Hereafter, all of the results shown for the asymmetries and the Monte
Carlo-calculated systematic corrections to the asymmetries were
obtained under the default Analysis Choice 3, the motivation for which
was noted previously in Section
\ref{sec:asymmetry_analysis_backscattering_choices_definitions}.

\subsection{Asymmetry Unblinding}
\label{sec:analysis_unblinding}

All of the asymmetry results shown hereafter constitute our final
(unblinded) results.  We note that during our actual analysis of the
blinded data, we did not unblind our asymmetries
until after all of the (already discussed) calibrations and cuts, the
systematic corrections and uncertainties now being discussed, and the
analysis energy window (discussed later in Section
\ref{sec:corrections_analysis_energy_window}) were finalized.  At that
time, the asymmetries were unblinded by removing the scale
factors applied to the detector rates which were used to blind the
data (as discussed previously in Section \ref{sec:analysis_blinding}).

We emphasize that our final results for $A_0$ presented later were
those obtained at the time the data were unblinded; no further data
analysis was conducted after the unblinding.

\subsection{Results for Measured Asymmetries}
\label{sec:corrections_asym_meas}


Our resulting measured asymmetries,
$A_{\text{meas}}(E_{\text{recon}})$, are shown in Fig.\
\ref{fig:spectra_asym_meas} as a function of $E_{\text{recon}}$ in 25
keV bins over the entire detectable energy range, 50--800 keV, for
each of the Geometries.  There, we also show for each Geometry the
measured background spectrum and the resulting bin-by-bin
background-subtracted $\beta$-decay spectrum.

\begin{figure*}
\includegraphics[scale=0.97,clip=]{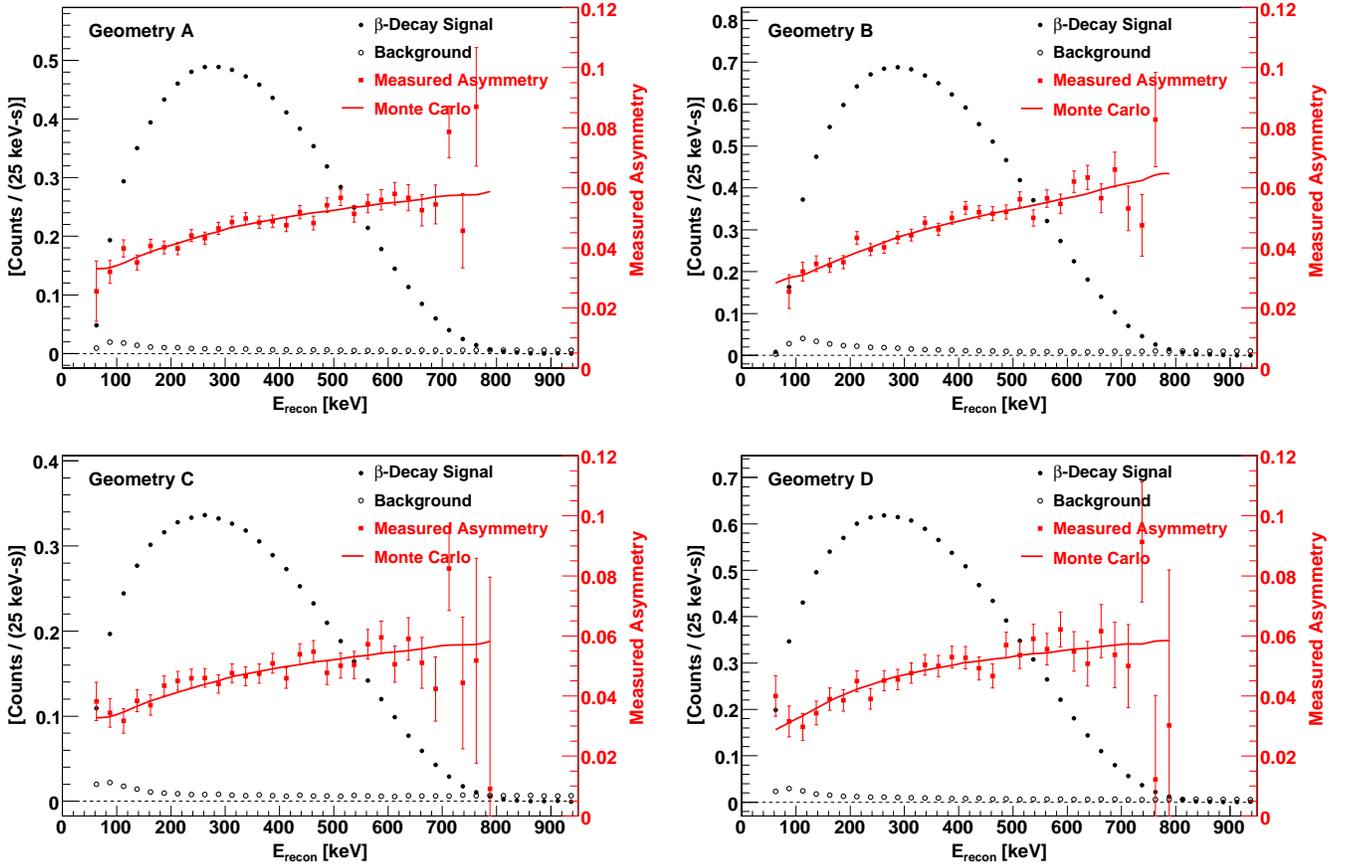}
\caption{(Color online) Results for
the measured background $E_{\text{recon}}$ energy spectra (open
circles) and the background-subtracted neutron $\beta$-decay
$E_{\text{recon}}$ energy spectra (filled circles) for each of the
Geometries.  The spectra shown here were summed over both detectors
and then averaged over the two neutron spin states.  The resulting
measured asymmetries $A_{\text{meas}}$ (filled red squares; see
vertical scale on the right) are shown as a function of
$E_{\text{recon}}$ for each of the Geometries.  These
are compared with the Monte Carlo calculated values for the measured
asymmetries (solid red lines).  The energy dependence of the measured
asymmetries is primarily due to the $\beta\cos\theta$
dependence of the measured asymmetries [see Eq.\
(\ref{eq:measured_asymmetry})]; there are also $\langle
\beta\cos\theta \rangle$ acceptance and backscattering effects (e.g.,
compare Geometry B).  All errors are statistical.}
\label{fig:spectra_asym_meas}
\end{figure*}

\subsection{Overview of Method for Systematic Corrections to Extract
$\bm{A_0}$}
\label{sec:corrections_overview}

For a given energy bin in ``true initial energy'', the true
energy-dependent physics asymmetry under the Standard Model would be
$P_n A_0 \langle \beta_{\text{true}} \cos\theta \rangle (1 +
f_{\text{RO}})$, where $P_n$ is the neutron polarization, $\langle
\beta_{\text{true}} \cos\theta \rangle$ denotes the average value of
the product of the true electron velocity in units of $c$ with the
true electron pitch angle $\cos\theta$ for that energy bin, and
$f_{\text{RO}}$ denotes the (energy-dependent) recoil-order physics
correction to $A_0$.  For now, we have omitted the small
$\mathcal{O}(0.1\%)$ radiative correction to the asymmetry discussed
in the Introduction to this article.

Therefore, it is clear that an extraction of the desired
$\beta$-asymmetry parameter $A_0$ from the (energy-dependent) measured
asymmetry $A_{\text{meas}}(E_{\text{recon}})$ in each
$E_{\text{recon}}$ bin requires a correct reconstruction of the mean
(initial) energy in each energy bin, a correction to the measured
asymmetry for missed backscattering, and a calculation of the mean
value of $\langle \beta_{\text{recon}} \cos\theta \rangle$ in that
energy bin.  For now, we will assume that in each
particular energy bin $\langle E_{\text{recon}} \rangle = \langle
E_{\text{true}} \rangle$, or $\langle \beta_{\text{recon}} \rangle =
\langle \beta_{\text{true}} \rangle$; later, we will explore the
implication for a systematic uncertainty to the asymmetry resulting
from a possible $E_{\text{recon}} \neq E_{\text{true}}$ error, such as
from an energy calibration error.  Note that even under the assumption
that $E_{\text{recon}} = E_{\text{true}}$ (i.e., ``perfect''
calibration), an acceptance correction must still be applied for the
average value of $\langle \beta \cos\theta \rangle$ in each energy
bin, because the electron energy loss (in the decay trap end-cap
foils, MWPC windows, etc.) is strongly angle dependent (hence the
designation ``angle effect''), with the energy loss a monotonically
increasing function of the pitch angle.  Therefore, each
$E_{\text{recon}}$ bin includes a distribution of events in initial
true energy and pitch angle.

We used our Monte Carlo simulation code to compute these required
corrections for missed backscattering and the $\langle \beta
\cos\theta \rangle$ angle effect.
After extraction of the measured asymmetry
$A_{\text{meas}}(E_{\text{recon}})$ in each $E_{\text{recon}}$ energy
bin, we then applied the following corrections in a sequential manner
in order to extract $A_0$ from $A_{\text{meas}}(E_{\text{recon}})$:

\subsubsection{Asymmetry Energy Dependence and $\beta$-Decay Spectra}
\label{sec:corrections_overview_energy_dependence}

(1) First, we made a first-order bin-by-bin correction for the
energy-dependence of the measured asymmetries,
$A_{\text{meas}}(E_{\text{recon}})$, by calculating an asymmetry
$A_1(E_{\text{recon}})$, defined by
\begin{equation}
A_1(E_{\text{recon}}) \equiv
A_{\text{meas}}/(\frac{1}{2}\beta_{\text{recon}}),
\end{equation}
where $\beta_{\text{recon}}$ was calculated for the central value of
each $E_{\text{recon}}$ bin for each event class.
Note that this first-order correction
simply assumed a uniform value (or distribution) of
$\beta_{\text{recon}}$ within each energy bin, and also a symmetric
value for $\langle \cos\theta \rangle$ of 1/2 in each energy bin.

\subsubsection{Backscattering Corrections}
\label{sec:corrections_overview_backscattering}

(2) Second, we applied a bin-by-bin correction for missed (or
misidentified) backscattering events, with the result of this
correction $A_2(E_{\text{recon}})$ defined to be
\begin{equation}
A_2(E_{\text{recon}}) \equiv A_1(E_{\text{recon}})(1 + \Delta_2),
\end{equation}
where $\Delta_2 \equiv \Delta_{2,0} + \Delta_{2,1} + \Delta_{2,2}$
represents the total backscattering correction in each
$E_{\text{recon}}$ bin, which we define to be the sum of the
individual corrections for events misidentified as Type 0, Type 1, and
Type 2/3 events, respectively.  The fractional
correction factors $\Delta_{2,i}$ were extracted from Monte Carlo.

\begin{figure*}
\includegraphics[scale=0.90,clip=]{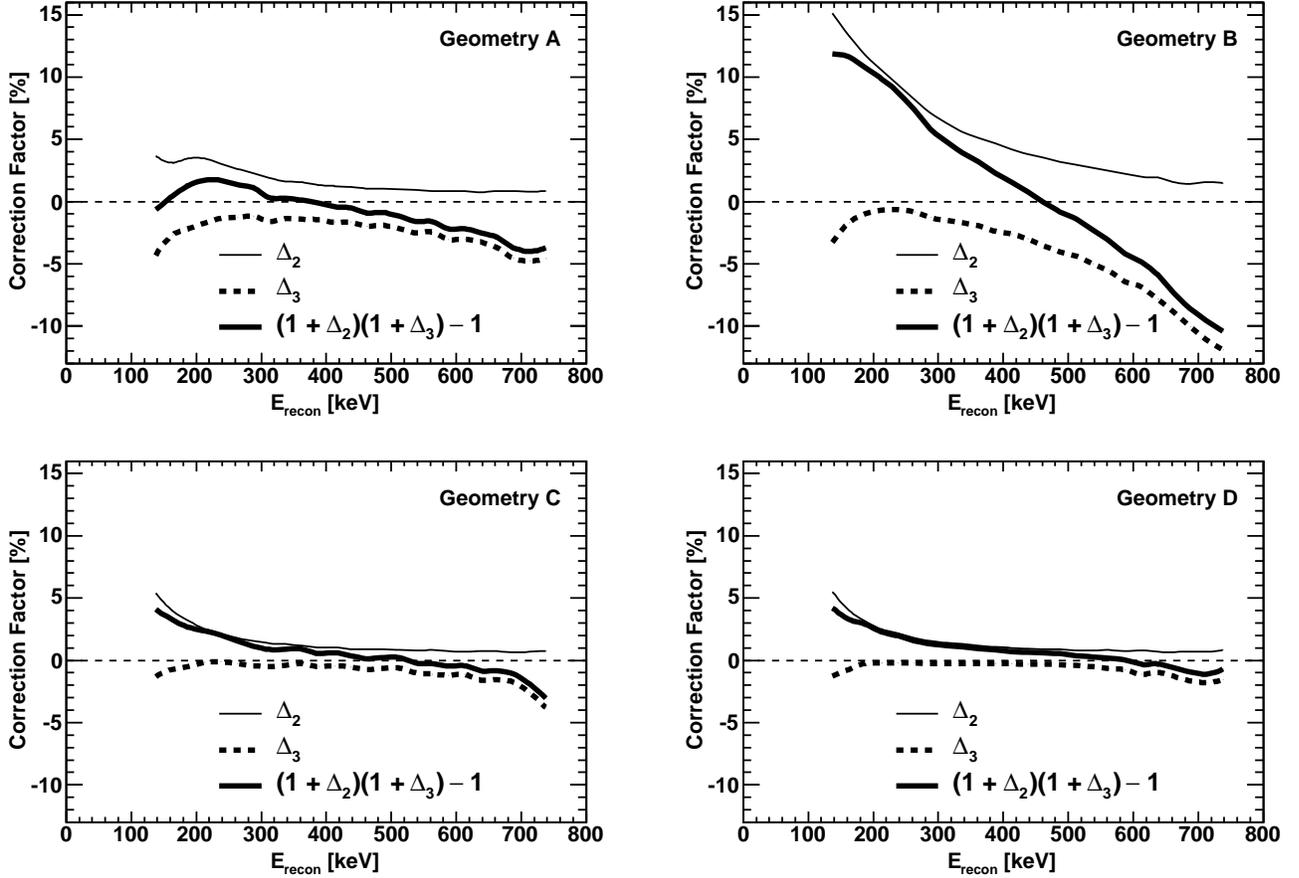}
\caption{Results from \texttt{GEANT4} Monte Carlo calculations of the
energy dependence of the relative sizes of the $\Delta_2$
backscattering (thin solid line) and $\Delta_3$ angle effect (thick
dashed line) systematic corrections.  The energy dependence of the
relative size of the combined correction $(1+\Delta_2)(1+\Delta_3)$ is
then shown as the thick solid line.  The systematic uncertainties in
these calculations are discussed in detail later in Section
\ref{sec:uncertainties_angle_effects_backscattering}.}
\label{fig:delta2_delta3_corrections}
\end{figure*}

\subsubsection{$\langle \beta\cos\theta \rangle$ Acceptance Corrections}
\label{sec:corrections_overview_betacostheta}

(3) Third, for each event class
we applied a bin-by-bin correction for the $\langle
\beta\cos\theta \rangle$ angle effect, with the result of this
correction $A_3(E_{\text{recon}})$ defined to be
\begin{eqnarray}
A_3(E_{\text{recon}}) &=& A_2(E_{\text{recon}}) \times
\frac{1}{2}\beta_{\text{recon}} / \langle
\beta_{\text{true}}\cos\theta \rangle \nonumber \\
&\equiv& A_2(E_{\text{recon}})(1 + \Delta_3).
\label{eq:delta3}
\end{eqnarray}
Again, the fractional correction factors $\Delta_3$
were extracted from Monte Carlo.

\subsubsection{Standard Model Corrections}
\label{sec:corrections_overview_standard_model}

(4) After application of the above three corrections, the resulting
asymmetry $A_3(E_{\text{recon}})$ is proportional to the product of
$P_n$, $A_0$, and the (energy-dependent) recoil-order and radiative
corrections (which are calculable under the Standard Model).  As
a final step, we extracted a value for $A_0$ in each
$E_{\text{recon}}$ bin
according to
\begin{eqnarray}
&&A_0(E_\text{recon})/P_n = \nonumber \\
&&~~~~~A_3(E_{\text{recon}}) [1 +
\Delta_{\text{RO}}(E_\text{recon})][1 +
\Delta_{\text{rad}}(E_\text{recon})], \nonumber \\
\end{eqnarray}
where $\Delta_{\text{RO}}(E_\text{recon})$ and
$\Delta_{\text{rad}}(E_\text{recon})$ denote the energy-dependent
recoil-order and radiative corrections, respectively, for that
particular $E_{\text{recon}}$ bin}.  In the absence of
new physics, there should be no residual energy dependence to the
values of the asymmetries $A_0(E_\text{recon})$ extracted from all of
the $E_{\text{recon}}$ bins after application of these Standard Model
corrections; therefore, the final energy-averaged value for $A_0$ will
be the statistically-weighted average of the bin-by-bin
$A_0(E_{\text{recon}})$ values,
\begin{equation}
A_0 = \langle A_0(E_\text{recon})\rangle.
\end{equation}

We now describe the results from our Monte Carlo calculations of the
$\Delta_{2,i}$ and $\Delta_3$ correction factors in more detail.

\subsection{Results for Monte Carlo Corrections}
\label{sec:corrections_monte_carlo_results}

\begin{figure*}
\includegraphics[scale=0.90,clip=]{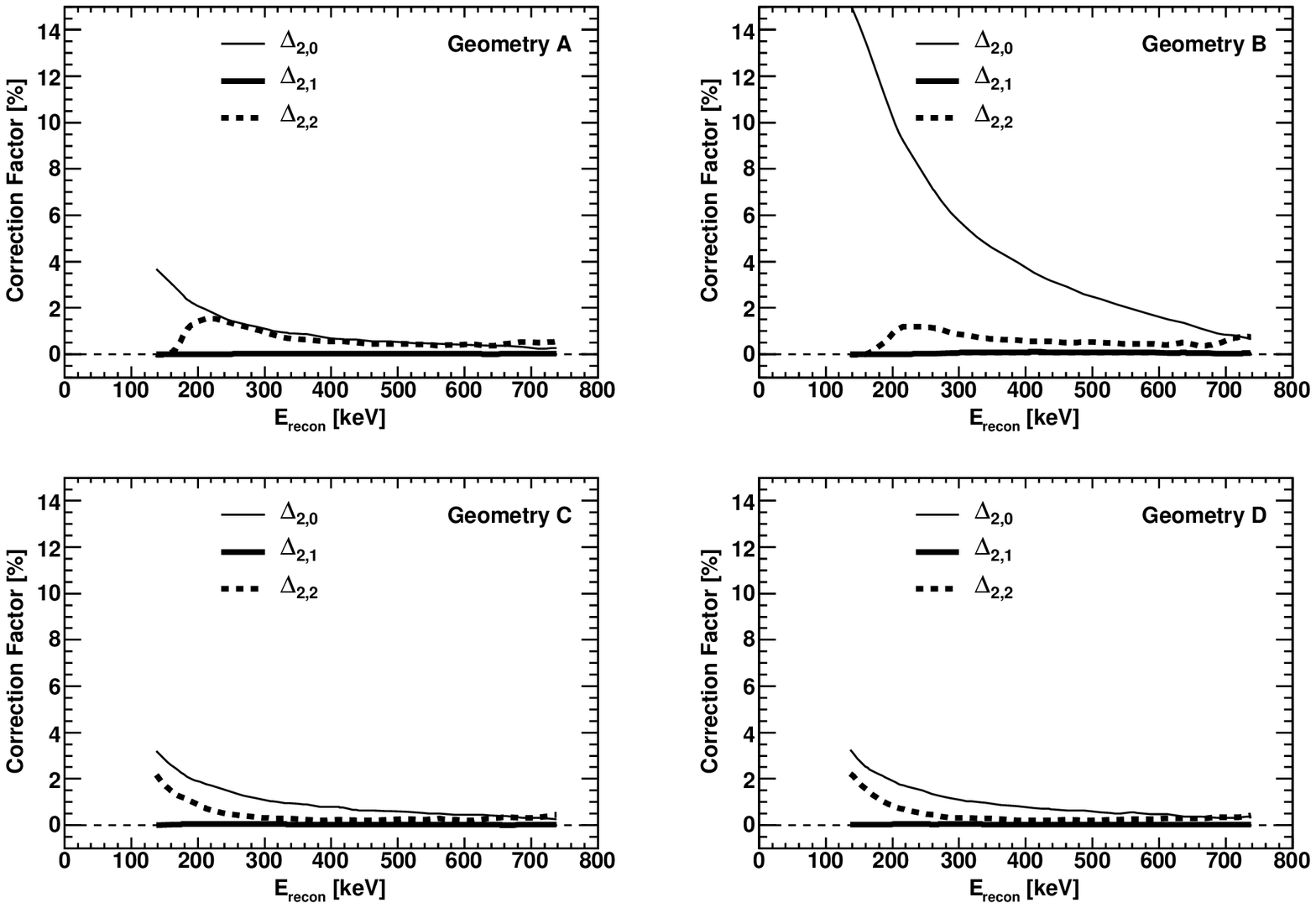}
\caption{Results from \texttt{GEANT4} Monte Carlo calculations of the
relative sizes of the individual $\Delta_{2,0}$ (thin solid line),
$\Delta_{2,1}$ (thick solid line), and $\Delta_{2,2}$ (thick dashed
line) contributions to the total $\Delta_2$ backscattering systematic
correction.  The systematic uncertainties in these
calculations are discussed in detail later in Section
\ref{sec:uncertainties_angle_effects_backscattering}.  [The
$\Delta_{2,1}$ backscattering correction is non-zero because some
small fraction of events will initially backscatter from the
decay-trap end-cap foils, and then undergo Type 1 backscattering
(however, the reconstructed initial direction will then be incorrect).
The probability for this type of event increases with decreasing
energy, and thus the acceptance for this type of events was suppressed
in Geometries A and B (i.e., thicker MWPC windows).]}
\label{fig:delta2_corrections_decomposed}
\end{figure*}

\begin{figure}
\includegraphics[scale=0.44,clip=]{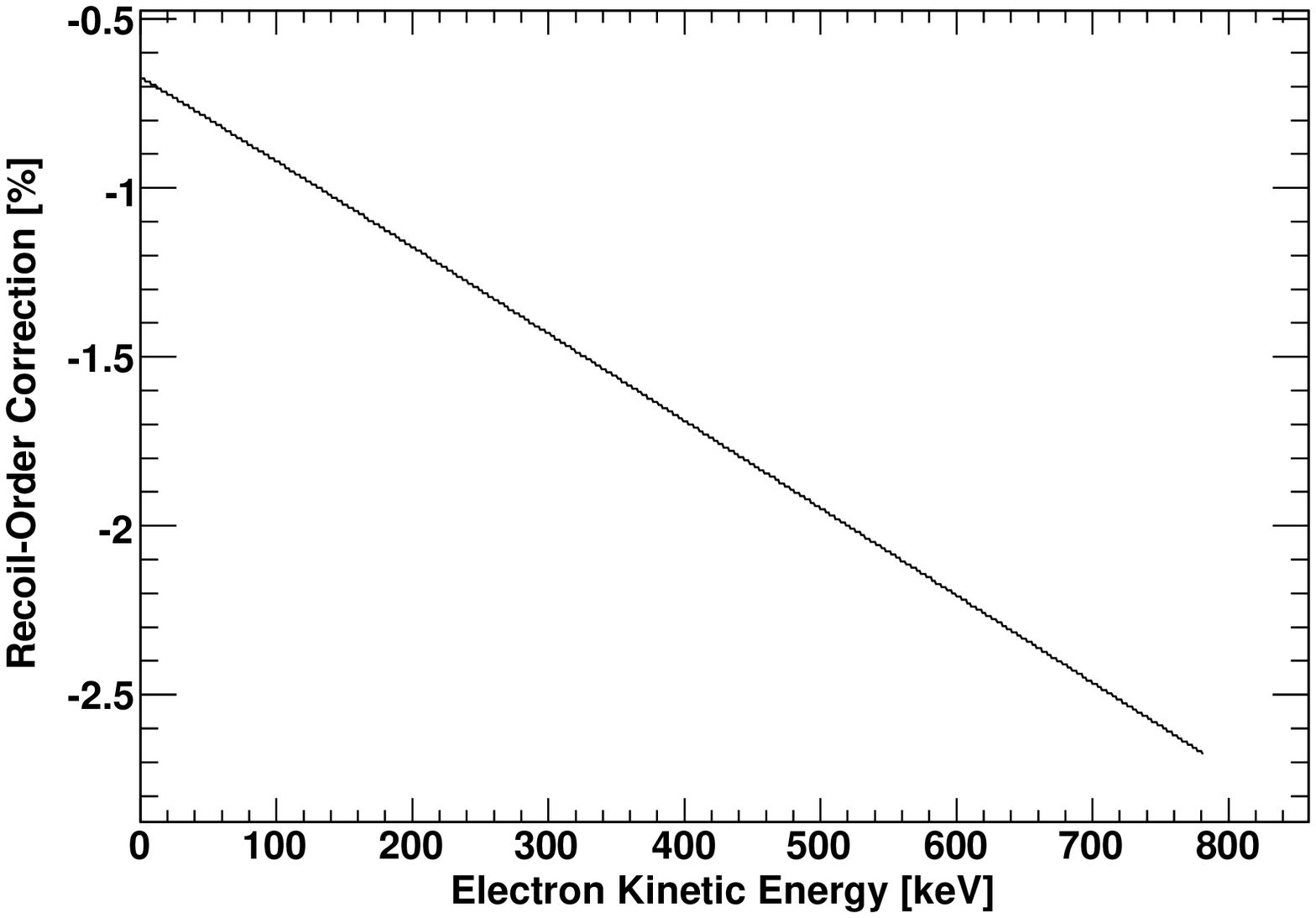}
\caption{Energy dependence of the Standard Model recoil-order
correction (calculated according to the formalism of
\cite{wilkinson82,gardner01}) for an extraction of $A_0$ from the
measured asymmetry $A_3$.}
\label{fig:recoil_order_correction}
\end{figure}

Monte Carlo calculations of the $\Delta_{2,i}$ and $\Delta_3$
correction factors were carried out for each of the Geometries, thus
accounting for their different foil thicknesses, their different
measured MWPC efficiencies (recall the discussion of such in Section
\ref{sec:analysis_mwpc_efficiency}), and their different two-fold PMT
trigger threshold functions (Section
\ref{sec:analysis_trigger_efficiency}).  To account for the
previously-discussed (Section
\ref{sec:asymmetry_data_monte_carlo_energy_backscattering}) known
deficit in the \texttt{GEANT4} backscattering strengths,
backscattering events in the Monte Carlo were re-weighted, with a
scaling factor of $f_{\text{bulk}}$ applied to Type 1 and Type 3
events, and $f_{\text{thin}}$ applied to Type 2 events and those
events misidentified as Type 0 events (i.e., the ``missed''
backscattering events discussed in Section
\ref{sec:analysis_event_type_definitions}).  However,
note that the $\Delta_3$ correction factors were evaluated in Monte
Carlo without the $f_{\text{bulk}}$ and $f_{\text{thin}}$ re-weighting
factors for the backscattering events.  After application of the
$\Delta_{2,i}$ backscattering corrections, the remaining $\Delta_3$
correction factor accounts for the distortion of the detected $\langle
\beta_{\text{recon}} \cos\theta \rangle$ acceptance from the ``true''
$\langle \frac{1}{2} \beta_{\text{true}} \rangle$ distribution.  Therefore,
employing these $f_{\text{bulk}}$ and $f_{\text{thin}}$ re-weighting
factors for the evaluation of the $\Delta_3$ correction factor in
Monte Carlo would have biased the angular distribution of the simulated
events, as the probablity for backscattering increases with angle.

The results of these calculations are plotted in Figs.\
\ref{fig:delta2_delta3_corrections} and
\ref{fig:delta2_corrections_decomposed} as a function of
$E_{\text{recon}}$ for each of the Geometries.  Figure
\ref{fig:delta2_delta3_corrections} shows the calculated results for
the $\Delta_2$ and $\Delta_3$ corrections, and the size of their
combined correction, $(1 + \Delta_2)(1 + \Delta_3)$.  For
completeness, Fig.\ \ref{fig:delta2_corrections_decomposed} then shows
the sources of the individual $\Delta_{2,i}$ contributions to
$\Delta_2$.  All of the corrections shown here were calculated for the
default Analysis Choice 3.  The systematic uncertainties in these
calculations are discussed in detail later in Section
\ref{sec:uncertainties_angle_effects_backscattering}.
Note that the results for $\Delta_2$ and $\Delta_3$
are shown over an $E_{\text{recon}}$ energy range of 150--750 keV.
Below 150 keV, the corrections become quite large, as the acceptance
is highly suppressed for large pitch angle events and the
probability for backscattering increases with decreasing energy.

Our sign convention for these corrections is such that if the
correction factor $\Delta_j > 0$, the resulting asymmetry calculated
according to $A_i (1 + \Delta_j)$ is larger in magnitude (i.e., a more
negative value for the asymmetry).  As can be seen in Fig.\
\ref{fig:delta2_delta3_corrections}, the $\Delta_2$ correction factors
are $>0$ for all energies and Geometries, whereas the $\Delta_3$
correction factors are $<0$.  The conceptual physical explanation for
this is as follows.

With regard to a physical explanation for $\Delta_2$,
the $\Delta_{2,i}$ correction factors correct the measured asymmetries
for events misidentified in data analysis as a Type $i$ event.  For
example, an electron incident initially on one of the detectors could
backscatter from either the decay-trap end-cap foil or the MWPC
entrance window and then be detected in the opposite side's MWPC and
scintillator (e.g., the ``Missed'' event illustrated in Fig.\
\ref{fig:event_types_schematic}); however, such an
event would subsequently be misidentified in data
analysis as a Type 0 event (with an incorrect initial
direction).  Because this type of event would dilute the magnitude of
the measured asymmetry for Type 0 events, the Monte-Carlo-calculated
$\Delta_{2,0}$ correction factor would be $>0$ (i.e., so as to
increase the magnitude of the asymmetry, per our sign convention) to
compensate for the dilution.  Further, as the probability for
backscattering from plastic scintillator decreases with energy over
the energy range of neutron $\beta$-decay \cite{martin06} (and also
for backscattering from thin Mylar films, as calculated in
\texttt{GEANT4} and verified via analytic integration of the
differential cross section for the Mott scattering of electrons from
atomic electrons), the $\Delta_{2,0}$ correction factor decreases with
energy.  Thus, conceptually, the $\Delta_{2,i}$ correction factors are
expected to be $>0$ for all energies and to decrease in magnitude with
energy, which is consistent with the results from our Monte Carlo
calculations shown in Fig.\ \ref{fig:delta2_delta3_corrections}.

With regard to a physical explanation for $\Delta_3$,
there are two potential sources of bias to the $\langle
\beta\cos\theta \rangle$ acceptance.  First, there is a bias to the
$\langle \cos\theta \rangle$ value, as the acceptance for large pitch
angles is suppressed (e.g., from ``Lost'' events such as those shown
in Fig.\ \ref{fig:event_types_schematic}), with the value of $\langle
\cos\theta \rangle > 1/2$ for detected events (as the acceptance is
biased towards small pitch angle events).  Thus, in general,
$\Delta_3$ is expected conceptually to be negative, given our
definition of $(1+\Delta_3) = \langle
\frac{1}{2}\beta_{\text{recon}}\rangle / \langle \beta_{\text{true}}
\cos\theta\rangle$ in Eq.\ (\ref{eq:delta3}).  Second, because the
energy loss in the decay-trap end-cap foils and the MWPC windows is
strongly angle-dependent, the $E_{\text{vis}}$ visible
energies in each visible energy bin, which then map to
$E_{\text{recon}}$ values via the (mostly linear) parametrizations
discussed in Section \ref{sec:analysis_initial_energy_reconstruction},
actually correspond to a distribution of initial true energies.
Therefore, there is a bias to the assumption that the value of
$\langle \beta_{\text{true}} \rangle = \langle \beta_{\text{recon}}
\rangle$ in each $E_{\text{recon}}$ bin, which is corrected for in our
Monte Carlo calculations of $\Delta_3$.  The variation of $\Delta_3$
with energy depends on the details and shape of the initial energy
distribution.

The differences between the values of the $\Delta_2$ and
$\Delta_3$ correction factors for Geometry C and Geometry D (despite
their identical foil thicknesses) is primarily because the measured
MWPC thresholds for Geometry D were higher than those for Geometry C.

Finally, it is interesting to note that because $\Delta_2$ and
$\Delta_3$ are of opposite signs, there is a ``zero crossing'' in the
total correction factor $(1 + \Delta_2)(1 + \Delta_3) \approx 1 +
\Delta_2 + \Delta_3$, which can be seen in Fig.\
\ref{fig:delta2_delta3_corrections}.

As discussed in the upcoming Section
\ref{sec:uncertainties_angle_effects_backscattering}, the systematic
uncertainties in these corrections were taken to be a relative
fraction of the magnitude of these correction factors.

\subsection{Analysis Energy Window}
\label{sec:corrections_analysis_energy_window}

Our final results for $A_0$ presented later in Section
\ref{sec:summary_final_results} were obtained over an analysis energy
window of 275--625 keV.  This optimized energy window minimized the
total absolute error in $A_0$ resulting from the quadrature sum of the
statistical error and the energy-dependent systematic errors in the
above-described Monte Carlo calculations of the $\Delta_2$ and
$\Delta_3$ backscattering and angle effects corrections (which varied
with energy as just shown in Figs.\
\ref{fig:delta2_delta3_corrections} and
\ref{fig:delta2_corrections_decomposed}).

\subsection{Recoil-Order and Radiative Corrections}
\label{sec:corrections_recoil_order_radiative}

\begin{table*}
\caption{Values of the $\Delta_2$ and $\Delta_3$ correction factors
over the analysis energy window of 275--625 keV for each of the
Geometries.}
\begin{ruledtabular}
\begin{tabular}{ccccccccc}
& \multicolumn{2}{c}{Geometry A}& \multicolumn{2}{c}{Geometry B}&
  \multicolumn{2}{c}{Geometry C}& \multicolumn{2}{c}{Geometry D} \\
$E_{\text{recon}}$ [keV]& $\Delta_2$ [\%]& $\Delta_3$ [\%]&
  $\Delta_2$ [\%]& $\Delta_3$ [\%]& $\Delta_2$ [\%]& $\Delta_3$ [\%]&
  $\Delta_2$ [\%]& $\Delta_3$ [\%] \\ \hline
275--300& 2.28& $-1.17$& 7.12& $-1.28$& 1.57& $-0.43$& 1.55& $-0.25$ \\
300--325& 1.89& $-1.58$& 6.32& $-1.52$& 1.36& $-0.49$& 1.41& $-0.24$ \\
325--350& 1.63& $-1.38$& 5.60& $-1.71$& 1.30& $-0.40$& 1.32& $-0.24$ \\
350--375& 1.54& $-1.43$& 5.12& $-1.94$& 1.18& $-0.28$& 1.19& $-0.25$ \\
375--400& 1.33& $-1.45$& 4.66& $-2.38$& 1.04& $-0.50$& 1.08& $-0.25$ \\
400--425& 1.25& $-1.67$& 4.19& $-2.61$& 1.05& $-0.43$& 1.00& $-0.28$ \\
425--450& 1.16& $-1.66$& 3.82& $-3.01$& 0.91& $-0.53$& 0.94& $-0.29$ \\
450--475& 1.04& $-1.96$& 3.51& $-3.56$& 0.89& $-0.76$& 0.90& $-0.29$ \\
475--500& 1.04& $-1.91$& 3.21& $-4.02$& 0.87& $-0.63$& 0.88& $-0.35$ \\
500--525& 0.97& $-2.16$& 2.96& $-4.39$& 0.85& $-0.63$& 0.80& $-0.42$ \\
525--550& 0.92& $-2.51$& 2.70& $-4.99$& 0.77& $-1.00$& 0.79& $-0.52$ \\
550--575& 0.85& $-2.44$& 2.46& $-5.57$& 0.83& $-1.10$& 0.83& $-0.65$ \\
575--600& 0.84& $-3.03$& 2.25& $-6.42$& 0.73& $-1.20$& 0.76& $-0.75$ \\
600--625& 0.84& $-3.02$& 1.99& $-6.86$& 0.70& $-1.13$& 0.80& $-1.13$ \\
\end{tabular}
\end{ruledtabular}
\label{tab:delta2_delta3_tabulated_values}
\end{table*}

\begin{figure}
\includegraphics[scale=0.44,clip=]{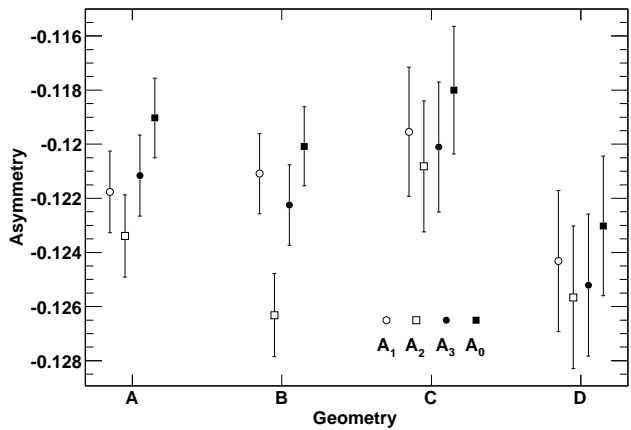}
\caption{Results for the asymmetries $A_1$ (open circles), $A_2$
(open squares), $A_3$ (filled circles), and $A_0$ (filled squares)
for each of the Geometries, integrated over the analysis energy
window of 275--625 keV.  All errors are statistical.}
\label{fig:asymmetries_geometries}
\end{figure}

\begin{figure*}
\includegraphics[scale=0.90,clip=]{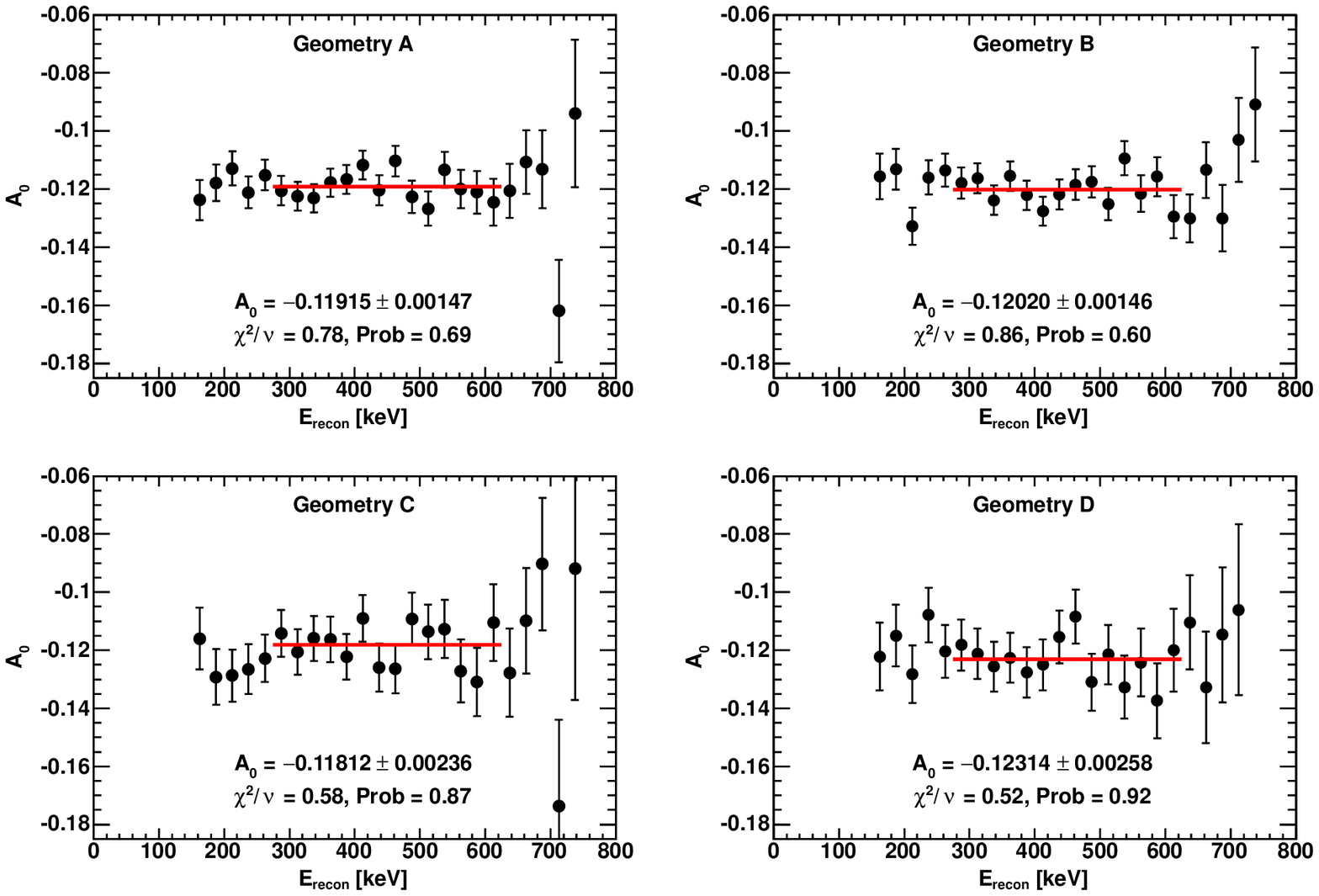}
\caption{(Color online) Geometry-by-Geometry results for $A_0$ values
extracted bin-by-bin.  The solid red lines indicate the analysis
energy window of 275--625 keV.  The quoted errors are statistical.}
\label{fig:A0_geometries}
\end{figure*}

Recoil-order corrections to the asymmetry were calculated within the
context of the Standard Model according to the formalism of
\cite{wilkinson82, gardner01}.  The numerical results from these two
parametrizations agree to better than $2 \times 10^{-5}$.  These
parametrizations for the asymmetry $A$, Eq.\
(\ref{eq:recoil_order_parametrization}), were then folded over the
$\beta$-decay energy spectrum (including contributions from the Fermi
function). In general, these recoil-order corrections increase the
magnitude of the measured asymmetry over that of $A_0$; thus, an
extraction of $A_0$ from the measured asymmetry requires (per our sign
convention) a negative correction (i.e., a decrease in the magnitude
of the asymmetry).  The energy-dependence of this correction is shown
in Fig.\ \ref{fig:recoil_order_correction}.  For our analysis
energy window of 275--625 keV, the integrated recoil-order correction
(assuming the Particle Data Group average value for $\lambda$) was
$(-1.79 \pm 0.03)$\% to $A_0$, where the $\pm 0.03$\% uncertainty
corresponds to the Particle Data Group's statistical uncertainty in
$\lambda$.

The value for the radiative correction was taken from the calculations
of \cite{gluck92}, who presented results for the absolute (as opposed
to relative) value of the radiative correction to the asymmetry.
These calculations were performed using the average value for
$\lambda$ available at that time, and were reported at six discrete
values of electron energy.  In general, these radiative corrections
increase the magnitude of the measured asymmetry by $\sim 0.1$\% over
the range of our 275--625 keV analysis energy window.
However, at the time of this analysis, we were not
aware of the functional form for the energy-dependence of these
radiative corrections which is presented in \cite{shann71}.
Therefore, we applied a $(0.10 \pm 0.05)$\% correction to our
asymmetries (again, per our sign convention), where our estimated $\pm
0.05$\% uncertainity accounts for differences between the value of
$\lambda$ available at the time of the calculation (1992,
\cite{gluck92}) and its present value, and for our incomplete
knowledge at the time of the energy-dependence for the radiative
correction.

Per the discussion in \cite{czarnecki04}, application of this
radiative correction to our reported value for $A_0$ then permits
extraction of a value for $g_A$ which can be compared with the
expression relating $G_F$, $\tau_n$, $g_A$, and $V_{ud}$, Eq.\
(\ref{eq:lifetime-Vud}), in which the (1 + RC) electroweak radiative
corrections have been factorized in the same way for both the vector
and axial-vector interactions \cite{czarnecki04}.

\subsection{Comparison of Geometry-by-Geometry Results}
\label{sec:corrections_comparison_geometry}

The values for the $\Delta_2$ and $\Delta_3$
corrections in Geometry over our analysis energy window of
275--625 keV are tabulated in
Table \ref{tab:delta2_delta3_tabulated_values} for all of the Geometries.
Figure \ref{fig:asymmetries_geometries} then compares the resulting
Geometry-by-Geometry asymmetries $A_1$, $A_2$, $A_3$,
and $A_0$ integrated over the analysis energy window.
The final values for $A_0$ from each of the Geometries are all seen to
be statistically consistent.  Geometry-by-Geometry results for the
$A_0$ values extracted bin-by-bin are shown in
Fig.\ \ref{fig:A0_geometries}.

We emphasize that the agreement between our results
from the different Geometries provides confidence in our Monte Carlo
treatment of the backscattering and $\langle \beta\cos\theta \rangle$
acceptance effects.  In particular, the result from Geometry B
(thickest decay trap end-cap and MWPC foils) agrees well with those
from the other Geometries, in spite of its relatively larger
$\Delta_2$ and $\Delta_3$ corrections (compare the differences in
$A_1$, $A_2$, and $A_3$ for Geometry B with those for the other
Geometries).

\begin{table}
\caption{Background-subtracted neutron $\beta$-decay rates integrated
over the complete energy range of 0--800 keV after all analysis cuts.}
\begin{ruledtabular}
\begin{tabular}{cccc}
& & East Detector& West Detector \\
Geometry& Spin State& Rate [s$^{-1}$]& Rate [s$^{-1}$] \\ \hline
A& $-$& 4.79& 4.90 \\
A& $+$& 3.82& 3.25 \\ \hline
B& $-$& 6.50& 6.87 \\
B& $+$& 5.43& 4.78 \\ \hline
C& $-$& 3.29& 3.42 \\
C& $+$& 2.81& 2.44 \\ \hline
D& $-$& 6.45& 6.61 \\
D& $+$& 4.80& 4.12 \\
\end{tabular}
\end{ruledtabular}
\label{tab:beta_decay_rates}
\end{table}

Finally, for completeness, our background-subtracted neutron
$\beta$-decay rates integrated over the complete energy range of
0--800 keV after all analysis cuts are listed in Table
\ref{tab:beta_decay_rates} for all the Geometries and for both spin
states.  As previously discussed in Section
\ref{sec:analysis_data_quality_cuts}, the data quality analysis cuts
related to DAQ electronics problems removed a significant fraction of
the Geometry A and Geometry B events (up to $\sim 30$\%).  Another
significant analysis cut included the 45 mm radius fiducial cut
discussed in Section \ref{sec:analysis_position_cuts}, which removed
$\sim 25$\% of the events.  Finally, under the 275--625 keV
analysis energy window, the rates from 0--800 keV listed in Table
\ref{tab:beta_decay_rates} would be reduced by another factor of
$\sim 40$\%.

\section{Systematic Uncertainties}
\label{sec:uncertainties}

\subsection{Summary}
\label{sec:uncertainties_summary}

\begin{table}
\caption{Summary of systematic corrections and uncertainties.  All
numbers quoted are fractional [\%] relative to $A_0$.  Upper Table:
Geometry-Independent systematic uncertainties.  No systematic
corrections were applied for these effects (with the exception of the
radiative corrections, already discussed in Section
\ref{sec:corrections_recoil_order_radiative}).  Lower Table:
Geometry-Dependent systematic corrections and uncertainties.  The
quoted value denotes the systematic correction, with the error the
systematic uncertainty.  As discussed in the text, $\Delta_2$
represents the correction for backscattering, and $\Delta_3$ the
correction for the angle effect.  $\epsilon_{\text{MWPC}}$ denotes the
systematic uncertainty associated with the MWPC efficiency.}
\begin{ruledtabular}
\begin{tabular}{lc}
Geometry-Independent Effect& Uncertainty [\%] \\ \hline
Dead Time& $\pm 0.01$ \\
Energy Reconstruction& $\pm 0.47$ \\
Fiducial Cut and Coordinate Systems& $\pm 0.24$ \\
Gain Fluctuations& $\pm 0.20$ \\
Live Time& $\pm 0.24$ \\
Magnetic Field Nonuniformity& $^{+0.20}_{-0.00}$ \\
Muon Veto Efficiency& $\pm 0.30$ \\ 
Neutron-Generated Backgrounds& $\pm 0.02$ \\
Polarization& $^{+0.52}_{-0.00}$ \\
Radiative Corrections& $\pm 0.05$ \\
Rate-Dependent Gain Shifts& $\pm 0.08$
\end{tabular}
\end{ruledtabular}
\vspace{0.25cm}
\begin{ruledtabular}
\begin{tabular}{lcccc}
\multicolumn{5}{c}{Geometry-Dependent Effects} \\ \hline
& A [\%]& B [\%]& C [\%]& D [\%] \\ \hline
$\Delta_2$& $1.34 \pm 0.40$& $4.32 \pm 1.30$& $1.07 \pm 0.32$&
  $1.08 \pm 0.32$ \\
$\Delta_3$& $-1.81 \pm 0.45$& $-3.22 \pm 0.81$& $-0.60 \pm 0.15$&
  $-0.36 \pm 0.09$ \\
$\epsilon_{\text{MWPC}}$& $0.00 \pm 0.02$& $0.00 \pm 0.01$& $0.00 \pm 0.16$&
  $0.00 \pm 0.50$
\end{tabular}
\end{ruledtabular}
\label{tab:systematic_corrections_uncertainties}
\end{table}

Our systematic corrections and uncertainties are summarized in Table
\ref{tab:systematic_corrections_uncertainties}, where we have
categorized the effects as either Geometry-Dependent (i.e., effects
which varied with the decay trap end-cap foil and MWPC window
thicknesses, measured detector thresholds, etc.), or
Geometry-Independent (e.g., UCN polarization, dead time effects, etc.)
In the rest of this Section we discuss each of these systematic
effects (in the order in which they appear in Table
\ref{tab:systematic_corrections_uncertainties}) in more detail.

\subsection{Geometry-Independent: Dead Time}
\label{sec:uncertainties_dead_time}

Nearly all dead time effects cancel in the super-ratio technique.
Indeed, in order for there to be any bias to the asymmetry resulting
from dead time effects in the background-subtracted $\beta$-decay
rates, there must be a difference in the two detectors' dead times,
and there must be a difference in a particular detector's dead time
for the two neutron spin states.  Thus, these effects are expected to
be quite small.

Nevertheless, as previously noted in Section
\ref{sec:experiment_electronics}, the dead time of the DAQ system was
monitored by counting, in scalers, the total number of detector
two-fold PMT triggers, including those that were vetoed by the DAQ
``busy logic'' during the $\sim 12$ $\mu$s readout gates for the PADC
modules.  However, to avoid spurious (and correlated) trigger chains
from scintillator afterpulses (as noted earlier in Section
\ref{sec:experiment_electronics}) distorting the determination of the
dead time, the dead time was determined only from the scaler counts of
detector triggers that occurred during triggers from the opposite-side
detector or from other experimental triggers, such as the UCN
monitors.  The dead time, as extracted from the
correlation between the DAQ trigger rate and the fraction of these
``missed triggers'', was found to be $\sim 13.5$ $\mu$s, which is
consistent with the nominal $\sim 12$ $\mu$s system dead time
(associated with the gate for the PADC readout).  Further, the
difference in the fraction of ``missed triggers'' for (up to) a 20
s$^{-1}$ trigger rate difference between the two spin states is no
larger than $\sim 0.03$\%.  Considered together, any possible bias to
the asymmetry was no greater than 0.01\%, which is the error we quote
in Table \ref{tab:systematic_corrections_uncertainties}.

Alternatively, another possible way dead time effects could bias the
asymmetry is in the background subtraction procedure, resulting from
differences in the DAQ total trigger rates during $\beta$-decay and
background runs.  However, these effects tend to cancel in the
super-ratio, as the four background-subtracted $\beta$-decay rates
appearing in the super-ratio would be expected to be biased in the
same direction.  Further, the effect is minimized as the
signal-to-background ratio increases.  Under the conservative
assumption of a 200 s$^{-1}$ DAQ trigger rate difference (e.g., from
differences in the scintillator trigger rates, UCN monitor trigger
rates, etc.) for $\beta$-decay versus background runs, and a
signal-to-background ratio greater than 5, any such systematic bias to
the asymmetry from dead time effects is $\ll 0.01$\%.

\subsection{Geometry-Independent: Energy Reconstruction}
\label{sec:uncertainties_energy_reconstruction}

Figure \ref{fig:energy_reconstruction_error_envelope} showed the error
envelope for the uncertainty in the visible energy calibration.  To
estimate the systematic error associated with possible errors in our
energy calibration, we generated a large number (200 per Geometry) of
random error curves that were constrained to fit within the limits of
this error envelope.  We then extracted from these error curves their
contributions to an error in the asymmetry, resulting from an
incorrect reconstruction of the electron energy, and hence $\langle
\beta\cos\theta \rangle$.  From these calculations we concluded that
the maximum (i.e., worst case) error, resulting from the case where
the error curves for the two detectors are identical, is a fractional
0.47\% uncertainty in the asymmetry for the analysis energy window of
275--625 keV.  As a conservative estimate of the systematic
uncertainty associated with our energy calibration, we then assign
this worst-case error of 0.47\% to be the systematic uncertainty
associated with possible errors in our energy calibration.

As discussed earlier in Section
\ref{sec:analysis_initial_energy_reconstruction}, the default
$E_{\text{recon}}$ parametrization we employed was based on a fit to
the scintillator visible energy $E_{\text{vis}}$ only; by constrast,
an alternative $E_{\text{recon}}$ fit included both $E_{\text{vis}}$
and the calibrated MWPC energy $E_{\text{MWPC}}$.  To study the
sensitivity of the reconstructed asymmetry to these two different
$E_{\text{recon}}$ parametrizations, we extracted values for the
energy-corrected asymmetry $A_1$ for these two different fits.  The
difference between these two methods, averaged over the entire data
set, was 0.2\%.  This is small relative to the 0.47\% systematic
uncertainty associated with the energy calibration, and we also noted
in Section \ref{sec:analysis_initial_energy_reconstruction} that this
alternative $E_{\text{recon}}$ parametrization based on both
$E_{\text{vis}}$ and $E_{\text{MWPC}}$ is subject to (uncorrectable)
overflow of the MWPC anode readout.

Another source of a systematic error resulting from the energy
reconstruction as discussed in detail in Section
\ref{sec:asymmetry_endpoint_distributions} (and shown in Fig.\
\ref{fig:kurie_fit_distributions}) was the observed systematic $\sim
10$--14 keV difference between the fitted endpoints and the Monte
Carlo prediction.  We investigated the systematic uncertainty to the
extracted asymmetry due to this systematic difference by extracting a
``stretching factor'', $f \equiv T_{0,\text{MC}}/T_{0,\text{fit}}$,
where $T_{0,\text{MC}}$ denotes the Monte Carlo predicted endpoint and
$T_{0,\text{meas}}$ the fitted endpoint, for each run.  The data were
then re-analyzed by applying on an event-by-event basis this
``stretching factor'' to the reconstructed energy $E_{\text{recon}}$,
thus effectively forcing the fitted endpoints to match the Monte Carlo
predicted endpoints.  The asymmetries extracted from the ``streched''
data differed by $<0.07$\% from the (original) ``unstretched'' data
which, again, is much less than the 0.47\% error associated with the
energy calibration.

Finally, to account for a slight mismatch ($\sim 2$ keV) between the
Monte Carlo and measured energy spectra (this is visible in the final
$E_{\text{recon}}$ spectrum later in Fig.\
\ref{fig:final_spectra_asymmetries}) we fitted the Monte Carlo visible
energy spectra to the measured visible energy spectra, and then
extracted values for the asymmetry assuming these modified values for
the visible energy.  The bias to the asymmetry was $0.13$\% averaged
over all four Geometries which, again, is much less than the 0.47\%
the energy calibration uncertainty.

\subsection{Geometry-Independent: Fiducial Cut and Coordinate Systems}
\label{sec:uncertainties_fiducial_cut}

As discussed in Section \ref{sec:analysis_position_cuts}, we required
backscattering events to satisfy a default vertex cut of $|\vec{x}_E -
\vec{x}_W| < 25$ mm.  We studied the impact of this cut on the
asymmetry by varying this cut from 10 mm to 40 mm; the effect on the
asymmetry was $<0.1$\%, indicating a negligible systematic effect.

As was also discussed there, our fiducial cut required the position
(radius) of the event on the primary triggering scintillator side to
satisfy $r_{\text{trigger}} < 45$ mm.  To examine whether there was
any position bias, we extracted the asymmetry in successive annuli via
cuts on $r^2$ in six different annular bins, ranging from [0,400]
mm$^2$ to [2025,2500] mm$^2$.  The asymmetries in all of these annuli
were in statistical agreement, with no statistical evidence for any
systematic difference with position.

Recall also in Section \ref{sec:analysis_position_cuts} we noted the
possibility for the definition of four different coordinate systems.
To determine whether there was any bias resulting from the choice of
the coordinate system (for example, a consideration could be whether
there were any systematic variations in the backscattering fractions
in the vicinity of the fiducial cut), we studied the variation of the
asymmetry with the choice of coordinate system, and for fiducial
volume radius cuts of 45 mm and 50 mm.  The RMS spread in the
asymmetries for the different coordinate system choices was 0.24\% for
the 45 mm radius cut and 0.21\% for the 50 mm radius cut.  Although
the RMS spread for the 50 mm radius cut was actually somewhat smaller
(suggesting that employing a larger fiducial volume would have
introduced no bias to the asymmetry), we nevertheless chose the 45 mm
radius cut as our (conservative) definition of the fiducial volume,
and thus assigned a 0.24\% systematic uncertainty to the definition of
the fiducial volume.


\subsection{Geometry-Independent: Gain Fluctuations}
\label{sec:uncertainties_gain_fluctuations}

As noted in Section \ref{sec:experiment_calibration_gain}, the PMT
gains were monitored on a run-to-run basis using the minimium-ionizing
peak from cosmic-ray muon events.  Nevertheless, any residual
uncompensated run-to-run gain fluctuations could bias the asymmetry on
a run-to-run basis; however, any such short-term run-to-run
fluctuations will average away according to the usual
$1/\sqrt{N_{\text{run}}}$ statistics assuming the long-term gain
corrections are accurate.  We estimated the level of any such
run-to-run residual gain errors by extracting the level of
fluctuations in the run-to-run fitted values for the $\beta$-decay
spectrum endpoint.  These were typically of order $\sim \pm 1.2$\% in
each detector, with the gain fluctuations in the two detectors only
slightly correlated relative to each other.  [Correlated gain
fluctuations are significantly more problematic than are
anti-correlated gain fluctuations.]  Conservatively assuming the
worst-case sensitivity for gain fluctuations in one of the Geometries
to be representative of the entire data set, we quote a systematic
uncertainty of 0.20\% for uncompensated gain fluctuations.

\subsection{Geometry-Independent: Live Time}
\label{sec:uncertainties_live_time}

The detector rates (and, hence, asymmetries) were ultimately
calculated from the number of events passing the analysis cuts
normalized to the detectors' respective live times; the concept of the
detector live time was discussed in detail previously in Section
\ref{sec:analysis_data_quality_cuts}.  As discussed there, we defined
a run's live time to be the fraction of that run surviving all of the
global data quality cuts.  However, as we noted there, it was
necessary to apply a correction for the Geometry B live times due to
the large fraction (up to $\sim 30$\%) of events suffering from an
event-by-event TDC event counter problem.  The
correction factors for each run were determined using events
identified as gamma rays, which were statistically independent of the
neutron $\beta$-decay events and also provided higher statistics
(event rates up to 100 s$^{-1}$ in each detector) than the neutron
$\beta$-decay events themselves for the calculation of the correction
factors.  The resulting correction factors, defined to be the ratio of
the number of gamma ray events surviving the event-by-event TDC event
counter cut to the total number of gamma ray events, were then
computed on a run-by-run basis for each detector.

Only the Geometry B live times were corrected
according to this procedure.  Nevertheless, to assess the systematic
error associated with our definition of and calculation of the live
time, we extracted values for the asymmetries for all four Geometries
with and without application of these live time correction factors
(the correction factors for Geometries A, C, and D were small, with
the values for the asymmetries differing by $< 0.1$\% under the two
scenarios).  Averaged over all four Geometries, the difference between
the asymmetries extracted under these two different scenarios was
0.24\%, which is the value for the systematic uncertainty associated
with this effect we quote in Table
\ref{tab:systematic_corrections_uncertainties}.

\subsection{Geometry-Independent: Magnetic Field Nonuniformity}
\label{sec:uncertainties_magnetic_field_nonuniformity}

Our Monte Carlo calculations of the corrections for
backscattering and the $\langle \beta\cos\theta \rangle$ acceptance
discussed in Section \ref{sec:corrections_overview} assumed a uniform
magnetic field in the decay trap region.  We studied the impact of the
actual measured nonuniformity in the spectrometer magnetic field shown
previously in Fig.\ \ref{fig:scs_field_uniformity} in Monte Carlo.
Qualitatively, the impact of the $\sim 30$ Gauss ``field dip'' in
the central decay-trap region is such that electrons from decays
occuring in this ``field dip'' region are either reflected (analogous
to backscattering) or are trapped (for large pitch angles).  We
studied these effects in our \texttt{GEANT4} Monte Carlo simulation
program by implementing the magnetic field profile shown in Fig.\
\ref{fig:scs_field_uniformity} directly in the simulation.  Neutron
$\beta$-decay events were then generated uniformly along the length of
the decay trap.

In the Monte Carlo, $\sim 0.3$\% of the events incident initially on
one of the two detectors were reflected from the field dip, with an
average $\langle \beta \cos\theta \rangle$ of $\sim 0.02$.
Because this small fraction of events carries little
$\langle \beta\cos\theta \rangle$ ``analyzing power'', the resulting bias
to the asymmetry is negligible.  The fraction of electrons trapped by
the field dip was $\sim 2.6$\%, again with an average $\langle
\beta\cos\theta \rangle$ of $\sim 0.02$.  The remaining 97.1\% of the
events were not impacted by the field dip.
Assuming that the electrons trapped by the field dip eventually
scatter from residual gas molecules, the impact is a dilution to the
asymmetry.  The calculated dilution to the asymmetry
was $-0.2$\%.  In lieu of applying a correction to the asymmetry, we
assigned a $^{+0.2\%}_{-0.0\%}$ systematic uncertainty to this
effect.  We also note that our Monte Carlo calculations found that
the time for a trapped electron to scatter from residual gas for a
vacuum pressure of $10^{-5}$ Torr is $\sim 4$ ms, with a small
distortion to their energy distribution of $\sim -8$ keV.

Note that our Monte Carlo results for the fraction of
events trapped by the field dip and their average value of $\langle
\beta\cos\theta \rangle$ are consistent with the following simple
estimates.  As discussed earlier in Section
\ref{sec:experiment_electron_spectrometer_SCS}, electrons emitted with
some momentum $p_0 = (p_{\perp,0}^2 + p_{\parallel,0}^2)^{1/2}$, with
$p_{\perp,0}$ ($p_{\parallel,0}$) the initial transverse
(longitudinal) momentum component, in some local field $B_0$ will be
reflected from higher field regions $B$ if $B > B_{\text{crit}} \equiv
(p_0^2/p_{\perp,0}^2) B_0$ (thus, only the pitch angle $\theta$ of the
emitted electron is relevant, not the magnitude of the momentum).
Taking $B_0 = 0.9925$~T and $B = 0.9955$~T for the measured 2009 field
profile (here, $B$ is taken to be the average of the local
maxima at $z=-100$ cm and $+50$ cm) shown in Fig.\
\ref{fig:scs_field_uniformity}, one finds electrons with pitch angles
$\theta > \theta_{\text{crit}} = 86.9^\circ$ will be trapped in the
field dip region.

Approximating the initial angular distribution of emitted
electrons as isotropic (reasonable, given that the $\beta$-asymmetry
is an $\mathcal{O}(10\%)$ effect), one finds that the fraction of
electrons emitted in the local field dip region $B_0 = 0.9925$~T which
will be trapped is
\begin{eqnarray}
f_\text{trap} &=& \frac{1}{4\pi} 2\times
{\int_{\theta_\text{crit}}^{\pi/2}}\sin\theta d\theta
{\int_0^{2\pi}d\phi} = \cos\theta_{\text{crit}} \nonumber \\
&=& 0.054.
\end{eqnarray}
Then, assuming a uniform distribution of events along
the 300-cm long decay trap, the fraction of events emitted in the
$\sim 150$-cm long field dip region is $\sim 0.5$, implying the total
fraction of events generated over the length of the decay trap which
will be trapped in the field dip region is $\sim 0.027$, which is
consistent with the Monte Carlo result of 2.6\%.  For a nominal value
of $\beta \sim 0.75$, $\langle \beta\cos\theta \rangle \approx
\beta\cos[(\theta_{\text{crit}} + \pi/2)/2] = 0.02$ for these trapped
events, again, consistent with the Monte Carlo result.

\subsection{Geometry-Independent: Muon Veto Efficiency}
\label{sec:uncertainties_muon_veto}

We estimated the effect of a possible systematic uncertainty resulting
from fluctuations in the muon-veto efficiency by extracting values for
the asymmetries with and without application of the muon-veto detector
cuts.  Averaged over Geometries, the variations in the asymmetry were
at the 0.3\% level.

We note that the assignment of this 0.3\% uncertainty
is quite conservative.  A linear drift in the muon veto cut efficiency
would be equivalent to a linear drift in the backgrounds, and as
discussed in Section \ref{sec:asymmetry_asymmetry_extraction},
linear background drifts cancel under the octet-based super-ratio
asymmetry structure.

\subsection{Geometry-Independent: Neutron-Generated Backgrounds}
\label{sec:uncertainties_neutron_generated_backgrounds}

As already dicussed in detail, ambient backgrounds were measured and
subtracted on a run-by-run basis.  However, a possible source of
irreducible backgrounds was neutron capture on materials near the
electron detectors, generating prompt gamma rays with energies up to
7.9~MeV, 7.1~MeV, 6.8~MeV, 4.9~MeV, or 8.2~MeV for
capture on $^{63}$Cu, $^{65}$Cu, $^9$Be, $^{12}$C, or $^{13}$C,
respectively, the elements of which the decay trap and end-cap foils
were primarily composed.  Such backgrounds cannot, of course, be
subtracted.

This background was expected to be significantly suppressed in the
UCNA experiment as compared to previous cold neutron beam experiments
because, as discussed earlier in Section
\ref{sec:experiment_overview}, the fraction of neutrons present in the
apparatus which contribute to the decay rate is orders of magnitude
larger in the UCNA Experiment than in previous cold neutron beam
experiments, and also because of the small probability for capture and
upscatter by UCN stored in the decay trap.

We carried out three different approaches to our
assessment of the contamination level from any such backgrounds.  The
idea of our first approach is as follows.  If a gamma ray emitted
from a neutron capture subsequently interacted with the scintillator,
the MWPC should not have recorded any energy deposition if the gamma
ray forward Compton scattered in the scintillator.
Further,
as calculated in simulations, there is
a factor of 10--20 suppression in the fraction of gamma ray
events incident on the electron detectors
triggering both the scintillator and MWPC as compared to
those triggering only the scintillator.  Therefore, any such
neutron-generated backgrounds should appear as non-zero residuals in a
comparison of background-subtracted scintillator spectra [i.e.,
($\beta$-decay run $-$ background run) spectra] formed with and
without application of a MWPC-scintillator coincidence cut.  In
particular, an excess would be expected in the background-subtracted
spectrum constructed without the requirement of a MWPC-scintillator
coincidence cut as compared to the background-subtracted spectrum
obtained with the requirement of a MWPC-scintillator coincidence cut.

Now consider the following model.
Under application of a MWPC coincidence cut, let
\begin{equation}
R_{\text{cut}} = \epsilon \left( S_{\text{cut}} - B_{\text{cut}} \right),
\end{equation}
where $R_{\text{cut}}$ denotes the resulting background-subtracted
scintillator event rate, $S_{\text{cut}}$ and $B_{\text{cut}}$ denote,
respectively, the underlying $\beta$-decay + background and background
event rates, respectively, and $\epsilon$ denotes the MWPC cut
efficiency.  We then write a similar expression for the
background-subtracted scintillator event rates obtained without
application of a MWPC cut as
\begin{equation}
R_{\text{no cut}} = \left( S_{\text{cut}} - B_{\text{cut}} \right) +
\left( S_\gamma - B_\gamma \right),
\end{equation}
where now $S_\gamma$ and $B_\gamma$ denote the signal and background
rates during the $\beta$-decay and background runs from gamma ray
events which would otherwise fail the MWPC cut.  Note that in the
absence of any neutron-generated gamma rays, the statistical averages
of $S_\gamma$ and $B_\gamma$ should be identical.

The difference between $R_{\text{no cut}}$ and $R_{\text{cut}}$ is
then
\begin{eqnarray}
\Delta R &=& R_{\text{no cut}} - R_{\text{cut}} \nonumber \\
&=& \left( S_\gamma - B_\gamma \right) +
\left(S_{\text{cut}} - B_{\text{cut}}\right)(1 - \epsilon) \nonumber \\
&=& \Gamma_\gamma + (1-\epsilon)\Gamma_n,
\end{eqnarray}
where $\Gamma_\gamma$ and $\Gamma_n$ denote, for gamma ray and neutron
$\beta$-decay events, respectively, the difference between the
background-subtracted scintillator rates with and without application
of a MWPC coincidence cut.  Thus, this model then
requires an estimate for the MWPC cut efficiency.

We extracted a value for our MWPC cut efficiency by examining the MWPC
anode spectrum for $^{113}$Sn source calibration data.  After placing
a FWHM cut on the scintillator visible energy spectrum, we then fitted
the resulting MWPC spectrum (such as shown, for example, in Fig.\
\ref{fig:anode_cathode_spectra}) to a Landau distribution, and its
pedestal to a Gaussian.  We then calculated the fraction of the Landau
distribution falling below the cut line, which should provide an
estimate of the MWPC cut efficiency.  These values were found to be
99.93(3)\% and 99.95(3)\% for the East and West detector,
respectively.  We do note that this provides for an
estimate of the MWPC efficiency only at the $^{113}$Sn energy.

The residual $\Delta R$ (no MWPC cut $-$ MWPC cut)
rates integrated over the analysis energy window (and after all
analysis cuts) ranged from $1.0 \times 10^{-3}$
s$^{-1}$ to $6 \times 10^{-3}$ s$^{-1}$ for the two detectors and two
spin states for all four Geometries, representing $\sim 10^{-3}$ of
the $\beta$-decay rates.  After accounting for the factor of 10--20
supression for the fraction of events which would trigger both the
scintillator and MWPC, the estimated contamination fractions for the
actual $\beta$-decay analysis employing the MWPC-scintillator
coincidence cut are then on the order of $10^{-4}$.  Propagation of
the measured contamination fractions (for each detector and spin
state) through the super ratio then led to a systematic bias to the
asymmetry of order $\sim 0.02$\%.


In our second approach, we extrapolated the residual background (i.e.,
after background subtraction) above the $\beta$-decay endpoint into
the signal region.  Above the endpoint, the residual
background-subtracted rates in the 25 keV $E_{\text{recon}}$ bins were
typically of order $10^{-4}$ s$^{-1}$ or less.  Under the assumption
that the neutron-generated background is independent of energy
(e.g., as was employed in the analysis of \cite{abele02}), we then
extrapolated these above-the-endpoint rates into the analysis energy
window.  The resulting contamination fractions, $\sim
10^{-4}$--$10^{-3}$, were similar to the analysis in our first
approach comparing background-subtracted scintillator spectra obtained
with and without a MWPC cut.  These contamination fractions were again
propagated through our super-ratio asymmetry analysis, and the
systematic bias was again found to be of order $\sim 0.02$\%.

\begin{figure}
\includegraphics[angle=270,scale=0.30,clip=]{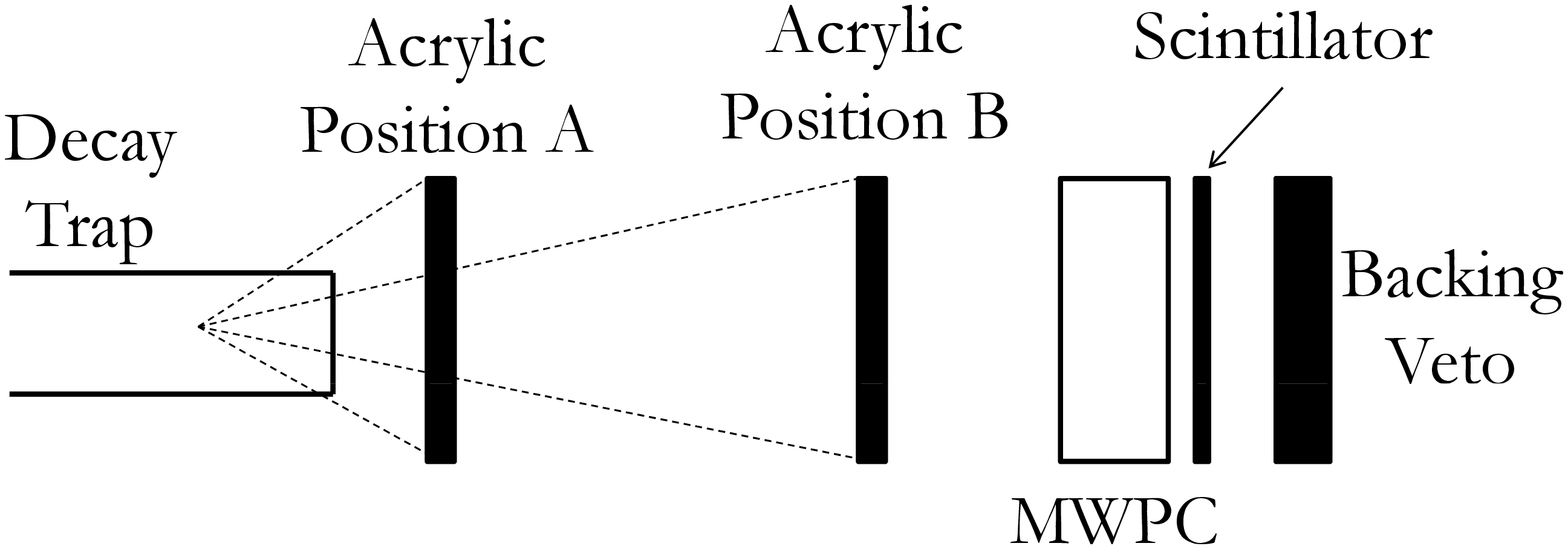}
\caption{Schematic diagram of the ``beta-blocker'' measurement of
neutron-generated backgrounds.  The dashed lines are meant to indicate
the solid angle for production of Compton-scattered electrons in the
acrylic at the two different positions.}
\label{fig:beta_blocker_schematic}
\end{figure}

Finally, in our third approach,
we carried out a ``beta-blocker'' measurement
in which a 6.35-mm thick piece of acrylic was placed between the decay
trap and the MWPC in the field-expansion region of the spectrometer,
as indicated schematically in Fig.\ \ref{fig:beta_blocker_schematic}.  The
idea for this measurement was two-fold: (1) the acrylic was
sufficiently thick to stop the endpoint $\beta$-decay electrons,
thereby ``blocking'' the $\beta$-decay signal of interest; and (2) the
acrylic then served as a ``source'' of Compton-scattered electrons,
resulting from interactions of neutron-generated gamma rays
with the acrylic.

Measurements were conducted with this piece of acrylic at two
different positions, A and B, as shown in Fig.\
\ref{fig:beta_blocker_schematic}, in front of one of
the detectors.  The motivation for doing so was that a comparison of
the results from positions A and B should, in principle, permit a
decomposition of the measured detector signal into contributions from
Compton-scattered electrons originating in this acrylic piece (the
solid angle for which was clearly smaller in position B as compared to
position A), and direct neutron-generated gamma-ray interactions in
the plastic scintillator detector (which should not have varied with
the position of the acrylic piece).

Note that this measurement is subject to some model
dependence, including an assumption for the source positions along the
decay trap of the neutron-generated backgrounds (which determines the
ratio of the solid angles for positions A and B).  Our
resulting estimates for the contamination fraction, as extracted from
our measurements of the residual (background-subtracted) rates with
the acrylic piece located at both positions A and B (under the
assumption that the ratio of the A and B solid angles for production
of Compton-scattered electrons was 20:1), were of order $10^{-3}$.
Then accounting for the factor of 10--20 suppression for gamma ray
events triggering both the scintillator and MWPC (in the actual
geometry) leads to an estimate for the contamination fraction on the
order of $10^{-4}$, consistent with the other two approaches.

\subsection{Geometry-Independent: Polarization}
\label{sec:uncertainties_polarization}

The UCN polarization systematic was discussed earlier in Section
\ref{sec:measurements_polarization}.  Because the measured
depolarization was consistent with zero at the $1\sigma$ level (i.e.,
$P > 0.9948$), we did not apply a correction for the polarization, and
instead quote a one-sided systematic uncertainty in $A_0$ of
$^{+0.52\%}_{-0.00\%}$ resulting from the constraint
$P>0.9948$.


\begin{figure*}
\includegraphics[scale=0.90,clip=]{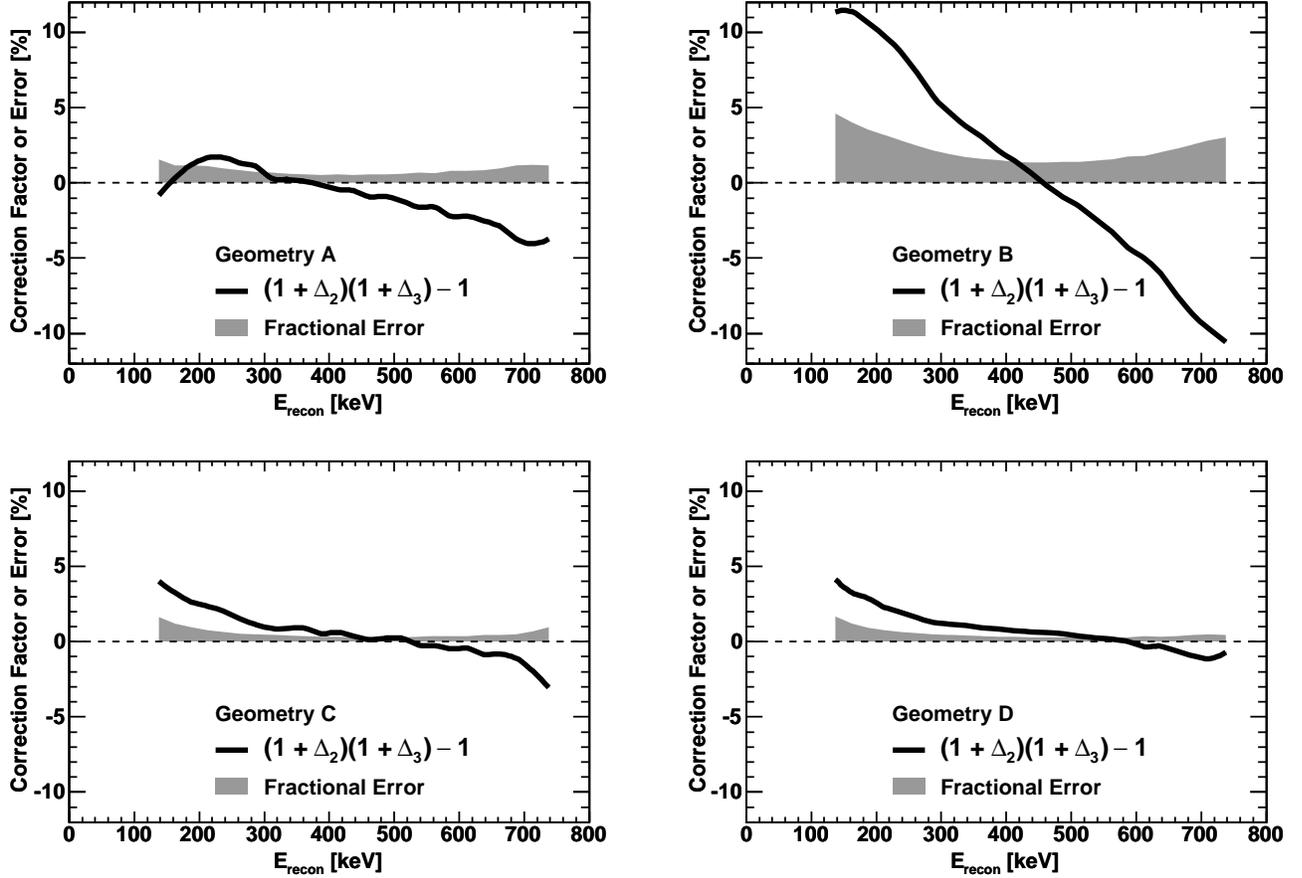}
\caption{Results from \texttt{GEANT4} Monte Carlo calculations of the
relative size of the combined $\Delta_2$ backscattering and $\Delta_3$
angle effect fractional systematic correction (thick solid lines) and
the total fractional systematic uncertainty assigned to these
corrections (gray bands).}
\label{fig:delta2_delta3_error_bands}
\end{figure*}

\subsection{Geometry-Independent: Rate-Dependent Gain Shifts}
\label{sec:uncertainties_rate_dependent_gain}

A potential systematic effect would arise from any
spin-state-correlated systematic gain shifts, such as from
rate-dependent gain shifts.  However, any such gain
shifts shared by both detectors cancel to first order in the super
ratio asymmetry and are thus expected to be small.  As noted earlier
in Section \ref{sec:experiment_source_guides}, during the operation of
the experiment, the total DAQ recorded data rates (i.e., the
``online'' $\beta$-decay rates integrated over all energies with few
cuts) during $\beta$-decay runs for the non-flipped spin state were
typically $\sim 10$ s$^{-1}$ greater than those recorded during
measurements of the flipped spin state.  By taking data with
calibration sources with different activities, we were able to bound
any such rate-dependent gain shifts to then be $< 0.02$\%/($\sim 10$
s$^{-1}$), which corresponds to a systematic uncertainty in the
asymmetry of $\alt 0.08$\%.

\subsection{Geometry-Dependent: Backscattering $\bm{\Delta_2}$ and
Angle Effect $\bm{\Delta_3}$ Corrections}
\label{sec:uncertainties_angle_effects_backscattering}

Our working assumption was that application of the $f_{\text{bulk}}$
and $f_{\text{thin}}$ scale factors to the \texttt{GEANT4}
backscattering distributions calibrated our Monte Carlo calculations
of our $\Delta_2$ backscattering corrections to the asymmetry.  To
estimate the uncertainty in these now-calibrated corrections, we
compared the \texttt{GEANT4} results for the $\Delta_2$ backscattering
correction with the \texttt{PENELOPE} results.  
[Note that the
\texttt{PENELOPE} calculations required $f_{\text{bulk}}$ and
$f_{\text{thin}}$ scale factors of 0.9--1.1 and 1.3, respectively,
somewhat smaller than those required by \texttt{GEANT4}.]
These agreed to better than 22\% for Geometry A, and better than 6\%
for Geometries B, C and D.  We also note that the RMS of the
$\Delta_2$ corrected asymmetries for Analysis Choices 1--5 was 0.10\%,
0.13\%, 0.30\%, and 0.27\% for Geometries A, B, C, and D,
respectively, providing a powerful check of the robustness of the
calculation of the $\Delta_2$ correction (indeed, consistent with the
robustness of the agreement between the measured and simulated
asymmetries for the various Analysis Choices demonstrated previously
in Fig.\ \ref{fig:asymmetries_analysis_choice_geometries}).  We have
taken a conservative approach to our estimate of the systematic
uncertainty in our backscattering corrections, and quote a 30\%
relative uncertainty in the $\Delta_2$ backscattering correction (and,
thus, in the asymmetry) for all of the Geometries.

To estimate the systematic uncertainty in the $\Delta_3$ angle effect
correction, we varied the thickness of the decay trap end-cap foil
thicknesses in the Monte Carlo.  Assuming 0.5 $\mu$m to be a
reasonable uncertainty in the foil thickness, the relative uncertainty
in $\Delta_3$ was found to be no larger than 25\%.  Further, we note
that the \texttt{GEANT4} and \texttt{PENELOPE} results for $\Delta_3$
agreed to better than $\sim 25$\% for all of the Geometries.
Therefore, we again quote a conservative 25\% relative uncertainty in
the $\Delta_3$ angle
effects correction for all of the Geometries.

Figure \ref{fig:delta2_delta3_error_bands} shows the resulting
energy-dependent error bands for the combined $\Delta_2$ and
$\Delta_3$ corrections for each of the Geometries.  The analysis
energy window of 275--625 keV referenced earlier in Section
\ref{sec:corrections_analysis_energy_window} was chosen such that the
total error in $A_0$ resulting from integration of these error bands
over some energy window combined with the statistical error within
that energy window was a global minimum.

\subsection{Geometry-Dependent: MWPC Efficiency}
\label{sec:uncertainties_mwpc_efficiency}

As discussed earlier in Section \ref{sec:analysis_mwpc_efficiency},
the ``standard'' MWPC cut for the separation of gamma rays and charged
particles was a cut on a fixed PADC channel number.  However, as shown
there, the MWPC exhibited a strong position-dependent response.
Therefore, for some particular energy deposition in the MWPC,
employing such a standard cut resulted in a position-dependent
efficiency for the identification of an event as either a gamma ray or
charged particle event.

We investigated the impact of this position-dependent efficiency on
the asymmetry by comparing results for the asymmetry extracted from
analyses employing the standard PADC channel number cut with those
obtained with fixed MWPC energy cuts ranging from 0.1--0.7 keV.  The
error bounds from this analysis appear as the MWPC efficiency
systematic uncertainties in Table
\ref{tab:systematic_corrections_uncertainties}.  Note that this effect
is largest for Geometry D, for which the differences between the two
MWPCs' efficiency curves was greatest, as shown previously in
Fig.\ \ref{fig:mwpc_efficiency_energy_deposition}.

\section{Summary of Final Results}
\label{sec:summary_final_results}

\subsection{Energy Spectra and Binned $\bm{A_0}$ Results}
\label{sec:summary_final_results_spectra_A0}

\begin{figure}
\includegraphics[scale=0.91,clip=]{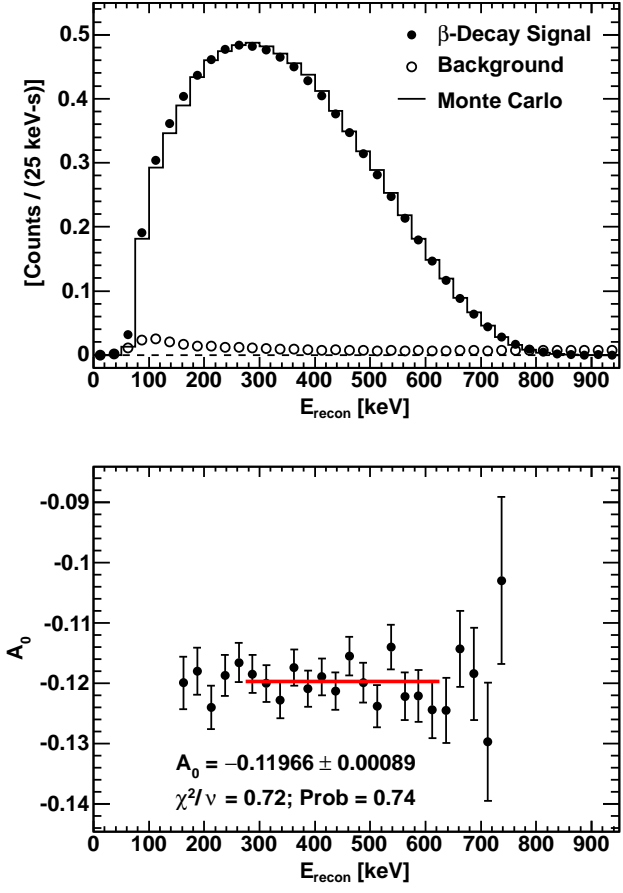}
\caption{(Color online) Upper plot: Final results for the measured
background $E_{\text{recon}}$ energy spectrum (open circles) and the
background-subtracted neutron $\beta$-decay $E_{\text{recon}}$ energy
spectrum (filled circles) summed over both detectors, averaged over
the two neutron spin states, and then averaged over all four
Geometries.  The Monte Carlo prediction for the $E_{\text{recon}}$
spectrum is shown as the solid line.  Lower plot: Final results for
$A_0$ extracted bin-by-bin and then averaged over all four Geometries.
The solid red line indicates the analysis energy window of 275--625
keV.  The quoted error is statistical.}
\label{fig:final_spectra_asymmetries}
\end{figure}

Our final results are shown in Fig.\
\ref{fig:final_spectra_asymmetries}, where we compare the
background-subtracted
$\beta$-decay $E_{\text{recon}}$ spectrum,
summed over both detectors, averaged over the two neutron spin states,
and averaged over all four of the Geometries, with the Monte Carlo
predicted $E_{\text{recon}}$ spectrum.  There, we also show our
Geometry-averaged energy-binned values for the $\beta$-asymmetry
$A_0$, and the final statistical result of $A_0 = -0.11966 \pm
0.00089$.  Note that the central value for $A_0$ was insensitive to
the choice of analysis energy window, with the variation less than
15\% of the statistical uncertainty for other windows between 150
and 750 keV.

\subsection{Final Combined Result}
\label{sec:summary_final_results_combined}

Each of the four Geometries yielded a data set with a statistical
error and a systematic error.  The statistical errors for each of
these Geometries were, of course, independent.  However, three of our
dominant systematic uncertainties (energy reconstruction,
backscattering, and angle effects) were all correlated.  For example,
a mistake in the decay trap end-cap window thickness would have biased
the angle effects correction for all of the data sets.  Therefore, we
assume all three of these systematic uncertainties are 100\%
correlated across the four different Geometries.

Under this assumption, we then combined the results from the four
Geometries according to the following procedure which incorporates
correlations properly in the construction of a global $\chi^2$
\cite{stump01}.  Each individual measurement $i$ gives a constraint of
\begin{equation}
A_0^i = A_0 \pm \sigma_i + {\sum_{k=1}^{3}} \pm \sigma(\Delta_k^i),
\end{equation}
where $A_0^i$ and $\sigma_i$ denote, respectively, the central value
and statistical uncertainty for the $i^{\text{th}}$ measurement (i.e.,
Geometry), and the $\sigma(\Delta_k^i)$ denotes the correlated
uncertainty due to the three systematic effects just discussed.  We
then constructed a $\chi^2$ as
\begin{equation}
\chi^2 = {\sum_{i}}{\sum_{j}} (A_0^i - A_0) (V^{-1})_{ij} (A_0^j - A_0),
\end{equation}
where $i$ and $j$ are the indices of the measurements, and $V$ is the
covariance matrix with $V_{ij} = \sigma_i^2 \delta_{ij} + \sum_{k=1}^{3}
\sigma(\Delta_k^i)\sigma(\Delta_k^j)$.  As was shown in
\cite{stump01}, the $\chi^2$ constructed this way satisfies a standard
$\chi^2$ distribution, in which the value of $A_0$ follows from
minimization of this $\chi^2$, and the one-sigma uncertainty is
obtained from the usual condition $\chi^2 - \chi^2_{\text{min}} = 1$.

Using standard minimization techniques, the final combined result for
$A_0$ we obtained is
\begin{eqnarray}
A_0 &=& -0.11966 \pm 0.00089~_{-0.00140} ^{+0.00123}, \\
&& [\chi^2/\nu = 2.4/3~(\text{Prob = 0.49})] \nonumber
\end{eqnarray}
where the first (second) error represents the statistical (systematic)
error.  From this, we extract the following value for $\lambda =
g_A/g_V$ under the Standard Model,
\begin{equation}
\lambda = \frac{g_A}{g_V} = -1.27590 \pm 0.00239~_{-0.00377}^{+0.00331}.
\end{equation}

\section{Summary and Conclusions}
\label{sec:conclusions}

In this article we have presented a comprehensive and detailed
description of the first precision result \cite{liu10} from the UCNA
Experiment, an experiment designed to perform the first-ever
measurement of the neutron $\beta$-asymmetry parameter $A_0$ with
polarized ultracold neutrons.  As demonstrated here,
the use of UCN in a neutron $\beta$-asymmetry experiment controls key
neutron-related systematic corrections and uncertainties, including
the neutron polarization and neutron-generated backgrounds.  Our
result for the neutron polarization was shown to only be statistics
limited, and our neutron-generated backgrounds were negligible to the
level of $<0.1$\% precision.
All of our results reported here are consistent
with our previously published proof-of-principle results obtained
during data-taking runs in 2007 \cite{pattie09}.

\begin{figure}
\includegraphics[scale=0.45,clip=]{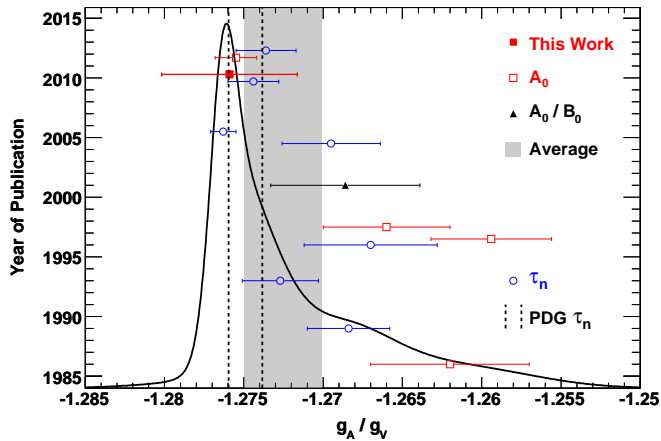}
\caption{(Color online) Ideogram of
values for $g_A/g_V$ extracted from measurements of the neutron
$\beta$-asymmetry parameter $A_0$ (this work \cite{liu10}, filled red
square; all other measurements of $A_0$ \cite{bopp86, yerozolimsky97,
liaud97, mund12}, open red squares), a simultaneous measurement of
$A_0$ and $B_0$ (black triangle, \cite{mostovoi01}), the current
average value of $g_A/g_V = -1.2726 \pm 0.0024$ from these
measurements (gray band; note that the current 2012 Particle Data
Group average value for $g_A/g_V$ has not yet incorporated the result
of \cite{mund12}), measurements of the neutron lifetime $\tau_n$ (blue
circles, \cite{serebrov05, pichlmaier10, mampe89, mampe93, byrne96,
nico05b, arzumanov12}) assuming the superallowed $0^+ \rightarrow 0^+$
value for $V_{ud}$ \cite{towner10}, and the current 2012 Particle Data
Group average value for the neutron lifetime of $880.1 \pm 1.1$ s
\cite{pdg} (between dashed lines).}
\label{fig:ideogram_ga}
\end{figure}

To evaluate the immediate impact of this work, we first note, as shown
earlier in Fig.\ \ref{fig:status_beta_asymmetry}, that our measurement
agrees well with the most recent (and most precise) published result
for the $\beta$-asymmetry $A_0$ from the PERKEO II experiment
\cite{abele97,abele02,mund12}, but is in poorer agreement with the
three other results
\cite{bopp86,erozolimskii91,yerozolimsky97,schreckenbach95,liaud97}
employed by the Particle Data Group in their averaging procedure.
Because of the differences between the PERKEO II and UCNA experimental
techniques, we believe this is a significant result.  Second, the
value for $g_A/g_V$ that we extract from our measurement is
$1.27590~_{-0.00445}^{+0.00409}$.  We compare our value for $g_A/g_V$
with results from a global fit under the Standard Model for $g_A/g_V$
values extracted from results for the $\beta$-asymmetry $A_0$
\cite{bopp86, yerozolimsky97, liaud97, liu10, mund12}, a simultaneous
measurement of $A_0$ and the neutrino asymmetry $B_0$
\cite{mostovoi01}, and from individual measurements of
the neutron lifetime \cite{serebrov05, pichlmaier10, mampe89, mampe93,
byrne96, nico05b, arzumanov12} and the current Particle Data Group
average value for the lifetime ($880.1 \pm 1.1$ s \cite{pdg}; recently
updated for the corrected result of \cite{arzumanov12} which
supersedes the original result of \cite{arzumanov00}), where the
lifetime results assume the superallowed $0^+\rightarrow 0^+$ value
for $V_{ud}$ of $0.97425\pm 0.00022$ \cite{towner10} in Eq.\
(\ref{eq:lifetime-Vud}).  The results of our global fit are displayed
as an ideogram in Fig.\ \ref{fig:ideogram_ga}, where it can be seen
that under the Standard Model the PERKEO II \cite{mund12} and UCNA
\cite{liu10} $\beta$-asymmetry experiments are in agreement with the
three most recent (or updated) results for the
neutron lifetime reported from experiments with
stored UCN \cite{serebrov05,pichlmaier10,arzumanov12}, but in poorer
agreement with the other results for the lifetime employed by the
Particle Data Group in their averaging procedure obtained in
experiments utilizing stored UCN \cite{mampe89,mampe93} and cold
neutron in-beam \cite{byrne96,nico05b} techniques.  Indeed, from our
result for $A_0$ alone of $-0.11966~^{+0.00152}_{-0.00166}$, we
extract, according to Eq.\ (\ref{eq:lifetime-Vud}), a value for the
neutron lifetime of
\begin{equation}
\tau_n = 879.0~^{+4.7}_{-5.1}~\text{s},~~~(\text{UCNA~} A_0~\text{and}~0^+
\rightarrow 0^+ V_{ud})
\end{equation}
in agreement with the measured values from the
three most recent results
reported from experiments using stored UCN
\cite{serebrov05,pichlmaier10,arzumanov12}.

Thus, we conclude that our $\beta$-asymmetry measurement already
provides significant impact to a self-consistent evaluation of the
landscape of neutron $\beta$-decay observables and the superallowed
$0^+ \rightarrow 0^+$ $V_{ud}$ data set.  The fact that the most
recent values for the $\beta$-asymmetry and the neutron lifetime
currently exhibit statistically significant deviations from their
respective world averages prepared by the Particle Data Group, but are
seen to be in agreement with each other (as shown in Fig.\
\ref{fig:ideogram_ga}), motivates further refinement of the UCNA
technique, with its novel approach to key neutron-related systematic
errors, in order to conduct a more precise evaluation of neutron
$\beta$-decay observables.  Indeed, recently demonstrated improvements
to our UCN source and refinements to the energy calibration and gain
monitoring systems will permit the future collection of a data set
with significantly improved statistical and systematic uncertainties.

\begin{acknowledgments}

This work was supported in part by the Department of Energy Office of
Nuclear Physics (Grant Number DE-FG02-08ER41557), the National Science
Foundation (Grant Numbers NSF-0555674, NSF-0855538, NSF-0653222,
NSF-1005233), and the Los Alamos National Laboratory LDRD program.  We
gratefully acknowledge the support of the LANSCE and AOT divisions of
Los Alamos National Laboratory.

\end{acknowledgments}



\begin{thebibliography}{999}

\bibitem{nico05} J.\ S.\ Nico and W.\ M.\ Snow,
         Annu.\ Rev.\ Nucl.\ Part.\ Sci.\ \textbf{55}, 27 (2005).

\bibitem{severijns06} N.\ Severijns, M.\ Beck, and O.\ Naviliat-Cuncic,
         Rev.\ Mod.\ Phys.\ \textbf{78}, 991 (2006).

\bibitem{abele08} H.\ Abele,
         Prog.\ Part.\ Nucl.\ Phys.\ \textbf{60}, 1 (2008).

\bibitem{nico09} J.\ S.\ Nico,
         J.\ Phys.\ G \textbf{36}, 104001 (2009).

\bibitem{dubbers11} D.\ Dubbers and M.\ G.\ Schmidt,
         Rev.\ Mod.\ Phys., \textbf{83}, 1111 (2011).

\bibitem{severijns11} N.\ Severijns and O.\ Naviliat-Cuncic,
         Annu.\ Rev.\ Nucl.\ Part.\ Sci.\ \textbf{61}, 23 (2011).

\bibitem{goldberger58} M.\ L.\ Goldberger and S.\ B.\ Treiman,
         Phys.\ Rev.\ \textbf{111}, 354 (1958).

\bibitem{weinberg58} S.\ Weinberg,
         Phys.\ Rev.\ \textbf{112}, 1375 (1958).

\bibitem{kaiser01} N.\ Kaiser,
         Phys.\ Rev.\ C \textbf{64}, 028201 (2001).

\bibitem{holstein74} B.\ R.\ Holstein,
         Rev.\ Mod.\ Phys.\ \textbf{46}, 789 (1974).

\bibitem{donoghue82} J.\ F.\ Donoghue and B.\ R.\ Holstein,
         Phys.\ Rev.\ D \textbf{25}, 206 (1982).

\bibitem{shiomi96} H.\ Shiomi,
         Nucl.\ Phys.\ A \textbf{603}, 281 (1996).

\bibitem{sasaki09} S.\ Sasaki and T.\ Yamazaki,
         Phys.\ Rev.\ D \textbf{79}, 074508 (2009).

\bibitem{webber11} D.\ M.\ Webber \textit{et al}.,
         Phys.\ Rev.\ Lett.\ \textbf{106}, 041803 (2011).

\bibitem{czarnecki04} A.\ Czarnecki, W.\ J.\ Marciano, and A.\ Sirlin,
         Phys.\ Rev.\ D \textbf{70}, 093006 (2004).

\bibitem{marciano06} W.\ J.\ Marciano and A.\ Sirlin,
         Phys.\ Rev.\ Lett.\ \textbf{96}, 032002 (2006).

\bibitem{jackson57} J.\ D.\ Jackson, S.\ B.\ Treiman, and H.\ W.\ Wyld, Jr.,
         Phys.\ Rev.\ \textbf{106}, 517 (1957).

\bibitem{bhattacharya12} T.\ Bhattacharya, V.\ Cirigliano, S.\ D.\ Cohen,
         A.\ Filipuzzi, M.\ Gonz\'{a}lez-Alonso, M.\ L.\ Graesser,
         R.\ Gupta, and H.-W.\ Lin,
         Phys.\ Rev.\ D \textbf{85}, 054512 (2012).

\bibitem{callan67} C.\ G.\ Callan and S.\ B.\ Treiman,
         Phys.\ Rev.\ \textbf{162}, 1494 (1967).

\bibitem{ando09} S.\ Ando, J.\ A.\ McGovern, and T.\ Sato,
         Phys.\ Lett.\ B \textbf{677}, 109 (2009).

\bibitem{wilkinson82} D.\ H.\ Wilkinson,
         Nucl.\ Phys.\ \textbf{A377}, 474 (1982).

\bibitem{gardner01} S.\ Gardner and C.\ Zhang,
         Phys.\ Rev.\ Lett.\ \textbf{86}, 5666 (2001).

\bibitem{gluck98} F.\ Gl\"{u}ck,
         Phys.\ Lett.\ B \textbf{436}, 25 (1998).

\bibitem{shann71} R.\ T.\ Shann,
         Nuovo Cimento A \textbf{5}, 591 (1971).

\bibitem{gluck92} F.\ Gl\"{u}ck and K.\ T\'{o}th,
         Phys.\ Rev.\ D \textbf{46}, 2090 (1992).

\bibitem{filippone02} B.\ W.\ Filippone and X.\ Ji,
         Adv.\ Nucl.\ Phys.\ \textbf{26}, 1 (2002).

\bibitem{bass05} S.\ D.\ Bass,
         Rev.\ Mod.\ Phys.\ \textbf{77}, 1257 (2005).

\bibitem{adelberger11} E.\ G.\ Adelberger \textit{et al}.,
         Rev.\ Mod.\ Phys.\ \textbf{83}, 195 (2011).

\bibitem{yamazaki08} T.\ Yamazaki \textit{et al}.,
         Phys.\ Rev.\ Lett.\ \textbf{100}, 171602 (2008).

\bibitem{choi10} K.-S.\ Choi, W.\ Plessas, and R.\ F.\ Wagenbrunn,
         Phys.\ Rev.\ C \textbf{81}, 028201 (2010).

\bibitem{gockeler05} M.\ G\"{o}ckeler \textit{et al}.,
         Phys.\ Rev.\ D \textbf{71}, 034508 (2005).

\bibitem{mathews05} G.\ J.\ Mathews, T.\ Kajino, and T.\ Shima,
         Phys.\ Rev.\ D \textbf{71}, 021302(R) (2005).

\bibitem{mention11} G.\ Mention \textit{et al}.,
         Phys.\ Rev.\ D \textbf{83}, 073006 (2011).

\bibitem{pdg} J.\ Beringer \textit{et al}.\ (Particle Data Group),
         Phys.\ Rev.\ D \textbf{86}, 010001 (2012).


\bibitem{towner10} I.\ S.\ Towner and J.\ C.\ Hardy,
         Rep.\ Prog.\ Phys.\ \textbf{73}, 046301 (2010);
         J.\ C.\ Hardy and I.\ S.\ Towner,
         Phys.\ Rev.\ C \textbf{79}, 055502 (2009).

\bibitem{gudkov06} V.\ Gudkov, G.\ L.\ Greene, and J.\ R.\ Calarco,
         Phys.\ Rev.\ C \textbf{73}, 035501 (2006).

\bibitem{konrad10} G.\ Konrad, W.\ Heil, S.\ Bae\ss ler, D.\ Po\v{c}ani\'{c},
         and F.\ Gl\"{u}ck,
         \texttt{arXiv:1007.3027}.

\bibitem{pattie09} R.\ W.\ Pattie, Jr.\ \textit{et al}.,
         Phys.\ Rev.\ Lett.\ \textbf{102}, 012301 (2009).

\bibitem{liu10} J.\ Liu \textit{et al}.,
         Phys.\ Rev.\ Lett.\ \textbf{105}, 181803 (2010).

\bibitem{bopp86} P.\ Bopp \textit{et al}.,
         Phys.\ Rev.\ Lett.\ \textbf{56}, 919 (1986);
         P.\ Bopp \textit{et al}.,
         Nucl.\ Instrum.\ Methods Phys.\ Res.\ A \textbf{267}, 436 (1988).

\bibitem{erozolimskii91} B.\ G.\ Erozolimskii \textit{et al}.,
         Phys.\ Lett.\ B \textbf{263}, 33 (1991).

\bibitem{yerozolimsky97} B.\ Yerozolimsky \textit{et al}.,
         Phys.\ Lett.\ B \textbf{412}, 240 (1997).

\bibitem{schreckenbach95} K.\ Schreckenbach \textit{et al}.,
         Phys.\ Lett.\ B \textbf{349}, 427 (1995).

\bibitem{liaud97} P.\ Liaud \textit{et al}.,
         Nucl.\ Phys.\ A \textbf{612}, 53 (1997).

\bibitem{abele97} H.\ Abele \textit{et al}.,
         Phys.\ Lett.\ B \textbf{407}, 212 (1997).

\bibitem{abele02} H.\ Abele \textit{et al}.,
         Phys.\ Rev.\ Lett.\ \textbf{88}, 211801 (2002).

\bibitem{mund12} D.\ Mund \textit{et al}.,
         \texttt{arXiv:1204.0013}.

\bibitem{ucn_book} R.\ Golub, D.\ Richardson, and S.\ K.\ Lamoreaux,
         \textit{Ultra-Cold Neutrons} (Adam Hilger, Bristol, 1991).

\bibitem{ito07} T.\ M.\ Ito \textit{et al}.,
         Nucl.\ Instrum.\ Methods Phys.\ Res.\ A \textbf{571}, 676 (2007).

\bibitem{plaster08} B.\ Plaster \textit{et al}.,
         Nucl.\ Instrum.\ Methods Phys.\ Res.\ A \textbf{595}, 587 (2008).

\bibitem{morris02} C.\ L.\ Morris \textit{et al}.,
         Phys.\ Rev.\ Lett.\ \textbf{89}, 272501 (2002).

\bibitem{saunders04} A.\ Saunders \textit{et al}.,
         Phys.\ Lett.\ B \textbf{593}, 55 (2004).

\bibitem{saunders11} A.\ Saunders \textit{et al}.,
         submitted to Rev.\ Sci.\ Instrum.

\bibitem{liu00} C.-Y.\ Liu, A.\ R.\ Young, and S.\ K.\ Lamoreaux,
         Phys.\ Rev.\ B \textbf{62}, R3581 (2000).

\bibitem{liu03} C.-Y.\ Liu \textit{et al}.,
         Nucl.\ Instrum.\ Methods Phys.\ Res.\ A \textbf{508}, 257 (2003).

\bibitem{morris09} C.\ L.\ Morris \textit{et al}.,
         Nucl.\ Instrum.\ Methods Phys.\ Res.\ A \textbf{599}, 248 (2009).

\bibitem{mammei-thesis} R.\ R.\ Mammei, Ph.D.\ thesis,
         Virginia Polytechnic Institute and State University (2010).

\bibitem{holley_afp} A.\ T.\ Holley \textit{et al}.,
         Rev.\ Sci.\ Instrum.\ \textbf{83}, 073505 (2012).

\bibitem{vladimirski61} V.\ V.\ Vladimirski,
         Sov.\ Phys.\ JETP \textbf{12}, 740 (1961).

\bibitem{wrede11} C.\ Wrede \textit{et al}.,
         Nucl.\ Instrum.\ Methods Phys.\ Res.\ B \textbf{269}, 1113 (2011).

\bibitem{rios11} R.\ Rios \textit{et al}.,
         Nucl.\ Instrum.\ Methods Phys.\ Res.\ A \textbf{637}, 105 (2011).

\bibitem{junhua_thesis} J.\ Yuan,
         Ph.D.\ thesis, California Institute of Technology (2006).

\bibitem{midas} \texttt{MIDAS} Data Acquisition System,
         \url{http://midas.psi.ch}~.

\bibitem{cernlib} CERN Program Library, \\
         \url{http://cernlib.web.cern.ch}~.

\bibitem{root} \texttt{ROOT} Data Analysis Framework,
         \url{http://root.cern.ch}~.

\bibitem{hogan99} G.\ E.\ Hogan, Proceedings of the 11th IEEE NPSS Real Time
         Conference, Santa Fe, NM (1999).

\bibitem{geant4} S.\ Agostinelli \textit{et al}.,
         Nucl.\ Instrum.\ Methods Phys.\ Res.\ A \textbf{506}, 250 (2003);
         \url{http://www.geant4.org}~.

\bibitem{penelope} J.\ Sempau \textit{et al}.,
         Nucl.\ Instrum.\ Methods Phys.\ Res.\ B \textbf{132}, 377 (1997).

\bibitem{martin03} J.\ W.\ Martin \textit{et al}.,
         Phys.\ Rev.\ C \textbf{68}, 055503 (2003).

\bibitem{martin06} J.\ W.\ Martin \textit{et al}.,
         Phys.\ Rev.\ C \textbf{73}, 015501 (2006).

\bibitem{hoedl_thesis} S.\ A.\ Hoedl,
         Ph.D.\ thesis, Princeton University (2003).

\bibitem{birks51} J.\ B.\ Birks,
         Proc.\ Phys.\ Soc.\ A \textbf{64}, 874 (1951).

\bibitem{yuan01} J.\ Yuan, B.\ W.\ Filippone, D.\ Fong, T.\ M.\ Ito,
         J.\ W.\ Martin, J.\ Penoyar, and B.\ Tipton,
         Nucl.\ Instrum.\ Methods Phys.\ Res.\ A \textbf{465}, 404 (2001).

\bibitem{stump01} D.\ Stump \textit{et al}.,
         Phys.\ Rev.\ D \textbf{65}, 014012 (2001).

\bibitem{mostovoi01} Yu.\ A.\ Mostovoi \textit{et al}.,
         Phys.\ Atm.\ Nucl.\ \textbf{64}, 1955 (2001).

\bibitem{serebrov05} A.\ Serebrov \textit{et al}.,
         Phys.\ Lett.\ B \textbf{605}, 72 (2005).

\bibitem{pichlmaier10} A.\ Pichlmaier, V.\ Varlamov, K.\ Schreckenbach,
         and P.\ Geltenbort,
         Phys.\ Lett.\ B \textbf{693}, 221 (2010).

\bibitem{mampe89} W.\ Mampe, P.\ Ageron, C.\ Bates, J.\ M.\ Pendlebury,
         and A.\ Steyerl,
         Phys.\ Rev.\ Lett.\ \textbf{63}, 593 (1989).

\bibitem{mampe93} W.\ Mampe, L.\ N.\ Bondarenko, V.\ I.\ Morozov,
         Yu.\ N.\ Panin, and A.\ I.\ Fomin,
         JETP Lett.\ \textbf{57}, 82 (1993).

\bibitem{byrne96} J.\ Byrne \textit{et al}.,
         Europhys.\ Lett.\ \textbf{33}, 187 (1996).

\bibitem{nico05b} J.\ S.\ Nico \textit{et al}.,
         Phys.\ Rev.\ C \textbf{71}, 055502 (2005).

\bibitem{arzumanov12} S.\ S.\ Arzumanov, L.\ N.\ Bondarenko, V.\ I.\ Morozov,
         Yu.\ N.\ Panin, and S.\ M.\ Chernyavsky,
         JETP Lett.\ \textbf{95}, 224 (2012).

\bibitem{arzumanov00} S.\ Arzumanov \textit{et al}.,
         Phys.\ Lett.\ B \textbf{483}, 15 (2000).

\end{thebibliography}

\end{document}